\documentclass[%
 reprint,
superscriptaddress,
showpacs,preprintnumbers,
nofootinbib,
 amsmath,amssymb,onecolumn,notitlepage,
 bm,
 aps,
 longbibliography,
]{revtex4-1}

\usepackage[T1]{fontenc}
\usepackage[dvipdfmx]{graphicx}
\usepackage{here}
\usepackage{amssymb}
\usepackage{comment}
\usepackage{physics}
\usepackage{epstopdf}
\usepackage{dcolumn}
\usepackage{booktabs}
\usepackage{enumitem}
\usepackage{bm}
\usepackage[dvipsnames]{xcolor}
\usepackage{subcaption}
\usepackage{mhchem}
\usepackage{tensor}
\usepackage[dvipsnames]{xcolor}
\usepackage{hyperref}
\hypersetup{
colorlinks=true,
linkcolor=blue,
citecolor=BlueViolet,
urlcolor=Aquamarine
}

\newcommand{\cD}{\mathcal{D}}

\newcommand{\cL}{\mathcal{L}}
\newcommand{\cO}{\mathcal{O}}
\newcommand{\cP}{\mathcal{P}}
\newcommand{\cG}{\mathcal{G}_2}
\newcommand{\bk}{\mathbf{k}}
\newcommand{\bq}{\mathbf{q}}

\usepackage[normalem]{ulem}

\usepackage[export]{adjustbox}

\newcommand{\mnras}{Mon. Not. R. Astron. Soc.}
\newcommand{\jcap}{J. Cosm. Astropart. Phys.}

\newcommand{\bPsi}{\boldsymbol{\Psi}}

\newcommand{\btheta}{\boldsymbol{\theta}}

\providecommand{\sorthelp}[1]{}
    
\begin{document}

\preprint{KEK-Cosmo-0386}
\preprint{RBI-ThPhys-2025-34}

\title{
Cosmology inference with perturbative forward modeling at the field level:
\\
a comparison with joint power spectrum and bispectrum analyses
}

\author{Kazuyuki Akitsu}
\email{kakitsu@post.kek.jp}
\affiliation{Theory Center, Institute of Particle and Nuclear Studies, High Energy Accelerator Research Organization (KEK), Tsukuba, Ibaraki 305-0801, Japan}

\author{Marko Simonovi\'c}
\affiliation{Dipartimento di Fisica e Astronomia, Universit{\`a} di Firenze; Via G. Sansone 1; I-50019 Sesto Fiorentino, Italy}
\affiliation{INFN, Sezione di Firenze; Via G. Sansone 1; I-50019 Sesto Fiorentino, Italy}

\author{Shi-Fan Chen}
\affiliation{School of Natural Sciences, Institute for Advanced Study, 1 Einstein Drive, Princeton, NJ 08540, USA}
\affiliation{Department of Physics, Columbia University, New York, NY, USA 10027}
\affiliation{NASA Hubble Fellowship Program, Einstein Fellow}

\author{Giovanni Cabass}
\affiliation{Division of Theoretical Physics, Ru{\dj}er Bo{\v s}kovi{\'c} Institute, Zagreb HR-10000, Croatia}

\author{Matias Zaldarriaga}
\affiliation{School of Natural Sciences, Institute for Advanced Study, 1 Einstein Drive, Princeton, NJ 08540, USA}

\date{\today}
    
\begin{abstract}
We extend field-level inference to jointly constrain the cosmological parameters~$\{A,\omega_{\rm cdm},H_0\}$, in both real and redshift space. 
Our analyses are based on mock data generated using a perturbative forward model, with noise drawn from a Gaussian distribution with a constant power spectrum. 
This idealized setting, where the field-level likelihood is exactly Gaussian, allows us to precisely quantify the information content in the nonlinear field on large scales. 
We find that field-level inference accurately recovers all cosmological parameters in both real and redshift space, with uncertainties consistent with perturbation theory expectations. 
We show that these error bars are comparable to those obtained from a joint power spectrum and bispectrum analysis using the same perturbative model. 
Finally, we perform several tests using the Gaussian field-level likelihood to fit the mock data where the true noise model is non-Gaussian, 
and find significant biases in the inferred cosmological parameters. 
These results highlight that the success of field-level inference critically depends on using the correct likelihood, 
which may be the primary challenge for applying this method to smaller scales even in the perturbative regime.
\end{abstract}

\maketitle

\tableofcontents

\section{Introduction}

Measurements of cosmological parameters in the Cosmic Microwave Background (CMB) and large-scale structure (LSS) surveys are traditionally done using~$n$-point correlation functions. 
An interesting alternative in which one uses the observed realization of fluctuations directly, without employing particular summary statistics, has recently become numerically feasible~\cite{Kitaura:2007pe,Jasche:2009hz,Jasche:2012kq,Wang:2013ep,Jasche:2014vpa,Ata:2014ssa,Lavaux:2015tsa,Alsing:2016hkh,Seljak:2017rmr,Ramanah:2018eed,Horowitz:2018tbe,Jasche:2018oym,Feng:2018for,Millea:2021had,Kostic:2022vok,Modi:2022pzm,Robnik:2022bzs,Bayer:2023rmj,Horowitz:2025kop,Simon-Onfroy:2025ziw}. 
In this so-called field-level inference (FLI), one assumes a likelihood for Fourier modes of the nonlinear field and then varies all initial conditions (with some prior given by the linear theory), jointly with cosmological and nuisance parameters. 
Marginalizing over all possible initial conditions and galaxy formation physics, it is possible to numerically find the posterior for cosmological parameters given the observations. 
One of the major advantages of this approach is that it is optimal since it uses the full available information in the observed field without any compression.

Besides the choice of the likelihood, in order to do the field-level inference one also has to specify the forward model. 
Both of these ingredients are highly nontrivial in the nonlinear regime. 
For this reason, most of the progress and practical applications in the context of galaxy clustering have been done on large scales, where nonlinearities are under perturbative control and both the likelihood and the forward model can be rigorously derived~\cite{Schmidt:2018bkr,Elsner:2019rql,Cabass:2019lqx,Schmidt:2020tao,Schmidt:2020ovm,Schmidt:2020viy,Cabass:2020jqo}. 
A lot of progress has been made in using the perturbative forward modeling for the field-level inference, with many interesting results and applications~\cite{Schmittfull:2018yuk,Schmittfull:2020trd,Millea:2020cpw,Millea:2020iuw,Porqueres:2020qgy,Tsaprazi:2021mft,Andrews:2022nvv,Babic:2022dws,Bayer:2022vid,Boruah:2022lsu,Zhou:2023ezg,Stadler:2023hea,Porqueres:2023drp,Beyond-2pt:2024mqz,Nguyen:2024yth,Stadler:2024fui,Stadler:2024aff,SPT-3G:2024atg,SPT-3G:2025bzu,Babic:2025fgv,Peron:2025lgh}.

Most of the field-level cosmological analyses done so far have focused on very simple examples, with biased tracers in real space and a single cosmological parameter---the amplitude of the linear density field~$A$ (equivalent to~$\sigma_8$ when all other cosmological parameters are fixed). 
While this setup is useful as a first test of the method and helps build intuition, it is important to explore generalizations that are closer to the real data. 
One of the main goals of this paper is to make extensions in two different directions. 
First, we extend the FLI to other cosmological parameters, besides the amplitude~$A$. 
In particular, we include the Hubble constant~$H_0$ and dark matter density~$\omega_{\rm cdm}$ in our analyses. 
Second, we apply the FLI in redshift space~\cite{Stadler:2023hea,Stadler:2024aff}.
We carefully examine several different setups in order to test the performance of the method and find that it robustly recovers the true cosmology with precision comparable to what is found in the previous studies. 
This is an important step forward towards applications to the real data. 

It is important to emphasize that in our analyses we do not use the data from simulations, but rather generate the mock data based on the same perturbative model used in the analysis. 
We make this choice for two reasons. 
First, the perturbative forward models on large scales (we focus on~$k_{\rm max}\leq 0.12\;h/{\rm Mpc}$) reproduce all features of the nonlinear evolution of biased tracers in real and redshift space~\cite{Schmittfull:2016jsw,Schmittfull:2020trd}. 
Therefore, any lesson about the FLI (related to the nonlinear dynamics) that we learn from the mock data will very likely apply equally well in reality. 
Second, one of the main issues in the FLI is the choice of the likelihood for the Fourier modes. 
While it has been shown that it approaches a simple Gaussian in the~$k\to 0$ limit~\cite{Schmidt:2018bkr,Cabass:2019lqx,Cabass:2020jqo,Cabass:2020nwf}, there are potentially significant deviations from Gaussianity for~$k_{\rm max}\sim 0.1\;h/{\rm Mpc}$, observed at the map level in comparisons of perturbation theory and simulations~\cite{Schmittfull:2016jsw,Schmittfull:2020trd}.
Since in the synthetic data we can freely choose the distribution of the noise in our forward model, this allows us to clearly separate this issue from the nonlinear evolution and information content in the nonlinear galaxy density field, which is the main focus of this paper. 
In most of our analyses we will use a Gaussian model for the noise field. 
However, we will also show that the results are very sensitive to this choice and that the mismatch in the noise model in the mock galaxy field and the field-level likelihood can lead to significant biases in cosmological parameter inference.
This remains an open problem and one of the main obstacles in applying the FLI to the real data, particularly going to smaller scales where the assumption of the Gaussian likelihood is less accurate. 

Our second major goal is to compare field-level inference with conventional joint power spectrum and bispectrum analyses, and we do it in all examples that we study in this paper. 
This comparison is crucial in order to establish the true potential of FLI and it is particularly relevant given recent contradictory claims about the performance of FLI compared to the conventional analyses. 
For example, results from Refs.~\cite{Beyond-2pt:2024mqz,Nguyen:2024yth} suggested that FLI has potential to constrain the amplitude~$A$ significantly better than the joint power spectrum and bispectrum analysis (see Fig.~22 and the discussion in Sec.~6 of Ref.~\cite{Beyond-2pt:2024mqz}). 
This was further explored in Ref.~\cite{Nguyen:2024yth} in a particularly simple setup, finding an even larger (factor of 3-5) difference in the error bars on~$A$ between the two approaches.\footnote{Note that the joint power spectrum and bispectrum analysis in Ref.~\cite{Nguyen:2024yth} was done using simulation-based inference (SBI) (relying on a cubic perturbative forward model), which in detail differs form the standard correlation function analysis in perturbation theory. 
For an argument of how these differences can potentially lead to significantly different error bars on cosmological parameters, see the discussion in Ref.~\cite{Spezzati:2025zsb}.} 
On the other hand, the expectation from perturbation theory is that FLI should be equivalent to the leading~$n$-point functions that can be reliably computed using a given forward model. 
This has been explicitly demonstrated in Refs.~\cite{Cabass:2023nyo,Schmidt:2025iwa}.
Given that in the perturbative regime each higher order~$n$-point function has a successively smaller signal (controlled by the same parameter as in the perturbative expansion) and that the strongest degeneracy is between the amplitude~$A$ and the linear bias~$b_1$, it is expected in practice that the constraints are dominated by the power spectrum and bispectrum. 
It has been recently shown in Ref.~\cite{Spezzati:2025zsb} that in a typical situations of interest, the higher-order correlation functions such as the trispectrum do not contribute significantly enough to the constraints on~$A$ in order to resolve the large discrepancy between the two inference methods found in Ref.~\cite{Nguyen:2024yth}. 

Our results provide a valuable input in this discussion. 
The fact that we perform all our analyses on mock data generated by a known forward model and with known noise is particularly well-suited to compare FLI and conventional~$n$-point function analyses. 
This setup allows us to use the exactly same model in both methods, with the exactly same number of cosmological and nuisance parameters. 
This is very similar to the approach of Ref.~\cite{Leclercq:2021ctr,Nguyen:2024yth}, except that we do not rely on the simulation based inference for the~$n$-point functions analysis. 
Our results in a wide range of examples confirm the theoretical expectation of Ref.~\cite{Cabass:2023nyo,Schmidt:2025iwa} and suggest that in a typical situation FLI and the joint power spectrum and bispectrum analysis lead to similar constraints on cosmological parameters. 
Of course, as expected from perturbation theory arguments, FLI is always a bit more optimal (by~$\lesssim 20\%$), in agreement with findings of Ref.~\cite{Spezzati:2025zsb} for the typical impact of the trispectrum. 
It is important to point out that we find similar results in a large range of different settings, for real and redshift space, different sets of nuisance and cosmological parameters and for Eulerian and Lagrangian perturbative forward models alike. 
This reaffirms our findings and sets the basic expectation for more general analyses, such as those at higher order in perturbation theory and with larger sets of nuisance and cosmological parameters.

This paper is organized as follows.
In Sec.~\ref{sec:model} we introduce the perturbative forward models that we will use in this paper and discuss some theoretical expectations.
Sec.~\ref{sec:numerical} summarizes the numerical implementation of the FLI pipeline.
We then compare the FLI and the joint power spectrum and bispectrum analysis in Sec.~\ref{sec:EPT} for a simple Eulerian model, 
and in Sec.~\ref{sec:LPT} for a LPT-based forward model in real and redshift space 
to discuss the similarity of these two approaches in terms of the constraining cosmological parameters.
Sec.~\ref{sec:resolution_mismatch} and Sec.~\ref{sec:non-Gaussian_noise} are devoted to the discussion of the differences between the FLI and the conventional analysis.
We conclude in Sec.~\ref{sec:conclusion} by summarizing the main results of this paper.

\section{Perturbative forward model}
\label{sec:model}

In this section we describe the perturbative forward models that we will use in this paper. 
We use results and methods from a large body of work where perturbation theory at the field level was developed to describe dark matter and biased tracers, in real and redshift space~\cite{Schmittfull:2018yuk,Schmittfull:2020trd}. 
We also comment on the expectations for the constraining power of joint analyses using~$n$-point functions and the comparison to optimal field-level inference. 

It is important to point out that in our forward models we do not use explicit cutoff for the initial conditions (there is always the Nyquist frequency). 
This is somewhat different from the implementation in Refs.~\cite{Kostic:2022vok,Nguyen:2024yth}. 
In other words, we allow for the nonlinear coupling of high-$k$ modes (up to the Nyquist frequency), beyond the maximum wavenumber~$k_{\rm max}$ used in the analysis. 
These modes can produce long-wavelength fields with~$k\leq k_{\rm max}$. Even though the high-$k$ couplings may not be properly computed in perturbation theory, this is not a problem. 
The counterterms and bias parameters serve exactly the purpose of absorbing consistently all small-scale dependence at each order in perturbation theory. 
Formally, this is equivalent to the usual calculation of loop diagrams where the cutoff is formally sent to infinity.

\subsection{Simple Eulerian model}
\label{subsec:simple_Eulerian}

Since our analyses will focus on large scales only, one of the forward models that we will use is based on Eulerian perturbation theory. 
This scheme is particularly intuitive, since the nonlinear fields are simple convolutions of the initial conditions, where one only has to specify perturbation theory kernels which couple in a particular way different Fourier modes. 
This is an ideal setup to test basic features of field-level inference and standard methods based on correlation functions and compare them. 

Besides building some intuition, we want to use this example to check at the field level the analytic derivation of the posterior for cosmological parameters (in the idealized zero noise limit) presented in Ref.~\cite{Cabass:2023nyo}. 
In particular, one of the key steps in computing the posterior is the construction of the inverse model, which gives the values of the initial linear field given some observed galaxy overdensity. 
While in the Eulerian perturbation theory it is always possible to find this inverse map, it is not obvious that the radius of convergence of the inverse model is the same as the forward model (see the discussion in Ref.~\cite{Cabass:2023nyo}). 
We will confirm that the inverse model works well, showing that the FLI posterior matches the joint power spectrum and bispectrum posterior up to the small corrections set by the perturbation theory expansion parameter.

In order to maximally simplify the analysis, we use the following Eulerian model
\begin{align}
    \delta_g (\bk) = b_1\big(\delta_1 (\bk) + {\cal G}_2 (\bk) \big) \;,
\end{align}
where~$\delta_1$ is the linear density field,~$b_1$ is the linear bias and the only nonlinearity is given by 
\begin{align}
    {\cal G}_2 (\bk) = \int_{\bq_1} \int_{\bq_2} \left( \frac{(\bq_1\cdot\bq_2)^2}{q_1^2q_2^2} -1 \right) (2\pi)^3 \delta^{(3)}_{\rm D}(\bk-\bq_1-\bq_2) \delta_1(\bq_1) \delta_1 (\bq_2) \;.
\end{align}
Here and throughout the paper we use the notation where~$\int_{\bq}\equiv \int \frac{{\rm d}^3\bq}{(2\pi)^3}$.
For simplicity, we neglect the time dependence, but it is implicitly assumed that all linear fields are evolved up to the given redshift~$z$. Note that all quadratic biases are fixed in this example. 
In particular, we choose~$b_2=0$ in order to avoid any confusion  with the constant low-$k$ contribution to the power spectrum which comes from the auto-correlation of~$\delta_1^2$ and~$b_{\mathcal G_2}=1$, in agreement with the typical values in the standard Eulerian kernels. 
The full nonlinear galaxy density field also has a stochastic component. 
Therefore, the data that we generate using the forward model always has the following form
\begin{align}
    \hat\delta_g (\bk) = \delta_g (\bk) + \epsilon (\bk) \;,
\end{align}
where~$\epsilon$ is a stochastic field. In most of our analyses, we will use a simple Gaussian realizations of~$\epsilon$ with a constant power spectrum, with the exception of Sec.~\ref{sec:non-Gaussian_noise}.

Even though it is very simple, this model is nontrivial. 
The linear bias and stochastic noise are degenerate with cosmological parameters. 
Most importantly,~$b_1$ is exactly degenerate with the amplitude of the linear field in linear theory. 
This degeneracy is broken only by the nonlinear term~${\cal G}_2$. 
This feature is similar to the more realistic setups where the measurement of the amplitude of the linear field has been used as a test of the field-level inference~\cite{Kostic:2022vok,Beyond-2pt:2024mqz,Nguyen:2024yth}.

One major advantage of this model is that one can easily compute all observables that can be used in the standard analyses. 
All correlation functions are one-loop exact. 
In other words, they do not have any corrections from two loops or higher orders in perturbation theory. 
For instance, assuming that the stochastic field~$\epsilon$ is Gaussian with the constant power spectrum~$P_\epsilon$, the full nonlinear power spectrum and bispectrum are given by
\begin{align}
    P_g(k) &= b_1^2 \big( P_\text{lin}(k) + P_{{\cal G}_2{\cal G}_2}(k) \big) + P_{\epsilon}\;, \\
    B_g(k_1, k_2, k_3) &=  b_1^3 \big( B_{\rm tree}(k_1, k_2, k_3) + B_{{\cal G}_2{\cal G}_2{\cal G}_2}(k_1, k_2, k_3) \big) \;,
    \label{eq:P_B_G2}
\end{align}
where~$P_{\rm lin}(k)$ is the linear matter power spectrum. 
The tree-level bispectrum has the standard form
\begin{align}
    B_{\rm tree}(k_1, k_2, k_3) = 2(\mu_{\bk_1,\bk_2}^2-1) P_\text{lin}(k_1) P_\text{lin}(k_2) +\text{2 perms.} 
    \label{eq:B_G2_tree}
\end{align}
and the one-loop power spectrum and bispectrum have only one contribution each
\begin{align}
    P_{{\cal G}_2{\cal G}_2}(k) &= 2\int_{\bq} (\mu_{\bk-\bq, \bq}^2 -1)^2 P_\text{lin}(|\bk - \bq|) P_\text{lin}(q)\;, \\
    B_{{\cal G}_2{\cal G}_2{\cal G}_2}(k_1, k_2, k_3) &= 8 \int_\bq (\mu_{\bq,\bk_1-\bq}^2-1) (\mu_{\bk_1-\bq,\bk_2+\bq}^2-1) (\mu_{\bk_2+\bq,-\bq}^2-1) P_{\rm lin}(q) P_{\rm lin}(|\bk_1-\bq|) P_{\rm lin}(|\bk_2+\bq|) \;.
    \label{eq:P_B_G2}
\end{align}
We checked that these equations exactly reproduce the measurements of the power spectrum and bispectrum on the grid, with the exact same forward model (see Fig.~\ref{fig:bispec} for the bispectrum comparison). 
Even though the one-loop bispectrum is subdominant on large scales, we keep it in order to estimate the impact of the next-to-leading order corrections in perturbation theory, particularly for setups with a few free parameters where the cosmological constraints can be very tight even in a relatively small volume. 
These corrections are controlled by the variance of the density field
\begin{align}
    \Delta^2 (k) = \frac{1}{2\pi^2} \int_0^{k} {\rm d}q \,  q^2 P_{\rm lin}(q) \;,
\end{align}
which is the only small parameter in this simple model.

Note that even though the model has only a single quadratic term, this is sufficient to generate all higher order~$n$-point functions (tree-level and one-loop). 
In this sense, despite its simplicity, this forward model is similar to the usual perturbation theory.
For this reason, we use it as the simplest possible setup for nontrivial tests of field-level inference and comparison to conventional analyses based on the~$n$-point functions.  
 
\subsection{LPT-based forward model}
\label{subsec:LPT_model}

Moving beyond the simple Eulerian model, a more realistic way to make perturbative maps for biased tracers in real and redshift space is to use some version of Lagrangian perturbation theory, including all relevant bias parameters and counterterms. 
This has been extensively studied and applied in practice, for example in Refs.~\cite{Schmittfull:2018yuk,Schmittfull:2020trd}. 

One practical implementation of LPT at the field level is in terms of the so-called shifted operators~\cite{Schmittfull:2018yuk,Schmittfull:2020trd}. 
Let us first focus on biased tracers in real space. In this formulation, all nonlinear terms in Eulerian Fourier space can be written as follows
\begin{align}
    \tilde{\mathcal O}(\bk) = \int {\rm d}^3\bq \ \mathcal O(\bq)\ e^{-i\bk\cdot \bPsi_1(\bq)}\ e^{-i\bk\cdot\bq} \;,
\end{align}
where~$\bq$ are initial Lagrangian coordinates,~$\mathcal O(\bq)$ are nonlinear terms built out of the linear fields in the initial conditions and~$\bPsi_1$ are the Zel'dovich displacements given by
\begin{align}
    \bPsi_1(\bq) = \int_{\bk} \frac{i\bk}{k^2}\ \delta_1(\bk) e^{i\bk\cdot\bq} \;.
\end{align}
In order to evaluate the nonlinear map at a given redshift~$z$, in all expressions it is implicitly assumed that all linear fields are evolved to the same~$z$. 

In the standard LPT one would first generate the displacement up to cubic order, compute the density field on the grid and then apply the bias expansion. 
Using shifted fields is a bit more convenient in practice. 
First, with shifted operators one only has to do a single linear displacement on the grid. As explicitly demonstrated in Ref.~\cite{Schmittfull:2018yuk}, all higher order displacements can be taken into account including the appropriate nonlinear terms in~$\mathcal O(\bq)$ (see also App.~\ref{app:2LPT}). 
Second, the fact that the nonlinear field is written directly in the final Eulerian coordinates allows to do a simple orthogonalization, where third-order operators can be decomposed into the part which correlates with linear terms and the part which is uncorrelated. Schematically, we can write
\begin{align}
    \tilde{\mathcal O}^{(3)} (\bk) = \frac{\langle \tilde\delta_1(\bk) \tilde{\mathcal O}^{(3)}(-\bk) \rangle'}{\langle \tilde\delta_1 (\bk) \tilde\delta_1(-\bk) \rangle '} \tilde\delta_1(\bk) + \left( \tilde{\mathcal O}^{(3)}(\bk) - \frac{\langle \tilde\delta_1 (\bk) \tilde{\mathcal O}^{(3)} (-\bk) \rangle'}{\langle \tilde\delta_1(\bk) \tilde\delta_1 (-\bk) \rangle '} \tilde\delta_1 (\bk) \right) \equiv \beta_1(k) \tilde\delta_1(\bk) + \tilde{\mathcal O}^{(3)}_\perp (\bk) \;,
\end{align}
where the prime on the correlation function indicates that the~$(2\pi)^3\delta^{(3)}_{\rm D}({\bf 0})$ is removed. 
The first term is only a~$k$-dependent rescaling of the shifted linear field. 
The transfer function~$\beta_1(k)$ has a fixed~$k$-dependence at the given order in perturbation theory and depends only on the bias parameters and counterterms (for details and explicit expressions for leading order corrections in~$\Delta^2$ see Ref.~\cite{Schmittfull:2018yuk}). 
The second term~$\tilde{\mathcal O}^{(3)}_\perp$ does not correlate with~$\tilde\delta_1$ by construction. 
In fact, this term does not appear in the one-loop power spectrum or the tree-level bispectrum and for the purposes of generating a nonlinear realization that is correct at leading nonlinear order for these two observables, one can safely neglect it. 
Therefore, the problem reduces to generating just quadratic operators in the initial conditions and shifting them with the Zel'dovich displacement. 
This is practically easier than doing third order LPT since one can use a larger range of scales for a fixed resolution of the grid. 

Motivated by this discussion, in order to maximally simplify the analysis and interpretation of the results, we use the following forward model to generate the field of biased tracers in real space
\begin{align}
\label{eq:LPT_model_field}
\delta_g(\bk) = \int {\rm d}^3\bq \left( 1 + b_1 \delta_1(\bq) + \frac{b_2}{2} \left(\delta_1^2(\bq) - \sigma^2 \right) + b_{\mathcal G_2} \mathcal G_2(\bq) \right) e^{-i\bk\cdot \bPsi_1(\bq) } e^{-i\bk\cdot\bq} \;.
\end{align}
This model has all complexities of the realistic nonlinear data, but it is easy do generate it on the grid and to compute all relevant correlation functions. 

Let us now turn to the forward model in redshift space. Everything we said for the real space applies in this case as well (for the detailed discussion and comparison to simulations see Ref.~\cite{Schmittfull:2020trd}). 
The only major difference is that the field of biased tracers now depends not only on the wavevector~$\bk$ but also the line of sight~$\hat {\boldsymbol n}$ that breaks isotropy. 
If we define~$\mu\equiv \hat{\bk}\cdot\hat{\boldsymbol{n}}$, our simplified model for the redshift space galaxy density field is given by
\begin{align}
    \delta_g(\bk,\mu) = \int {\rm d}^3\bq \left( 1 + b_1 \delta_1(\bq) + \frac{b_2}{2} \left(\delta_1^2(\bq) - \sigma^2 \right) + b_{\mathcal G_2} \mathcal G_2(\bq)   \right) e^{-ik_a \left( \delta^K_{ab}+f\hat{n}_a \hat{n}_b \right) \Psi_{1b}(\bq) } e^{-i\bk\cdot\bq} \;,
\end{align}
where~$f$ is the growth rate. Compared to the real space, the linear displacement in the exponent has an additional contribution which corresponds to the additional displacement in redshift space along the line of sight due to the peculiar linear velocity. 

In addition to the galaxy density field, we also have to include the counterterms. We have three free parameters~$c_0,\;c_2$ and~$c_4$ which contribute to long-wavelength fluctuations in the following way 
\begin{align}
    \delta_c(\bk,\mu) = (c_0+c_2\mu^2 + c_4\mu^4) \tilde\delta_1(\bk,\mu) \;.
\label{eq:counterterm}
\end{align}
Note that in real space~$c_2=c_4=0$. Besides the usual role in the EFTofLSS to absorb the large momentum contributions in the loop integrals coming from small scales, the counterterms here plays another important role. 
They are there to absorb the dependence on the maximum wavenumber available in the grid, given by the resolution of the simulated data.  
We will come back to this issue in Sec.~\ref{sec:resolution_mismatch} and discuss it in detail.

\subsection{Power spectrum and bispectrum in the LPT-based forward model}
\label{subsec:PBinLPT_model}

Starting with the LPT-based forward model, it is possible to derive expressions for the correlation functions. In particular, in this paper, we are interested in the one-loop power spectrum and the tree-level bispectrum. These correlation functions can be computed directly in LPT~\cite{Carlson:2012bu,Vlah15,Schmittfull:2018yuk,Chen:2020fxs,Chen:2020zjt}. 
For example, the power spectrum has the following schematic form
\begin{align}
    P_g(k,\mu) = \sum_{i,j} b_i b_j P_{\tilde{\mathcal O}_i \tilde{\mathcal O}_j}(k,\mu) + 2k^2(c_0+c_2\mu^2+c_4\mu^4) P_{\tilde{\delta}_1\tilde{\delta}_1}(k) + P_\epsilon \;.
\end{align}
Here indices~$i$ and~$j$ run over four different building blocks of the model. In particular,~$b_i\in \{1,b_1,b_2/2,b_{\mathcal G_2} \}$ and~$\tilde{\mathcal O}_i \in \{ 1, \tilde\delta_1, \tilde \delta_1^2, \tilde{\mathcal G}_2 \}$. Each auto- or cross-spectrum is given by
\begin{align}
    P_{\tilde{\mathcal O}_i \tilde{\mathcal O}_j} (k,\mu) \equiv \int {\rm d}^3\bq \, e^{-i\bk\cdot\bq} \langle \mathcal O_i(\bq) \mathcal O_j({\bf 0}) e^{-ik_a \left( \delta^K_{ab}+f\hat{n}_a \hat{n}_b \right) \left( \Psi_{1b}(\bq) - \Psi_{1b}({\bf 0}) \right) } \rangle \;. 
\end{align}
These correlation functions are the standard building blocks of the LPT in redshift space and can be computed using the standard techniques in LPT~\cite{Carlson:2012bu,Vlah15,Schmittfull:2018yuk,Chen:2020fxs,Chen:2020zjt}. 
Note that we include the three free parameters~$c_0,\;c_2$ and~$c_4$ as the counterterms. 
Finally,~$P_\epsilon$ is the constant power spectrum of the stochastic noise. 
While we write down the model in redshift space, the real space counterpart can be easily obtained by setting~$f=0$,~$c_2=0$ and~$c_4=0$ in the equations above.

We can proceed in a similar way and compute the tree-level bispectrum. 
While this can be also done in Lagrangian perturbation theory~\cite{Chen:2024pyp}, it is computationally simpler and theoretically justified to use the Eulerian framework in this case. 
First, on large scales the Eulerian and Lagrangian description do not differ significantly even at the map level~\cite{Taruya:2018jtk,Taruya:2021ftd,Spezzati:2025zsb}. 
Second, the difference at the map level is mainly due to the nonlinear treatment of large displacements. 
However, these displacements cancel in the correlation functions due to the equivalence principle~\cite{Jain:1995kx,Peloso:2013zw,Kehagias:2013yd,Creminelli:2013mca,Creminelli:2013poa}.\footnote{This is not true in the case of a feature such as the BAO peak~\cite{Senatore:2014via,Baldauf:2015xfa,Vlah:2015zda,Blas:2016sfa,Senatore:2017pbn,Ivanov:2018gjr}. 
However, on large scales used in this paper this exception is irrelevant.} 
For this reason, even the relatively small mismatch between the two perturbative schemes at the field level is even smaller for the~$n$-point functions. 

In order to write down the tree-level bispectrum, we first have to derive the second order Eulerian kernels for our forward models, defined by expanding the nonlinear galaxy density field in linear modes
\begin{align}
    \delta_g(k,\mu) = \sum_{n=1}^\infty \int_{\bq_1} \cdots \int_{\bq_n} (2\pi)^3\delta^{(3)}_{\rm D}(\bk-\bq_1-\cdots -\bq_n) Z_n(\bq_1,\ldots,\bq_n) \delta_1(\bq_1)\cdots \delta_n(\bq_n) \;.
\end{align}
Expanding the redshift space forward model to leading order in perturbation theory, we get the following linear kernel 
\begin{align}
    Z_1(\bk) = (b_1+1)+f\mu^2 \;,
\end{align}
which is the well-known Kaiser formula. At second order in perturbation theory, we get the following quadratic kernel
\begin{align}
\label{eq:Z2redshiftspace}
&Z_2(\bq_1,\bq_2) = (b_1+1) \left[ \frac12 + \frac 12\frac{\bq_1\cdot\bq_2}{q_1q_2} \left( \frac{q_1}{q_2} + \frac{q_2}{q_1} \right) + \frac12 \frac{(\bq_1\cdot\bq_2)^2}{q_1^2q_2^2} \right] + \frac{b_2}{2} + \left( b_{\mathcal G_2} - \frac{b_1}{2} \right) \left[ \frac{(\bq_1\cdot\bq_2)^2}{q_1^2q_2^2} -1 \right] \nonumber \\
& \qquad + f \mu^2 \left[ \frac12 \frac{\bq_1\cdot\bq_2}{q_1q_2}\left( \frac{q_1}{q_2} + \frac{q_2}{q_1} \right) + \frac{(\bq_1\cdot\bq_2)^2}{q_1^2q_2^2} \right] + \frac{1}{2} f \mu k \left( \frac{\mu_1}{q_1} ((b_1+1)+f\mu_2^2) + \frac{\mu_2}{q_2} ((b_1+1)+f\mu_1^2) \right) \;,
\end{align}
where~$\mu_a \equiv \hat{\bq} \cdot \hat{\boldsymbol n}$ and $\bk = \bq_1+\bq_2$. Note that $\mu=\tfrac{q_1}{k}\mu_1 + \tfrac{q_2}{k}\mu_2$, which is important when computing the permutations in the bispectrum. 
This result is similar to the usual SPT kernel, but it is missing all gravitational evolution contributions beyond the linear displacement. For this reason, the factors~$(b_1+1)$ and~$f\mu^2$ multiply the Zel'dovich second order kernels and not the matter density and velocity SPT kernels~$F_2$ and~$G_2$. Note that the standard Eulerian bias parameters are given by
\begin{align}
    b_1^{\rm E} = b_1+1\;, \qquad b_2^{\rm E} = b_2\;, \qquad b_{\mathcal G_2}^{\rm E} = b_{\mathcal G_2} - \frac{b_1}{2} \;. 
\end{align}
The real space linear and quadratic kernels can be easily obtained from these expressions by setting~$f=0$. 

The second order kernel is sufficient to compute the tree-level bispectrum. In redshift space we get
\begin{align}
    B_{g,{\rm tree}}(\bk_1,\bk_2,\bk_3,\hat{\boldsymbol n}) = 2Z_1(\bk_1)Z_1(\bk_2)Z_2(\bk_1,\bk_2) P_{\rm lin}(k_1) P_{\rm lin}(k_2) + 2\; {\rm perms.} \;,
\end{align}
and the corresponding real space counterpart can be easily obtained by setting~$f=0$. The full redshift space bispectrum has complicated momentum dependence given the projections along the line of sight~$\hat{\boldsymbol{n}}$. However, this can be simplified by expanding the bispectrum in multipoles. By far the most relevant contribution on large scales is the bispectrum monopole, which can be written as
\begin{align}
    B_{g,{\rm tree}}^0(\bk_1,\bk_2,\bk_3) = 2 Z_2^0 (\bk_1,\bk_2) P_{\rm lin}(k_1) P_{\rm lin}(k_2) + 2\; {\rm perms.}
\end{align}
The new {\em monopole} quadratic kernel can be computed easily starting from Eq.~\eqref{eq:Z2redshiftspace} and we give its explicit form in Appendix~\ref{app:bispectrum}. 
We will use this model for the joint power spectrum and bispectrum analyses in redshift space and the comparison to field-level inference on large scales.

\subsection{Expectations for the signal-to-noise ratio in perturbation theory}
\label{subsec:expectationsPT}

Before moving on to the practical implementation of the forward models and analyses in various setups, it is instructive to estimate the expected signal-to-noise ratio (SNR) for different observables and the main cosmological parameters. 
This can be done in perturbation theory and it sets expectations for concrete analyses and comparisons to field-level inference. 

For a given {\em connected}~$n$-point function, estimated in a simulation or survey of volume~$V$ and using wavenumbers up to some specified~$k_{\rm max}$, we can estimate its SNR as follows 
\begin{align}
    ({\rm SNR})^2_n \approx \int^{k_{\rm max}}_{\bk_1} \cdots \int^{k_{\rm max}}_{\bk_n} \frac{\langle \delta_g(\bk_1) \cdots \delta_g(\bk_n) \rangle_c^2}{P_g(k_1) \cdots P_g(k_n)} = \mathcal{O}(1) \times N_{\rm pix}\, \Delta^{2(n-2)}(k_{\rm max})  \;, 
\label{eq:SNR_for_n_point}
\end{align}
where we have defined the effective number of voxels  (or equivalently, the total number of Fourier modes) $N_{\rm pix}\equiv V\int^{k_{\rm max}}_{\bk}$. 
This estimate is already rather uncertain up to a factor of order one which depends on details of the nonlinear evolution, bias parameters etc. How this SNR converts to errors on different cosmological parameters is even harder to estimate. 
However, we can still get some intuition from Eq.~\eqref{eq:SNR_for_n_point}. 

As expected, we can see that each higher order~$n$-point function has SNR suppressed by an additional power of the variance of the density field at the given maximum wavenumber~$\Delta^2(k_{\rm max})$. 
This means that in the perturbative regime most of the signal will be in a few leading-order correlation functions. 
Naively, most of the signal should be just in the power spectrum, given that the field is mildly non-Gaussian for small~$\Delta^2$. 
As we are going to see, this is indeed the case when all nuisance parameters describing the galaxy density field are fixed. 
However, in real world examples, bias parameters can lead to strong degeneracies with cosmological parameters. 
The best-known one is the degeneracy between the linear bias~$b_1$ and the amplitude of the linear density field~$A$. 
These two parameters are exactly degenerate in linear theory and can be disentangled only by looking at the nonlinearities. 
This means that the linear power spectrum, even though it has the highest SNR, cannot tell us anything about~$b_1$ and~$A$ individually, just about their product~$b_1A$. 
It is then interesting to ask where the leading nonlinear information on~$A$ is. 
Given the structure of the loop expansion in the power spectrum, it is easy to see that (for fixed~$b_1A$) that the signal for~$A$ in the power spectrum comes only from loops and that it is given by
\begin{align}
    ({\rm SNR})^2_{A,P_\text{1-loop}} \approx V\int^{k_{\rm max}}_{\bk} \frac{P_\text{1-loop}^2(k)}{P_g^2(k)} = \mathcal{O}(1) \times N_{\rm pix} \,\Delta^{4}(k_{\rm max})  \;. 
\label{eq:SNR_A_P1loop}
\end{align}
This is very suppressed, since~$\Delta^4\approx 0.01$ (for~$k_{\rm max}\approx 0.1\;h/{\rm Mpc}$ at low redshift), compared to the total SNR of the power spectrum. 
At the same time, it is also parametrically smaller than the SNR for~$A$ in the bispectrum.
Assuming again that~$b_1A$ is fixed by the power spectrum on large scales, we can estimate
\begin{align}
    ({\rm SNR})^2_{A,B_{\rm tree}} \approx V\int^{k_{\rm max}}_{\bk_1} \int^{k_{\rm max}}_{\bk_2} \frac{B^2_{g,{\rm tree}}(\bk_1,\bk_2,-\bk_1-\bk_2)}{P_g(k_1) P_g(k_2) P_g(|\bk_1+\bk_2|)} = \mathcal{O}(1) \times N_{\rm pix}\, \Delta^2 (k_{\rm max})  \;. 
\label{eq:SNR_A_Btree}
\end{align}
Assuming the absence of large numbers that can significantly modify the factors of order one in the two expressions, it is clear that the SNR for the amplitude~$A$ in the bispectrum is approximately 10 times larger than in the one-loop power spectrum for the scales and redshifts of interest. 
We confirm this expectation in the explicit examples that we analyze in this paper.

The estimate for the SNR for other cosmological parameters is far more subtle for two reasons.
First, they do not appear as a simple multiplicative factor in front of the linear power spectrum, but modify the shape of~$P_{\rm lin}$ instead. 
Second, since they affect both the linear theory and the loops in a nontrivial way, the estimate for the SNR from the power spectrum has ``mixed'' terms in the numerator, where a part of the signal comes from~$P_{\rm lin}$ and the other part from~$P_\text{1-loop}$. 
This would lead to the~${\rm SNR}\sim N_{\rm pix}\,\Delta^2(k_{\rm max})$, comparable to the tree-level bispectrum. 
However, this does not mean that the other cosmological parameters can be measured with similar precision form the one-loop power spectrum and the tree-level bispectrum. 
The reason is that~$A$ is relatively strongly degenerate with other cosmological parameters. 
For example, this is the case for both~$H_0$ and~$\omega_{\rm cdm}$, the two additional cosmological parameters that we vary in this paper. For this reason, in the absence of the bispectrum to break the dominant~$b_1-A$ degeneracy, the one-loop power spectrum always leads to very weak constraints on all cosmological parameters.

Once the tree-level bispectrum is added and the main~$b_1-A$ degeneracy broken, it is interesting to ask how much more signal remains in the higher order~$n$-point functions. 
Following the Eq.~\eqref{eq:SNR_for_n_point}, a simple estimate for the connected trispectrum gives 
\begin{align}
    ({\rm SNR})^2_{A,T_{\rm tree}} \approx V \int^{k_{\rm max}}_{\bk_1} \int^{k_{\rm max}}_{\bk_2} \int^{k_{\rm max}}_{\bk_3} \frac{T^2_{g,{\rm tree}}(\bk_1,\bk_2,\bk_3-\bk_1-\bk_2-\bk_3)}{P_g(k_1) P_g(k_2) P_g(k_3) P_g(|\bk_1+\bk_2+\bk_3|)} = \mathcal{O}(1) \times N_{\rm pix}\, \Delta^4 (k_{\rm max})  \;. 
\label{eq:SNR_A_Ttree}
\end{align}
Parametrically, the trispectrum signal is suppressed and comparable to the one-loop power spectrum. 
For this reason we expect that in a generic situation the optimal constraints from the field-level analysis are similar to the joint power spectrum and bispectrum analysis. 
This has been recently argued in Ref.~\cite{Spezzati:2025zsb} at the level of a forecast and we will confirm this in explicit examples. 
However, note that in the absence of free cubic bias parameters, the trispectrum may help break the residual degeneracies (such as~$A-b_2-b_{\mathcal G_2}$). 
For particular choices of higher order biases, one can in principle design a setup in which the trispectrum has sufficiently high SNR to significantly improve the joint power spectrum and bispectrum analysis (for some examples, see Ref.~\cite{Spezzati:2025zsb}). 

Finally, let us comment on some expectations for the error bars in redshift space. 
In the absence of any selection effects that can change this simple picture~\cite{Hirata:2009qz,Krause:2010tt,Zheng:2010jf,Wyithe:2011mt,Behrens:2017xmm,Desjacques:2018pfv,Martens:2018uqj}, the~$b_1-A$ degeneracy in redshift space is already broken by the power spectrum monopole and quadrupole. 
As before, we can imagine that the combination~$b_1A$ is measured very precisely from the power spectrum monopole. 
Then the constraint on~$A^2$ is simply given by the SNR in the power spectrum quadrupole. 
This SNR is roughly smaller by a factor of $(P_2/P_0)^2/5$ compared to the SNR of the monopole, at all scales. 
On the other hand, the SNR in the bispectrum grows with~$k$. 
The quadrupole and the bispectrum have comparable SNR when 
\begin{align}
    \frac15 \left(\frac{P_2}{P_0}\right)^2 N_{\rm pix} \sim N_{\rm pix} \Delta^2(k_{\rm max}) \;.
\label{eq:SNR_RSD}
\end{align}
This happens approximately at~$k_{\rm max}\sim 0.1\;h/{\rm Mpc}$ where~$\Delta^2(k_{\rm max})\sim 0.1$ at redshifts of interest. 
Therefore, we expect that the joint power spectrum monopole and quadrupole analysis in redshift space leads to comparable constraints on~$A$ as the combination of power spectrum monopole and bispectrum monopole.  

In conclusion, the basic arguments of perturbation theory suggest that on large scales there is a hierarchy in the SNR among different correlation functions. In the absence of free nuisance parameters, almost all information is in the power spectrum. 
Once the bias parameters and the counterterms are allowed to vary, one has to carefully examine different degeneracies in order to estimate which statistics dominate the constraining power for a given cosmological parameter. 
However, for a generic situation, we expect that the joint power spectrum and bispectrum analysis to achieve similar constraints to optimal field-level inference. 

\section{Numerical implementation}
\label{sec:numerical}
In this section, we describe the numerical implementation of the field-level forward model, the sampling method, and the measurement and computation of the power spectrum and bispectrum.

\subsection{Field-level forward model}
\label{subsec:field_level_forward_model}
Our forward model translates the theoretical framework of Sec.~\ref{sec:model} into an efficient, differentiable code.
First, a Gaussian linear field $\delta(\bk)$ is generated on an $N_g^3$ grid by drawing independent complex amplitudes with variance $P_{\rm lin}(k)/2V$.
Next, we embed the resulting linear density field into a larger grid of size $(N_g^\text{L})^3$ (with $N^\text{L}_g > N_g$)
by zero-padding higher-frequency modes (those beyond the Nyquist frequency of the original grid).
This procedure minimizes aliasing effects in the subsequent generation of non-linear fields. 
Following Orszag's rule \cite{Orszag_rule}, and given that our model includes contributions up to quadratic order, we set $N_g^\text{L} = 1.5 N_g$ by default.

In the Eulerian variant the non-linear galaxy field is the direct weighted sum of relevant operators evaluated on that lattice.
In the Lagrangian implementation, each operator is attached as a ``weight'' to a Lagrangian particle that is then displaced with the Zel'dovich, or optional 2LPT, displacement~\cite{Schmittfull:2018yuk,Schmittfull:2020trd}.
After displacement, the particles are deposited with the weight $\cO(\bq)$ onto an Eulerian grid of size $(N^\text{E}_{g})^3$ using the Cloud-In-Cell (CIC) assignment scheme.
To further mitigate aliasing effects, we optionally apply an interlacing technique, in which fields from two interlaced grids are combined to cancel aliasing contributions~\cite{Sefusatti:2015aex}.  
Additionally, we have implemented the Triangular-Shaped-Cloud (TSC) scheme; however, our primary analyses employ CIC with interlacing as it provides an optimal balance between accuracy and computational efficiency.  
Our framework can also be readily extended to a Hybrid-EFT (HEFT) scheme \cite{Modi:2019qbt} by substituting the LPT displacement with a fully non-linear displacement (see e.g. Refs.~\cite{Kokron22,Sullivan25} for HEFT analyses at the field level)
obtained from a differentiable $N$-body simulation (e.g., using \texttt{pmwd}~\cite{Li:2022bsu,Li:2022qlf} or \texttt{JaxPM}\footnote{\url{https://github.com/DifferentiableUniverseInitiative/JaxPM}}).

All of the procedures described above are implemented using \texttt{JAX}~\cite{jax2018github}, 
which facilitates just-in-time (JIT) compilation and reverse-mode automatic differentiation throughout the entire field-level pipeline. 
\footnote{The code for perturbative forward modeling at the field level is publicly available at \url{https://github.com/kazakitsu/PT-field-modeling}.}
This framework supports differentiation not only with respect to the global parameters, $\btheta$ (including cosmological and bias parameters), 
but also for each Fourier mode of the initial conditions, $\delta(\bk)$.  
To ensure that the linear power spectrum is differentiable with respect to the cosmological parameters (particularly $\omega_\text{cdm}$), 
we compute it using \texttt{CosmoPower-JAX}~\cite{Piras:2023aub, SpurioMancini2022}.
\footnote{For differentiable Einstein-Boltzmann solvers see also e.g. \texttt{DISCO-EB} \cite{Hahn:2023nvb} based on \texttt{JAX} and \texttt{Bolt
} (\url{https://github.com/xzackli/Bolt.jl}) based on \texttt{Julia}.}
Consequently, even in high-dimensional parameter space involving every Fourier mode of the initial conditions, we can leverage \texttt{JAX}'s efficient GPU execution for rapid forward model evaluation and gradient computation.  
For instance, generating a single realization of the galaxy density field on a grid with $N^3_g = (N_g^\text{E})^3 = 128^3$ and $(N_g^{\rm L})^3 = 192^3$ takes just under 30 milliseconds on a single NVIDIA A100 GPU.
This efficiency is essential for employing gradient-based sampling methods such as Hamiltonian Monte Carlo (HMC), which we discuss in the following section.

\subsection{Field-level likelihood and sampling}
\subsubsection{The field-level posterior}
The cornerstone of Bayesian cosmological inference is the posterior distribution of the cosmological parameters, $\btheta$, 
given an observed galaxy density field, $\hat{\delta}_g$: $\cP [\btheta| \hat{\delta}_g]$. 
According to Bayes' theorem, this posterior can be formally expressed as
\begin{align}
    \cP[\btheta| \hat{\delta}_g] = & \int \cD \delta ~\cP[\btheta, \delta| \hat{\delta}_g]
    \\
    \propto & \int \cD \delta ~ \cL [\hat{\delta}_g|\btheta, \delta] \pi[\delta|\btheta] \pi[\btheta],
    \label{eq:bayes}
\end{align}
where $\cL [\hat{\delta}_g|\btheta, \delta]$ is the likelihood of the observed galaxy density field given the model prediction $\delta_g$, 
which in principle is a function of model parameters $\btheta$ and a realization of the initial conditions $\delta$,
$\pi [\delta | \btheta]$ is the prior distribution of the initial conditions given the cosmological parameters, 
and $\pi[\btheta]$ is the prior distribution of the model parameters including the cosmological parameters.
A prediction of the galaxy density field depends on a realization of the initial conditions, which we do not know a priori.
Therefore, we need to marginalize over realizations of the initial conditions to obtain the posterior distribution of the cosmological parameters.
Mathematically, this marginalization corresponds to the path integral over the initial conditions, denoted as $\int \cD \delta$.

Basically, what we can compute in cosmology is summarized in the prior term of the initial conditions, $\pi [\delta|\btheta]$,
and the likelihood term, $\cL [\hat{\delta}_g|\btheta, \delta]$, in the above.
In other words, early universe models can only predict the statistical properties of the primordial perturbations;
that is, they specify the probability distribution of the initial conditions given the cosmological parameters, which is precisely the role of the prior. 
The standard inflationary model predicts that the primordial fluctuations follows a Gaussian distribution with diagonal covariance in Fourier space:
$\langle \delta(\bk) \delta(\bk') \rangle = (2\pi)^2 \delta_\text{D}(\bk + \bk')P_{\rm lin}(k)$.
In our context the explicit form of the prior is given by
\begin{align}
    \pi[\delta({\bf k})|\btheta_{\rm cosmo}]
    = 
    \left( \prod_{\bf k}\sqrt{\frac{1}{2\pi P_{\rm lin}(k;\btheta_{\rm cosmo})}} \right)
    \exp\left( -\frac12 \int_{\bf k}\frac{|\delta({\bf k})|^2}{P_{\rm lin}(k;\btheta_{\rm cosmo})} \right).
    \label{eq:prior}
\end{align}

On the other hand, our understanding about the subsequent evolution of the fluctuations is encapsulated in the likelihood $\cL [\hat{\delta}_g|\btheta, \delta]$.
If we had a perfect model for the observables (e.g., the galaxy density field), and if we were in the cosmic variance limit, 
the likelihood would reduce to a Dirac delta function: 
$\cL [\hat{\delta}_g|\btheta, \delta] \to \delta_\text{D}^{(\infty)}[\hat{\delta}_g - \delta_g(\btheta, \delta)]$ 
where $\delta_g(\btheta, \delta)$ denotes the model.
In reality, however, model imperfections and the discrete nature of the observables lead to a likelihood with finite width.
To obtain an analytical form amenable to sampling, we assume a simple Gaussian noise model in this paper.
Although we note that such a simplistic Gaussian noise is insufficient even at relatively large scales in practice~\cite{Schmittfull:2018yuk},
this yields a Gaussian likelihood with a diagonal covariance, $P_{\rm err}(k)$,
\begin{align}
    \cL [\hat{\delta}_g|\btheta,\delta({\bf k})]
    =
    \left( \prod_{\bf k}^{k_\text{max}}\sqrt{\frac{1}{2\pi P_{\rm err}(k)}} \right)
    \exp \left(
    -\frac12 \int_{\bf k}^{k_\text{max}} \frac{|\hat{\delta}_g({\bf k}) - \delta_g({\bf k};\btheta, \delta)|^2}{P_{\rm err}(k)}
    \right),
    \label{eq:likelihood}
\end{align}
where $k_\text{max}$ denotes the maximum wavenumber used in the analysis.
This cutoff is imposed spherically in Fourier space, i.e., $|\bk| \leq k_\text{max}$, unless otherwise stated.
We study the outcome of non-Gaussian noise when using a Gaussian likelihood in field-level inference in Sec.~\ref{sec:non-Gaussian_noise}.

Putting everything together, the log-posterior of the global parameters $\btheta$, and the initial conditions $\delta(\bk)$, given the observed galaxy density field $\hat{\delta}_g$, can be written as:
\begin{align}
    \ln \cP [\btheta, \delta({\bf k})| \hat{\delta}_g]
    = & \ln \cL [\hat{\delta}_g|\btheta, \delta({\bf k})] + \ln \pi[\delta({\bf k})|\btheta] \\
    = & 
    -\frac12 \int_{\bf k}^{k_\text{max}}
    \left[ \ln(2\pi P_{\rm err}(k)) + \frac{|\hat{\delta}_g({\bf k}) - \delta_g({\bf k};\btheta)|^2}{P_{\rm err}(k)}\right]
    -\frac12 \int_{\bf k}^{\Lambda}
    \left[\ln(2\pi  P_{\rm lin}(k;\btheta_{\rm cosmo}))  + \frac{|\delta({\bf k})|^2}{P_{\rm lin}(k;\btheta_{\rm cosmo})} \right].
    \label{eq:log_posterior}
\end{align}
Here, constant terms independent of the parameters are omitted.  
In principle, the prior term is defined as an integral of the continuous field over the entire $\bk$-space
(hence the notation $\cD \delta$ in Eq.\eqref{eq:bayes}).
However, in practice we need to discretize the field on a finite grid to perform a numerical evaluation.
This discretization naturally introduces a cutoff in the $\bk$-integral of the initial conditions, which we denote by $\Lambda$.
This cutoff is determined by the grid size, $N_g$, and corresponds to the Nyquist frequency, $\Lambda =  \pi N_g/L$, 
leading to a cubic cutoff in Fourier space, satisfying $|k_i| \leq \Lambda$ with $i=x,y,z$.
\footnote{Another option is to choose the spherical cutoff in Fourier space, satisfying $|\bk| \leq \Lambda$.}
The two cutoffs, $\Lambda$ and $k_\text{max}$, can be different, although $\Lambda$ should be greater than $k_\text{max}$.
Note that due to the reality condition, $\delta^*(\bk) = \delta(-\bk)$, we take the summation over the $k_z \geq 0$ modes only.
Special care is taken for the $k_z = 0$ and $k_z = k_\text{Nyq}$ planes to avoid double-counting Fourier modes that are equal to their own complex conjugates.

Field-level inference aims to estimate the above posterior, $\cP [\btheta, \delta({\bf k})| \hat{\delta}_g]$,
by jointly sampling $\btheta$ and $\delta(\bk)$ using a Markov chain Monte Carlo (MCMC) approach.
The challenge here is that the posterior is a quite high-dimensional distribution and thus a naive random-walk Metropolis-Hastings sampler would mix extremely slowly.
To efficiently explore this high-dimensional posterior, we rely on the HMC method~\cite{Duane:1987de,Neal1996}.\footnote{This sampling technique was first proposed in the context of lattice QCD in Ref.~\cite{Duane:1987de}, 
where it is called Hybrid Monte Carlo 
(One can notice the similarity of Eq.~\eqref{eq:path_integral} to the partition function in field theory, $Z = \int {\cal D}\phi~e^{-S_\text{E}(\phi)}$).
Later it was imported to the statistics community as Hamiltonian Monte Carlo~\cite{Neal1996}. In either case the abbreviation is HMC.
}
It leverages Hamiltonian dynamics to propose distant yet appropriately likely samples, substantially reducing random-walk behavior, as we briefly review below.

\subsubsection{Hamiltonian Monte Carlo (HMC)}

Let $ \boldsymbol{q} = (\btheta, \delta(\bk))$ denote the full set of parameters we wish to sample,\footnote{Here $\boldsymbol{q}$ is not Lagrangian coordinates, which we denote $\bq$.}
and let $\boldsymbol{p}$ be an auxiliary momentum vector of the same dimension.
HMC augments the target posterior $\cP(\boldsymbol{q})$ with a kinetic term, forming the joint distribution $\propto \exp[- H(\boldsymbol{q},\boldsymbol{p})]$
with $H(\boldsymbol{q},\boldsymbol{p}) = U(\boldsymbol{q}) + K(\boldsymbol{p})$, 
where the potential energy is $U(\boldsymbol{q}) = -\ln \cP(\boldsymbol{q})$ and the kinetic energy is chosen as
\begin{align}
    K(\boldsymbol{p})
    =
    \frac12\,\boldsymbol{p}^{\mathsf T}
    \boldsymbol{M}^{-1}\,
    \boldsymbol{p}.
\end{align}
Here $\boldsymbol{M}$ is a symmetric, positive-definite \emph{mass matrix}.  
Drawing $\boldsymbol{p}$ from the Gaussian $\mathcal{N}(\mathbf{0},\boldsymbol{M})$ 
and evolving the pair $(\boldsymbol{q},\boldsymbol{p})$ according to Hamilton's equations
\begin{align}
    \frac{\dd\boldsymbol{q}}{\dd\tau} =
    \frac{\partial K}{\partial \boldsymbol{p}}
    =
    \boldsymbol{M}^{-1}\boldsymbol{p},
    \qquad
    \frac{\dd\boldsymbol{p}}{\dd\tau}=
    -\frac{\partial U}{\partial\boldsymbol{q}},
\end{align}
preserves $H(\boldsymbol{q},\boldsymbol{p})$.
These equations are integrated with a time-reversible, symplectic leapfrog scheme of step size $\epsilon$ over $L$ steps,
and then the resultant proposal $(\boldsymbol{q}',\boldsymbol{p}')$ is accepted or rejected with probability $\min[1, \exp (-\Delta H )]$,
where \(\Delta H\) denotes the change in the Hamiltonian resulting from numerical integration errors.
Hence, ideally the acceptance rate would be the unity.

The mass matrix governs the kinetic term and therefore the typical momentum drawn for each parameter.
A larger diagonal entry in $\boldsymbol{M}$ results in smaller typical momenta and hence smaller updates in that parameter, 
whereas smaller entries allow larger steps. 
Furthermore, correlations encoded in off-diagonal elements let HMC move efficiently along degenerate directions of the posterior.
In our implementation we partition $\boldsymbol{M}$ into blocks,
\begin{align}
    \boldsymbol{M}
    =
    \begin{pmatrix}
        \boldsymbol{M}_{\btheta} & 0\\[2pt]
        0 & \boldsymbol{D}_{\delta(\bk)}
    \end{pmatrix},
\end{align}
where $\boldsymbol{M}_{\btheta}$ is a dense sub-matrix corresponding to the global parameters \(\boldsymbol{\theta}\) (the cosmological and bias parameters) and
$\boldsymbol{D}_{\delta(\bk)}$ is a diagonal block for each Fourier modes of the initial conditions.
This reflects two empirical observations:
(1) The global parameters tend to exhibit appreciable posterior correlations (e.g.\ the well-known amplitude-bias degeneracy),
so a dense sub-matrix accelerates exploration along these directions, and
(2) Different Fourier modes of the initial conditions are nearly independent \emph{a priori};
a diagonal block avoids the memory cost of storing a full matrix of dimension $N_{\text{modes}}\times N_{\text{modes}}$.

During the warm-up phase of the HMC algorithm, the mass matrix is tuned based on the local curvature of the log-posterior.
One important observation here is that the gradient with respect to the global parameters $\btheta$ is generally larger than that with respect to the individual Fourier modes $\delta(\bk)$. 
Although the gradient does not directly determine the mass matrix values, it reflects a higher local curvature for $\btheta$. 
To compensate, the mass matrix is set with larger entries for 
$\btheta$ (thereby reducing the effective step size through the inverse mass in the integrator) and smaller entries for $\delta(\bk)$, 
whose smoother variation permits larger steps. 
This adaptive adjustment ensures that the integration step size is balanced across all parameters, thereby enhancing sampling efficiency.

\subsubsection{Inference setup and mock-data construction}
Our full sampling procedure is built upon \texttt{NumPyro}~\cite{phan2019composable}, a Pythonic probabilistic programming library that leverages \texttt{JAX}'s automatic differentiation, and just-in-time compilation via XLA. 
We employ the No-U-Turn Sampler (NUTS)~\cite{Hoffman:2011ukg}, a variant of HMC that adaptively adjusts the trajectory length, step size, and mass matrix during the warm-up phase, 
to jointly sample the global parameters and the initial conditions, $\cP[\btheta, \delta|\hat{\delta}_g]$.
\footnote{The code for field-level inference is publicly available at \url{https://github.com/kazakitsu/field-level-inference}.}
The posterior distribution of the cosmological parameters $\btheta$ given the observed galaxy density field $\hat{\delta}_g$ can be obtained by marginalizing over the initial conditions:
\begin{align}
    \cP[\btheta| \hat{\delta}_g] 
    \propto 
    \int \cD \delta 
    \exp \left(
        -\frac12 \int_{\bf k}^{k_\text{max}} \frac{|\hat{\delta}_g({\bf k}) - \delta_g({\bf k};\btheta)|^2}{P_{\rm err}(k)}
    \right)
    \times
    \exp\left( -\frac12 \int_{\bf k}^{\Lambda}\frac{|\delta({\bf k})|^2}{P_{\rm lin}(k;\btheta_{\rm cosmo})} \right).
    \label{eq:path_integral}
\end{align}

Our implementation allows us to vary the five cosmological parameters: $\{A, n_s, \omega_\text{cdm}, \omega_\text{b}, H_0\}$.
However, as galaxy clustering does not competitively constrain $n_s$ and $\omega_\text{b}$, we fix these two parameters at their input values 
and only vary $\{A, \omega_\text{cdm}, H_0\}$.
The true input values and the prior distributions for the global parameters are summarized in Tab.~\ref{tab:params}.
For all inference runs presented in this paper, we use eight independent MCMC chains and collect at least $10^3$ effective samples.
Details regarding the convergence of the MCMC chains for the field-level inference are provided in Appendix~\ref{app:mcmc}.

As noted in the introduction, we construct synthetic data to isolate the impact of the likelihood itself.
The mock data is generated with the \textit{exact} perturbative forward model described in Sec.~\ref{sec:model}, and the Gaussian white noise is subsequently added.
\footnote{Although we inject the Gaussian white noise into the mock data, 
we expect that the residual noise in the field-level likelihood has some scale-dependence due to the resolution mismatch, which scales as $k^4$ for the matter case~\cite{Peebles:1980yev,Baldauf:2015zga}.
Thus, in principle, one should include these scale-dependent corrections in $P_\text{err}(k)$, 
while we omit these scale-dependent terms in the noise given that they are negligibly small compared to the constant term.
The resolution mismatch also induces non-Gaussianity in the residual field-level noise,  whose implications will be discussed in Sec.~\ref{sec:resolution_mismatch} and Sec.~\ref{sec:non-Gaussian_noise}.
}
In other words, the mock and inference share the same forward model, but differ in the resolution of the initial grid:
the mock data is built on a $N^3_g = 512^3$ grid, whereas inference employs a coarser $N^3_g = 128^3$ grid.
Note that the Eulerian grid is kept the same for both the mock and the inference: $(N_g^\text{E})^3 = 128^3$.
The consequence of this resolution mismatch is examined in Sec.~\ref{sec:resolution_mismatch}.

\begin{table}[t]
  \centering
  \begin{subtable}[t]{0.45\textwidth}
    \centering
    \begin{tabular}{c|c|c}
      Parameters & Input & Prior \\ \hline\hline
      $A$             & $1.0$          & ${\cal U}[0.5, 1.5]$  \\
      $~\omega_{\rm cdm}~$ & $~0.11933~$   & $~{\cal U}[0.05, 0.255]~$\\
      $H_0$           & $73.0$         & ${\cal U}[64, 82]$    \\
      $A b_1$           & $1.0$          & ${\cal U}[0, 4]$      \\
      $A^2 b_2$           & $-0.5$         & ${\cal N}[0, 2^2]$    \\
      $A^2 b_{{\cal G}_2}$& $-0.5$         & ${\cal N}[0, 2^2]$    \\
      $c_0$, $c_2$, $c_4$   & $0.0$          & ${\cal N}[0, 20^2]$   \\
      $\log P_{\rm err}$ & $8.0$      & ${\cal N}[8, 1^2]$\\
      \bottomrule
    \end{tabular}
  \end{subtable}
  \hfill
  \begin{subtable}[t]{0.45\textwidth}
    \centering
    \begin{tabular}{c|c|c}
      &~Mock data~&~Inference~\\ \hline\hline
      $~L~[\rm {Mpc}/h]~$ & \multicolumn{2}{c}{$2000$}  \\
      $~z~$ & \multicolumn{2}{c}{$0.5$}  \\
      $N^3_{g}$ & $512^3$ & $128^3$ \\
      $(N^{\rm L}_{g})^3$ & $768^3$ & $192^3$ \\
      $(N^{\rm E}_{g})^3$ & \multicolumn{2}{c}{$128^3$}  \\
      \bottomrule
    \end{tabular}
  \end{subtable}
  \caption{%
    \textit{Left:} The parameters we vary in the inference and their priors. $A$ is the normalized amplitude for the initial density field: $A \equiv \sqrt{A_s/A^{\rm input}_s}$. The uniform prior ranges for $\omega_\text{cdm}$ and $H_0$ are adopted from  the prior range of \texttt{CosmoPower-JAX}.
    \textit{Right:} The choice of the grid sizes in the mock data and the inference.
  }
  \label{tab:params}
\end{table}

\subsection{Power spectrum and bispectrum analysis}
One of the main goals of this paper is to compare the field-level inference with the joint power spectrum and bispectrum analysis. 
Given the synthetic data generated by one of the perturbative forward models we described in Sec.~\ref{sec:model}, 
the theoretical predictions are also made in terms of the perturbation theory as we described in Sec.~\ref{subsec:simple_Eulerian} and \ref{subsec:PBinLPT_model}.
All one-loop contributions are computed using the FFTlog method of Refs~\cite{Schmittfull:2016jsw,Simonovic:2017mhp,Chen:2020fxs,Chen:2020zjt,Kokron:2022iok,DeRose:2022zfu} and implemented with \texttt{JAX}~\cite{JAX_oneloop}, 
such that we can utilize the HMC for the joint power spectrum and bispectrum analysis.

We measure the power spectrum and bispectrum from the synthetic data by using the following estimators~\cite{Sefusatti:2015aex}
{\footnote{The code for measuring the power spectrum and bispectrum with \texttt{JAX} is publicly available at \url{https://github.com/kazakitsu/lss_utils}, (although this part is not necessarily differentiable).}
}
:
\begin{align}
    \hat{P}_\ell(k_i) & = \frac{1}{N_P(k_i)}\sum_{\bq \in k_i} |\hat{\delta}_g(\bk)|^2 \cL_\ell(\mu), \\
    \hat{B}_0(k_1, k_2, k_3) &= \frac{1}{N_B(k_1,k_2,k_3)}
    \sum_{\bq_1 \in k_1} \sum_{\bq_2 \in k_2} \sum_{\bq_3 \in k_3} \hat{\delta}_g(\bq_1) \hat{\delta}_g(\bq_2) \hat{\delta}_g(\bq_3) \delta_{\rm K}(\bq_1 + \bq_2 + \bq_3),
\end{align}
where $N_P(k_i)$ is the number of the independent Fourier modes in the bin and $N_B(k_1,k_2,k_3)$ is the number of fundamental triangles in the bin
with binning $\Delta k = 0.01~h/\text{Mpc}$.
Note that these spectra are measured from the exactly same mock used in the field-level inference for each model, 
i.e., $\hat{\delta}_g(\bk)$ appeared above is exactly the same field as the one in Eq.~\eqref{eq:path_integral}.
In measuring the bispectrum, we do not include the open triangle bins, where the bin centers do not satisfy the triangle condition $|k_2 - k_3|< k_1 < k_2 + k_3$, since it is hard to compute the theoretical prediction for that.

We use a Gaussian likelihood for the joint power spectrum and bispectrum analysis:
\begin{align}
    \ln {\cal L}_{P+B} \propto -\frac12 \sum_{ij} \left( \hat{d}_i - m_i \right) C^{-1}_{ij} \left( \hat{d}_j - m_j \right),
\end{align}
where $\hat{d}_i$ is the data vector in the $i$-th bin, $m_i$ is the model prediction in the $i$-th bin, and $C_{ij}$ is the covariance matrix.
Given that it is easy to generate PT-based mock realizations,
the covariance matrix is estimated from $10^4$ realizations of the mock catalogs:
\begin{align}
    C_{ij} = \frac{h^{-1}_f}{N_{\rm mock}-1} \sum_{n=1}^{N_{\rm mock}} \left(d^{(n)}_i - \langle d_i \rangle \right) \left(d^{(n)}_j - \langle d_j \rangle \right) ,
\end{align}
with $h_f = (N_{\rm mock}- N_{\rm data}-2)/(N_{\rm mock}-1)$ being the Hartlap factor~\cite{Hartlap:2006kj}.
For each analysis case, we estimate a separate covariance from $10^4$ PT-based mocks generated with the same forward model and parameter settings as the mock data under study.
Note that on large scales one can also use the simple Gaussian covariance as a good approximation to the truth. In the mildly nonlinear regime we expect differences on inferred cosmological parameters to be small and of order~$\Delta^2(k_{\rm max})$. 
We explicitly test this in Appendix~\ref{app:cov_NG}. 
Finally, the priors on the parameters we vary in the joint power spectrum and bispectrum analysis are the same as in the field-level inference.

\section{Comparison with the joint power spectrum and bispectrum analysis I: Simple Eulerian model}
\label{sec:EPT}
In this section, we perform the field-level inference and the joint power spectrum and bispectrum analysis for the simple Eulerian model, where all nonlinearities are well under control. 
We use this setup to test the FLI algorithm, test predictions for the SNR from Sec.~\ref{sec:model} for different sets of free parameters and test the prediction of Ref~\cite{Cabass:2023nyo} by directly comparing the posteriors for the FLI and the standard inference approach. 
For all analyses presented in this section, we choose a maximum wavenumber of $k_\text{max} = 0.12~h/\text{Mpc}$.

\subsection{Linear problem: Gaussian random field}
\label{subsec:Gaussian_case}

\begin{figure}[t]
    \centering
    {$k_\text{max} = 0.12~h/\text{Mpc}$\par\vspace{1ex}}
    \begin{subfigure}{0.49\textwidth}
        \centering
        \includegraphics[width=\textwidth]{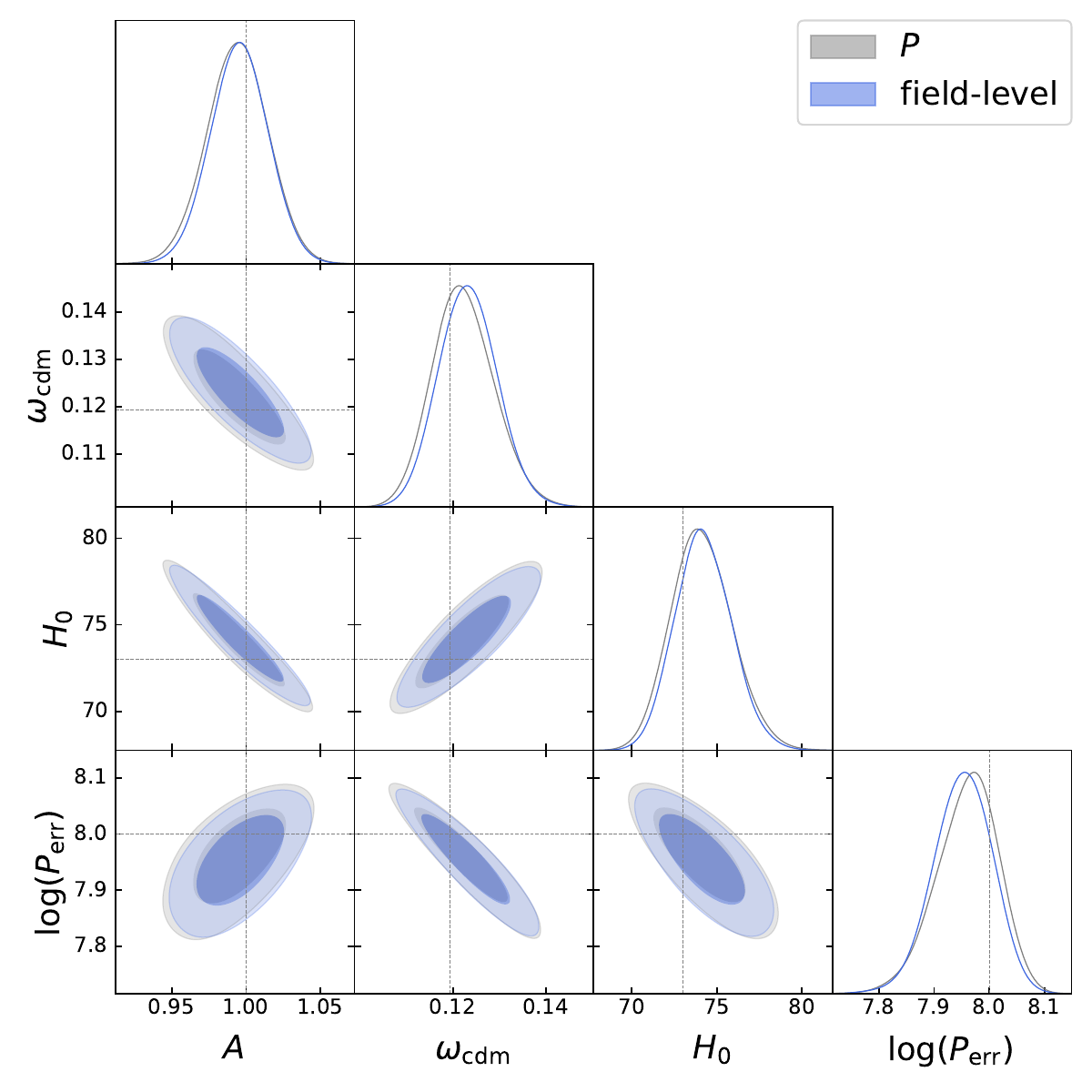}
    \end{subfigure}
    \begin{subfigure}{0.49\textwidth}
        \centering
        \includegraphics[width=\textwidth]{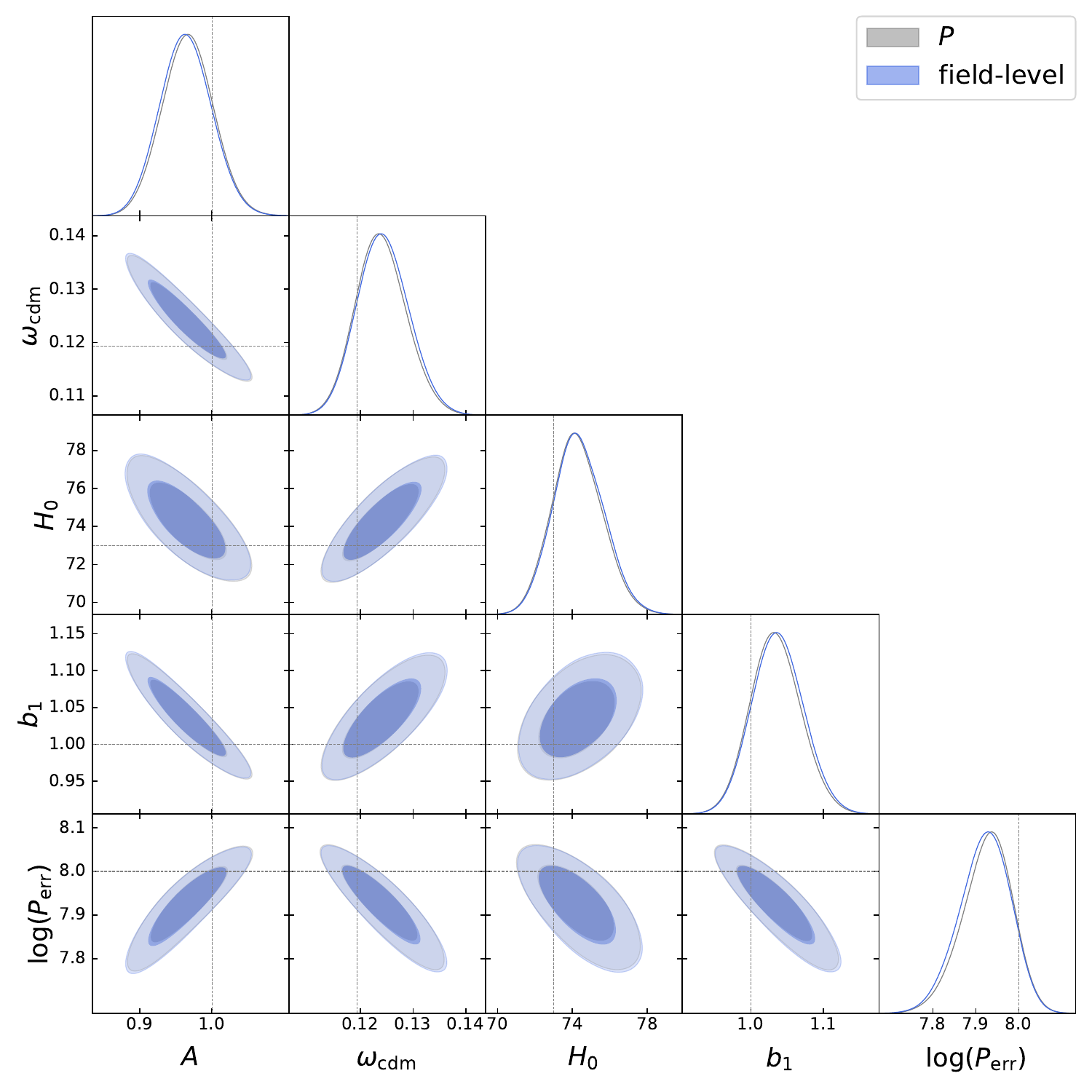}
    \end{subfigure}
    \caption{The 2D marginalized posterior distributions for the Gaussian random field. 
    The left panel shows the result in real space and the right panel shows the result in redshift space. 
    The power spectrum analysis is shown in gray and the field-level analysis in blue.
    The contours indicate 68\% and 95\% credible intervals.
    The dashed lines indicate the true input values.}
    \label{fig:Gaussian}
\end{figure}

Before we start with the nonlinear models, let us begin with the simplest scenario: a Gaussian random field. In this case, it is known that the compression of the full field into the power spectrum is optimal. In other words, the posterior obtained from the power spectrum analysis should coincide with that obtained from the full field-level analysis. This direct comparison serves as a sanity check for our implementation.

The first model we consider consists solely of the linear density field with Gaussian noise added, i.e.,
\begin{align}
    \hat{\delta}_g = \delta_1 + \epsilon.
\end{align}
In this analysis, we vary the three cosmological parameters $\{A, \omega_{\rm cdm}, H_0\}$ and the noise power spectrum amplitude $P_\text{err}$.
The comparison between the power spectrum posterior and FLI is displayed in the left panel in Fig.~\ref{fig:Gaussian}.
The resultant posterior distributions obtained from both the power spectrum analysis (the gray contour) and the field-level analysis (the blue) are nearly identical, as expected.
Here we observe that they are in agreement, including the random variations in the mean value shifts relative to the true input values, 
because we use the same mock data for both analyses.

As a further sanity check, we repeat the same analysis but with the linear density field in redshift space:
\begin{align}
    \hat{\delta}_{\rm g} = (b_1 + f \mu^2)\delta_1 + \epsilon,
\end{align}
Due to the anisotropy introduced in redshift space, the power spectrum analysis is based on its multipoles~$P_0$,~$P_2$, and~$P_4$. In this case, the parameter set we vary is~$\{A, \omega_{\rm cdm}, H_0, b_1, P_\text{err}\}$, while the growth rate~$f$ is computed as a derived parameter in  both the field-level and power spectrum analyses.
The right panel in Fig.~\ref{fig:Gaussian} shows that the two approaches yield nearly identical posterior distributions,
although we note that taking into account the binning effect and the discreteness effects are crucial for the power spectrum analysis to get the this level of agreement.

\subsection{Slightly non-linear (non-Gaussian) examples}
\label{subsec:G2_case}
\subsubsection{Varying the cosmological parameters only}

\begin{figure}[t]
    \centering
    \begin{subfigure}{0.49\textwidth}
        \centering
        {$k_\text{max} = 0.12~h/\text{Mpc}$\par\vspace{1ex}}
        \includegraphics[width=\textwidth]{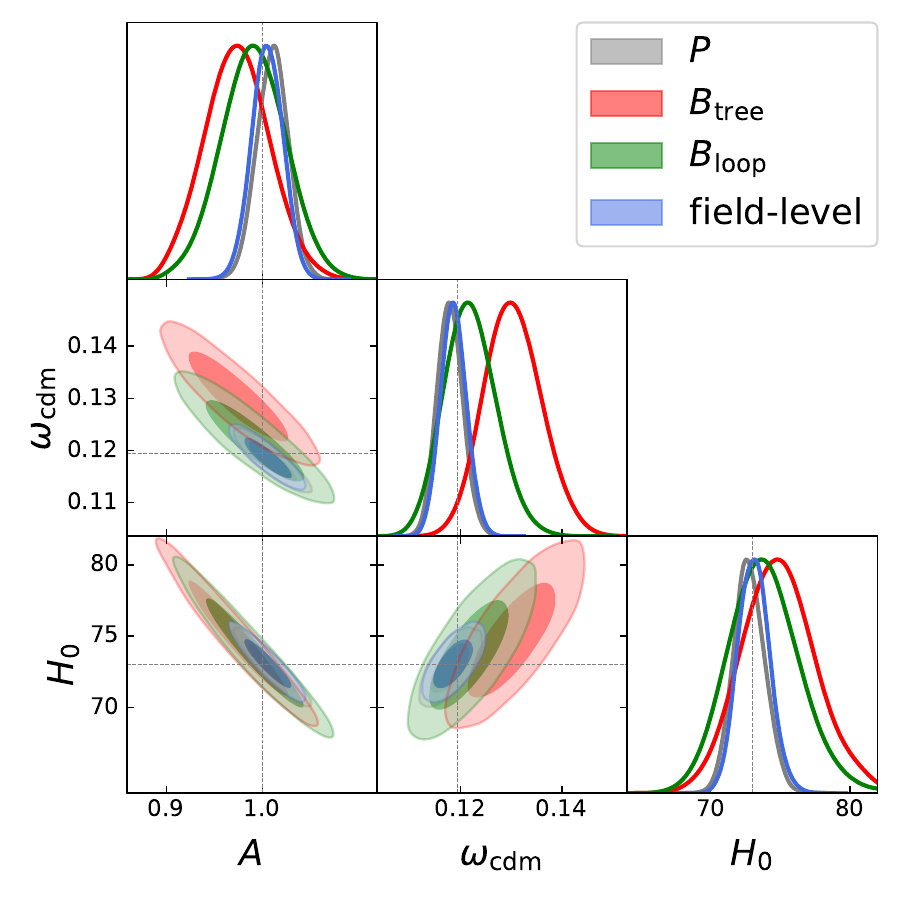}
    \end{subfigure}
    \begin{subfigure}{0.49\textwidth}
        \centering
        \includegraphics[width=\textwidth]{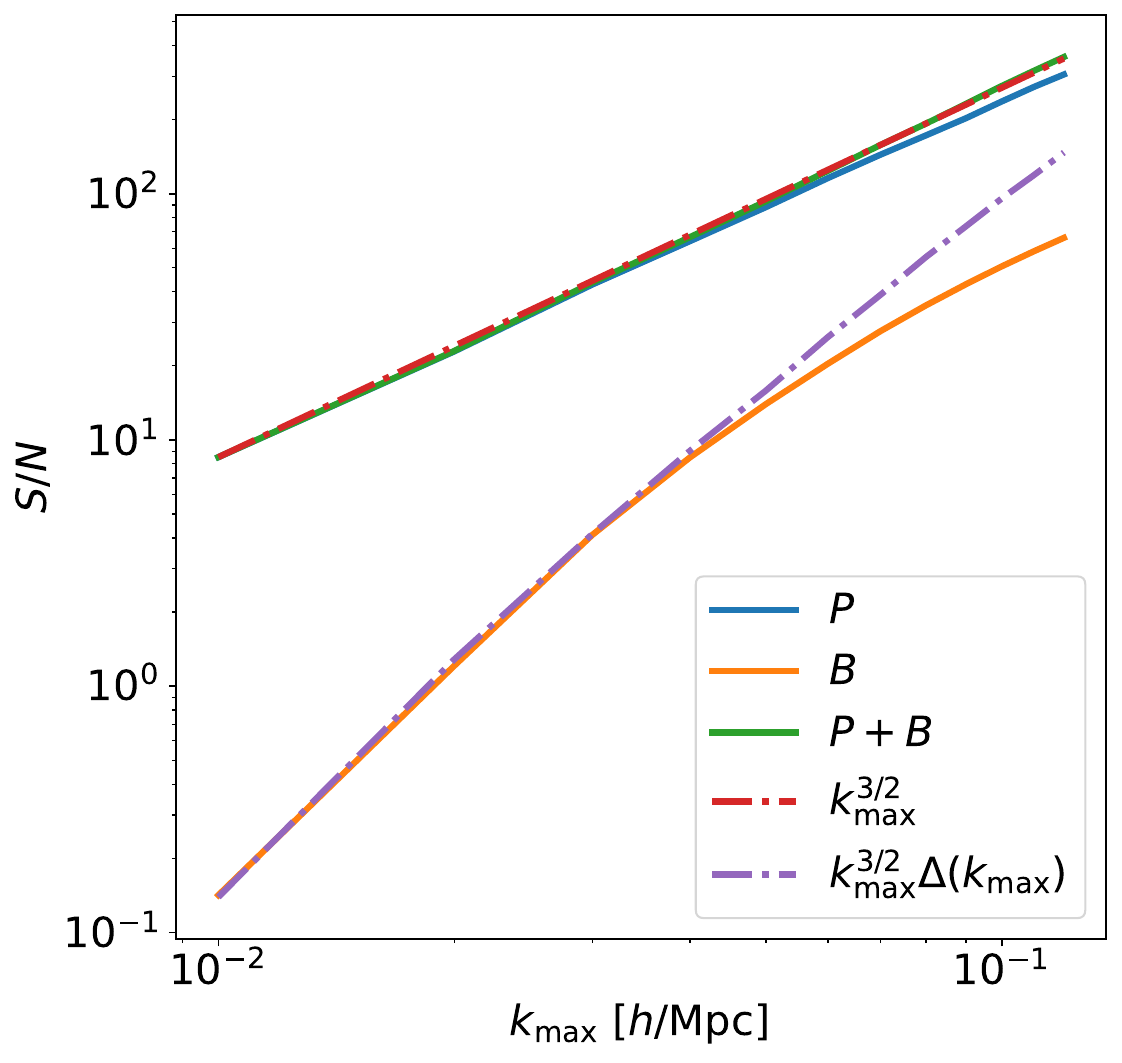}
    \end{subfigure}
    \caption{\textit{Left panel}:The 2D marginalized posterior distributions for the simple non-Gaussian example when varying the cosmological parameters only, 
    fixing the linear bias and the noise power spectrum amplitude.
    The power spectrum constraints are shown in gray and the field-level constraints in blue, which are almost overlapped with each other.
    The red contours show the constraints from the tree-level bispectrum only and the green contours show the constraints when including one-loop bispectrum.
    \textit{Right panel}: The cumulative SNR as a function of $k_\text{max}$ for the power spectrum (blue), the bispectrum (orange), and the joint power spectrum and bispectrum (green) in this model.
    The red dash-dotted line indicates the Gaussian limit, which is proportional to $k_\text{max}^{2/3}$, while the purple dash-dotted line represents $k_{\rm max}^{3/2}\Delta(k_{\rm max})$ estimate.
    }
    \label{fig:G2}
\end{figure}

We now move on to a slightly more complex scenario by introducing a small non-Gaussian component into the field.
In this simple model, the observed galaxy density field in real space is given by:
\begin{align}
    \hat{\delta}_{\rm g} = b_1(\delta_1 + {\cal G}_2) + \epsilon,
\end{align}
where $\cG$ represents the quadratic non-Gaussian term (see Sec.~\ref{sec:model} for details).
We first vary only the cosmological parameters $\{A, \omega_{\rm cdm}, H_0\}$, while keeping the linear bias $b_1$ and the amplitude of the noise power spectrum fixed for a moment.

In the absence of the nuisance parameters, as argued in Sec.~\ref{sec:model}, most of the information is in the power spectrum which has the highest SNR. 
We confirm this expectation in practice. As illustrated in the left panel of Fig.~\ref{fig:G2}, the posteriors obtained from the power spectrum analysis (displayed in gray) are in excellent agreement with those from the full field-level analysis (shown in blue).  
This indicates that, even in the presence of a small non-Gaussian component but with no free nuisance parameters, the power spectrum analysis remains nearly optimal.

Since the nonlinearities generate all higher-order correlation functions, they carry a part of the information about cosmological parameters. The leading higher-order~$n$-point function is the bispectrum. The results from the tree-level bispectrum (red) and the one-loop bispectrum (green) analyses are presented in the left panel of Fig.~\ref{fig:G2}. There are three key observations that merit further discussion.  

First, the resulting constraints from the bispectrum are significantly weaker than those obtained from the power spectrum.  
Such a result is expected given the fact that the signal-to-noise ratio (SNR) of the bispectrum is inherently lower than that of the power spectrum, as illustrated in the right panel of Fig.~\ref{fig:G2}. This hierarchy is compatible with Eq.~\eqref{eq:SNR_A_Btree} and suggests that the relative suppression of the bispectrum SNR is approximately given by~$\Delta(k_{\rm max})$. For~$k_{\rm max}\sim0.1\;h/{\rm Mpc}$ and the redshift of interest, this estimate gives roughly a factor of 3 larger error bars in the bispectrum analysis compared to the constraints from the power spectrum. This is in agreement with the results in Fig.~\ref{fig:G2}.

Second, the parameter degeneracy directions in the bispectrum are the same as those in the power spectrum, implying that no additional degeneracies are broken by adding bispectrum data.  
An important lesson from these two observations is that incorporating higher-order statistics does not automatically lead to improved constraints even for non-Gaussian fields,
unless these additional statistics serve to break parameter degeneracies that persist in the power spectrum analysis.\footnote{
Related to the first point, one can think the following way: The Fisher information matrix can be expressed as 
$F_{\alpha \beta} = \sum_{ij} m_i (C^{-1})_{ij} m_j \frac{\partial \ln m_i}{\partial\theta_\alpha} \frac{\partial \ln m_j}{\partial\theta_\beta}$, 
while the SNR$^2$ is given by $\sum_{ij} m_i (C^{-1})_{ij} m_j$,
where $m_i$ is the signal in $i$-th bin, $C$ is the covariance matrix, and $\theta_\alpha$ is the parameter of interest.
Thus, the Fisher information matrix can be viewed as an SNR$^2$, weighted by the derivatives of the signal with respect to the parameters.
This suggests that for an observable with low SNR to contribute to the improving parameter constraints, 
it must exhibit either a large gradient or a gradient in a different direction.
Conversely, if the gradient is similar, as in the present example, an observable with low SNR does not improve parameter constraints.
}
This is also true for more non-Gaussian fields such as the non-linear matter density field, as we show in Sec.~\ref{sec:resolution_mismatch}.

Finally, let us comment on the difference in the results depending on the bispectrum model (see also Fig.~\ref{fig:beyond_2pt_comparison}).
We can see from Fig.~\ref{fig:G2} that the simple tree-level bispectrum leads to slightly biased results (particularly for~$\omega_{\rm cdm}$). 
This is in line with what was found in other similar analyses when varying all cosmological parameters~\cite{Ivanov:2021kcd,DAmico:2022osl}, even though the analyses which vary~$A$ only are usually unbiased in real space at these scales~\cite{Eggemeier:2021cam,Oddo:2021iwq,Rizzo:2022lmh}.\footnote{It is important to keep in mind that this is a toy example, with a particular type of the nonlinear term. This makes connection to the real world examples nontrivial.} 
Inclusion of the higher order terms in perturbation theory solves this problem. 
As shown in Fig.~\ref{fig:G2}, the one-loop bispectrum leads to completely unbiased results. 
We will find similar behavior in several analyses below (particularly in redshift space). 
However, since the main goal of our power spectrum and bispectrum analyses is to compare the error bars on cosmological parameters to FLI, we do not have to worry about the exact match of the data and perturbation theory predictions. 
Crucially, the error bars from the tree-level bispectrum and one-loop bispectrum are almost identical. 
This is not surprising, given that in our simple models there are no new free parameters in the one-loop model. 
In reality, the situation would be different and one would have to include all cubic fields with new cubic bias parameters in order to compute the one-loop bispectrum. 
However, in that case the same cubic term would have to be added to the field-level analysis as well. 

\begin{figure}[t]
    \centering
    {$k_\text{max} = 0.12~h/\text{Mpc}$\par\vspace{1ex}}
    \begin{subfigure}{0.49\textwidth}
        \centering
        \includegraphics[width=\textwidth]{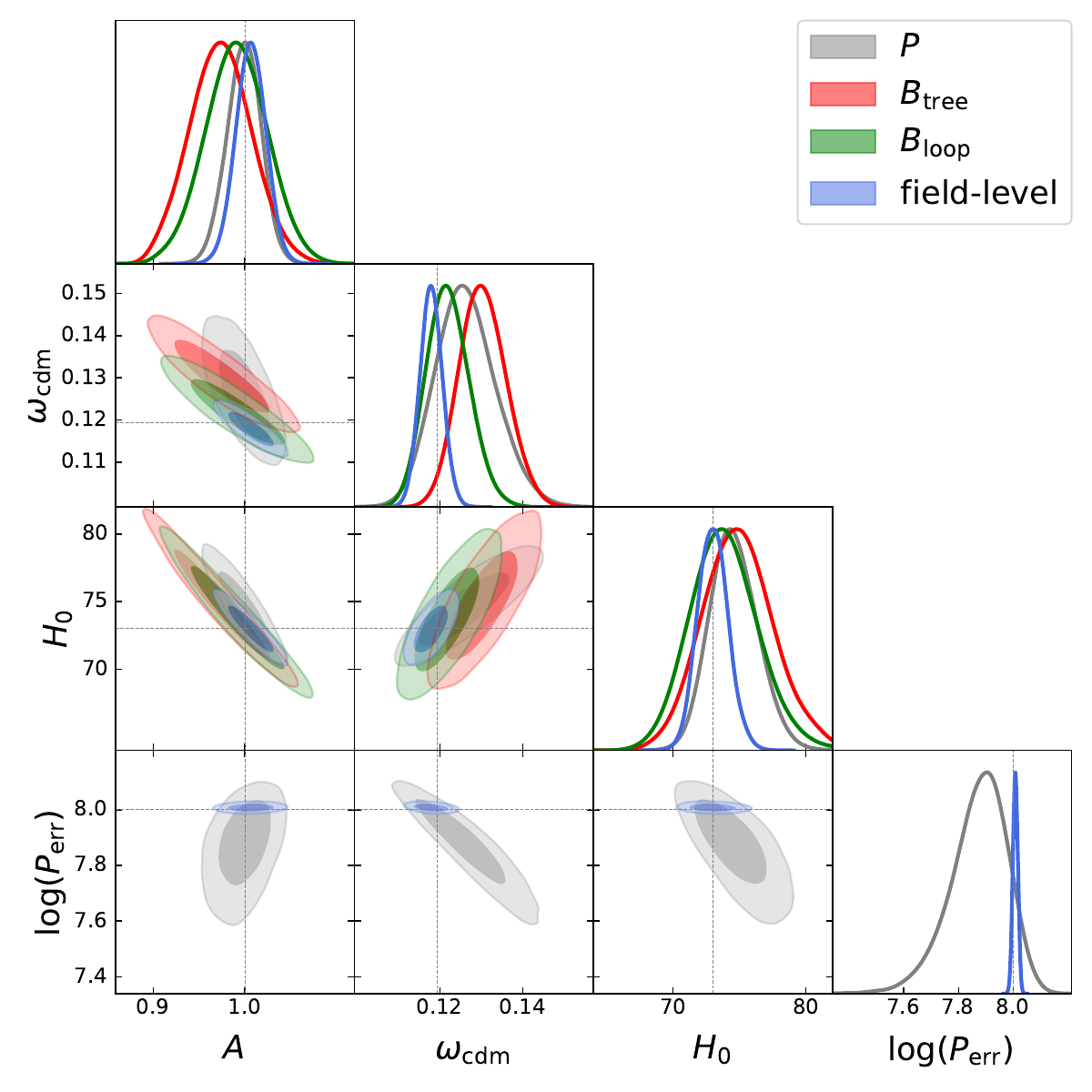}
    \end{subfigure}
    \begin{subfigure}{0.49\textwidth}
        \centering
        \includegraphics[width=\textwidth]{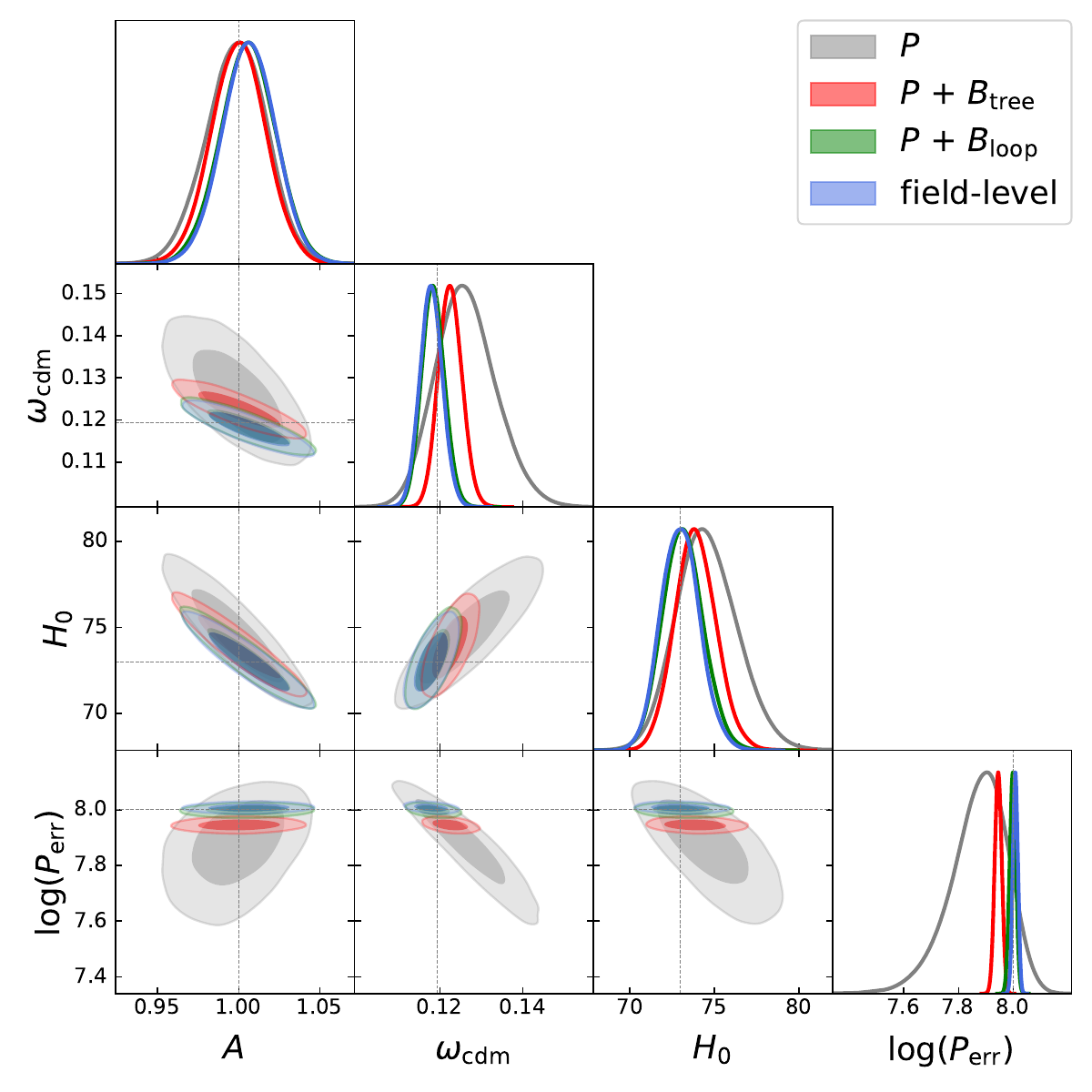}
    \end{subfigure} 
    \caption{The both panels show the 2D marginalized posterior distributions for the simple non-Gaussian example when varying the cosmological parameters and the noise amplitude, 
    fixing the linear bias. 
    In the both panels,  the constraints from the power spectrum are shown in gray, and the field-level in blue.
    In the left, the red and green constours show the constraints from the tree-level bispectrum alone and the one-loop bispectrum alone, respectively.
    In the right, the red and green contours correspond to the joint power spectrum with the tree-level bispectrum and with the one-loop bispectrum, respectively (although the green contours almost completely overlap with the blue ones).
    }
    \label{fig:G2_noise}
\end{figure}

\subsubsection{Varying the cosmological parameters and the noise amplitude, fixing the linear bias}

Next we ask what happens when varying the noise power spectrum amplitude $P_\text{err}$ while still fixing the linear bias $b_1$.
Fig.~\ref{fig:G2_noise} illustrates the resulting posterior constraints, and it is evident that the power spectrum analysis (depicted by the gray contours) is no longer optimal in this case.
A comparison between the field-level posterior (shown in blue) and the power spectrum posterior reveals that the loss of optimality in the power spectrum analysis is driven by a degeneracy between the noise amplitude and the other cosmological parameters, particularly~$\omega_{\rm cdm}$ and~$H_0$ (note that in this example the power spectrum and the field-level constraints on~$A$ are almost identical).  
This behavior is expected because the power spectrum contains contributions from both the signal and the noise, whereas the bispectrum is sensitive only to the signal, as described in Eq.~\eqref{eq:P_B_G2}.  
Consequently, by including the bispectrum information, the field-level inference is able to break the degeneracy between the signal and the noise, thereby yielding improved cosmological constraints.  
In fact, the left panel of Fig.~\ref{fig:G2_noise} demonstrates that the field-level constraints effectively represent the intersection of the power spectrum results and the one-loop bispectrum results (illustrated by the green contours).

This result implies that the joint power spectrum and bispectrum analysis could give constraints equivalent to those obtained from the full field-level analysis.  
To verify this, we perform a joint analysis of the power spectrum and bispectrum, the results of which are shown in the right panel of Fig.~\ref{fig:G2_noise}. There are two important observations that deserve a closer discussion. 

First, the combination of the one-loop power spectrum and tree-level bispectrum leads to the posterior that is very similar to the one in the field-level analysis. As discussed in Sec.~\ref{sec:model} (see Ref.~\cite{Cabass:2023nyo} for more details), this is expected in perturbation theory at leading order in the variance of the density field. 
The nontrivial assumption in the derivation of this result is the validity of the inverse perturbative model which gives the linear density field~$\delta_1$ in terms of the observed galaxy density field~$\hat\delta_g$. 
If we assume that the inverse model has similar radius of convergence as the forward model, given that all computations are done in perturbation theory, we expect that the equivalence of the two posteriors is correct up to~$\Delta^2(k_{\rm max})$. 
This is exactly what we see at the right panel of Fig.~\ref{fig:G2_noise}. 
The shifts in the best fit parameters and the size of the error bars are within~$5-10\%$. 
This is an important confirmation of the theoretical expectations and reinforces the simple connection between the information in the field and the leading~$n$-point functions. 

The second important point is that shifts in the best-fit parameters between the field-level and the joint power spectrum and tree-level bispectrum posteriors can be comparable to the estimated error bars, even though they are small (controlled by~$\Delta^2(k_{\rm max})$). 
This can lead to 1-2 sigma biased results, as can be seen for~$\omega_{\rm cdm}$ and~$\log\,P_{\rm err}$ in the $P+B_{\rm tree}$ analysis in the left panel of Fig.~\ref{fig:G2_noise}.
\footnote{The shift of the best-fit values can be approximated as $\delta\theta_\alpha = \sum_\beta (F^{-1})_{\alpha\beta}\sum_{ij}\frac{\partial m_i}{\partial\theta_\beta} (C^{-1})_{ij} \Delta m_j$,
where $\Delta m_i$ represents the model mismatch, which in our case is zero for the power spectrum bins and equal to $B_{\rm loop}$ for the bispectrum bins.
One can show that $|\delta \theta_\alpha| \leq \sqrt{(F^{-1})_{\alpha\alpha}}\cdot{\rm FoB}$ with ${\rm FoB} \equiv \sqrt{\sum_{\alpha\beta}\delta\theta_\alpha F_{\alpha\beta}\delta\theta_{\beta}}$ defined as the figure of bias.
Furthermore, the FoB is bounded by ${\rm FoB}\leq {\rm SNR}(B_{\rm loop})$, since
${\rm SNR}^2(B_{\rm loop}) = \sum_{ij} B^i_{\rm loop}(C^{-1})_{ij}B^j_{\rm loop} = \sum_{ij} r_i (C^{-1})_{ij}r_j + {\rm FoB}^2$ with $r_i$ is the residual, defined as $r_i \equiv \Delta m_i - \frac{\partial m_i}{\partial \theta_\alpha}\delta\theta_\alpha = B^i_{\rm loop} - \frac{\partial B^i}{\partial \theta_\alpha}\delta\theta_\alpha$.
Thus, the fractional shift of the best-fit parameter is bounded by ${\rm SNR}(B_{\rm loop})$: $|\delta\theta_\alpha|/\sigma_{\alpha} \leq {\rm SNR}(B_{\rm loop})$.
Given that the SNR of the total bispectrum is $\sim {\mathcal O}(10)$, we expect ${\rm SNR}(B_{\rm loop})$ to be of order unity, due to the suppression of $\Delta^2(k_{\rm max})$.
The shift of the best-fit values observed in Fig.~\ref{fig:G2_noise} is consistent with this upper bound.
\label{footnote}
}
Clearly, including higher order corrections such as one-loop bispectrum resolves these biases. Indeed, the joint~$P+B_{\rm loop}$ posterior in Fig.~\ref{fig:G2_noise} almost perfectly agrees with the field-level analysis. 
It is important to stress again that the error bars, which is the main focus or our paper, remain the same since the one-loop bispectrum does not have new parameters in our simple model. 
This is similar to the analysis in the previous section. 
Therefore, we can conclude that in this example the FLI and the standard joint power spectrum and bispectrum analysis lead to almost indistinguishable results, both for the best fit parameters and the error bars. This also indicates that the information in the higher-order~$n$-point functions is very small,~$\mathcal O(10\%)$, in agreement with our estimate of the SNR in Eq.~\eqref{eq:SNR_A_Ttree}.

\begin{figure}[t]
    \centering
    {$k_\text{max} = 0.12~h/\text{Mpc}$\par\vspace{1ex}}
    \begin{subfigure}{0.49\textwidth}
        \centering
        \includegraphics[width=\textwidth]{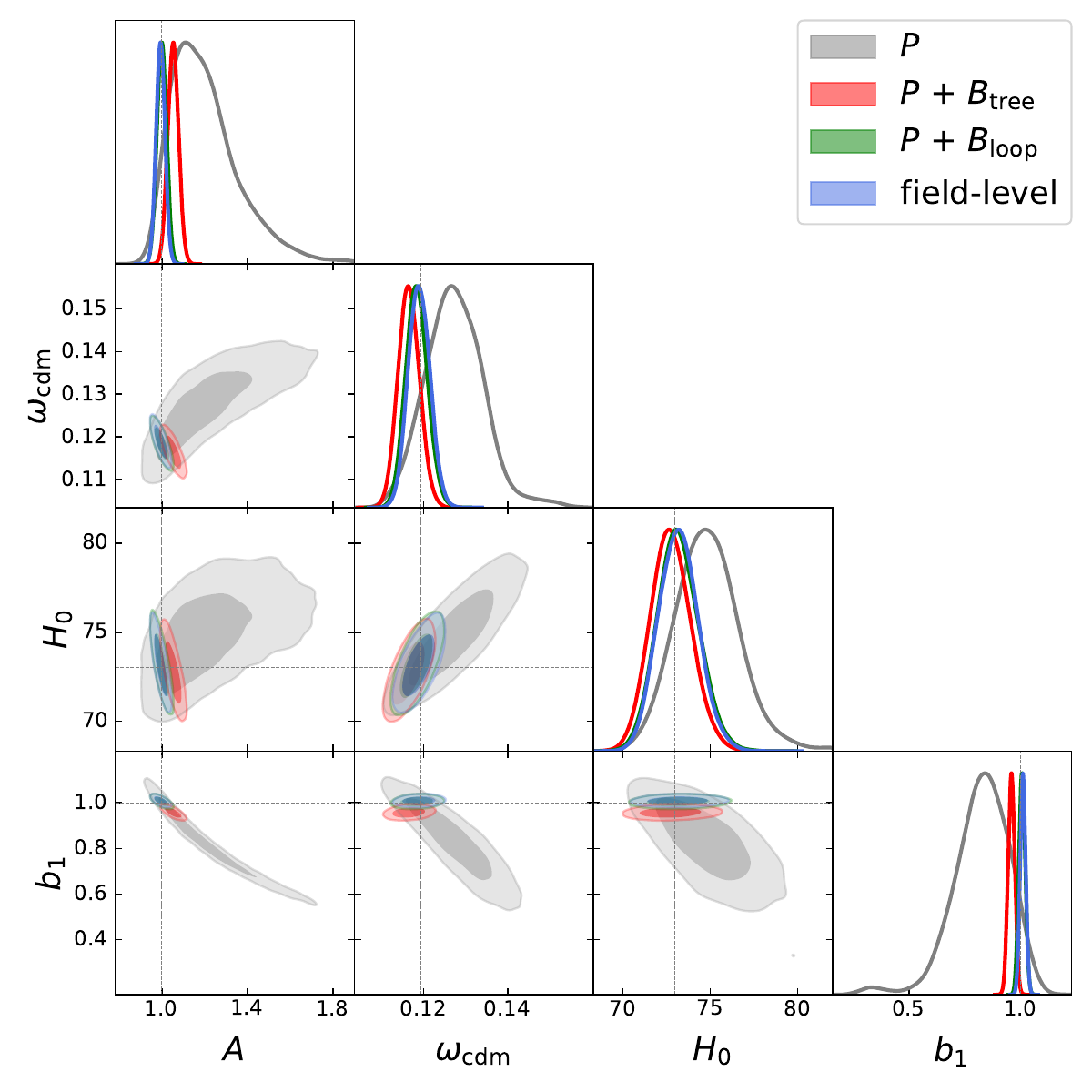}
    \end{subfigure}
    \begin{subfigure}{0.49\textwidth}
        \centering
        \includegraphics[width=\textwidth]{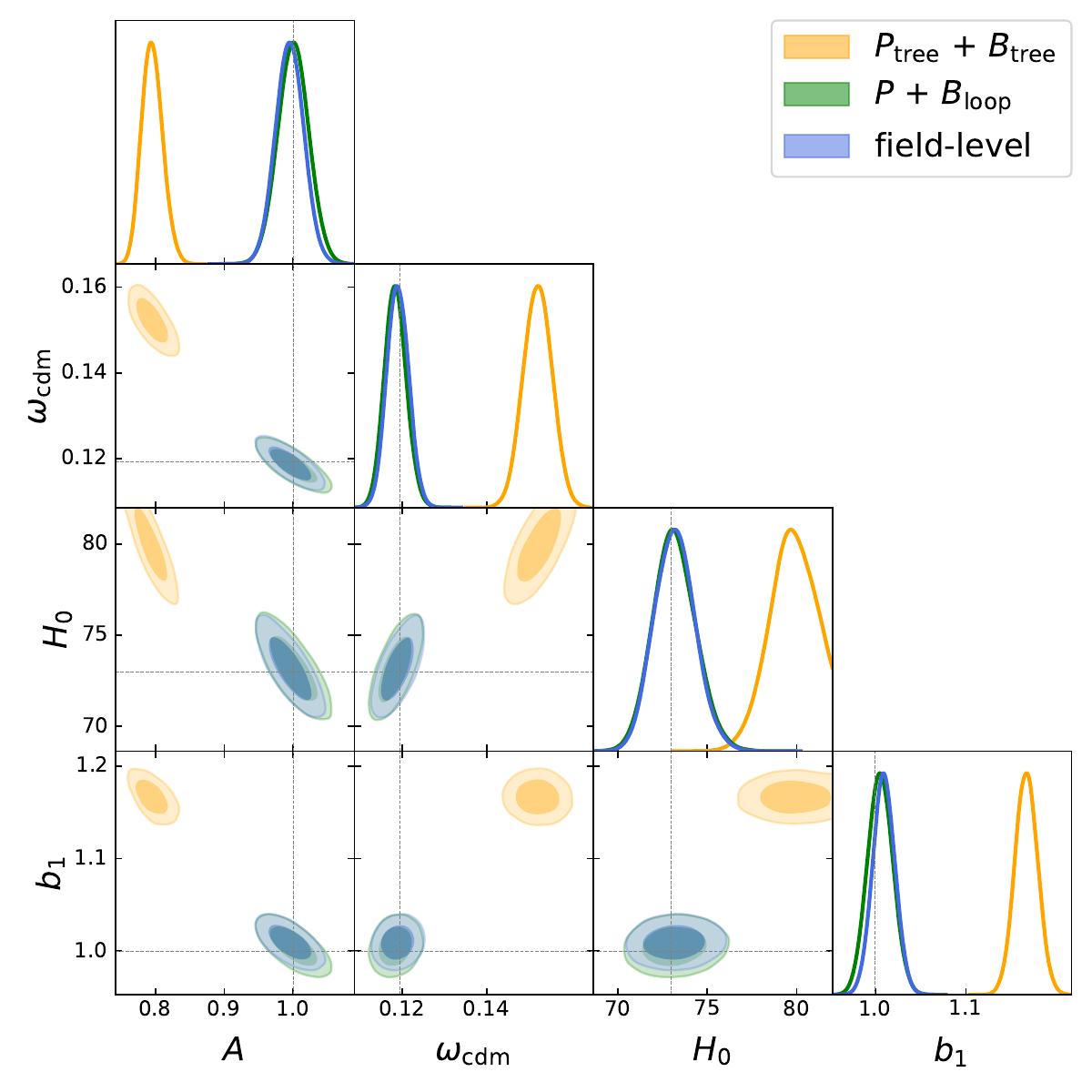}
    \end{subfigure} 
    \caption{The both panels show the 2D marginalized posterior distributions for the simple non-Gaussian example when varying the cosmological parameters and the linear bias, 
    fixing the noise amplitude.
    The gray contours show the constraints from the power spectrum alone, the red from the joint power spectrum and tree-level bispectrum, the green from the joint power spectrum and one-loop bispectrum,
    and the blue from the field-level.
    The orange contours in the right panel show the constraints from the joint tree-level power spectrum plus the tree-level bispectrum, which is quite biased, yet the error bars are similar to the field-level constraints.
    }
    \label{fig:G2_b1}
\end{figure}

\subsubsection{Varying the cosmological parameters and the linear bias, fixing the noise amplitude}
\label{subsubsec:G2_all}

\begin{figure}[t]
    \centering
    {$k_\text{max} = 0.12~h/\text{Mpc}$\par\vspace{1ex}}
    \includegraphics[width=0.5\textwidth]{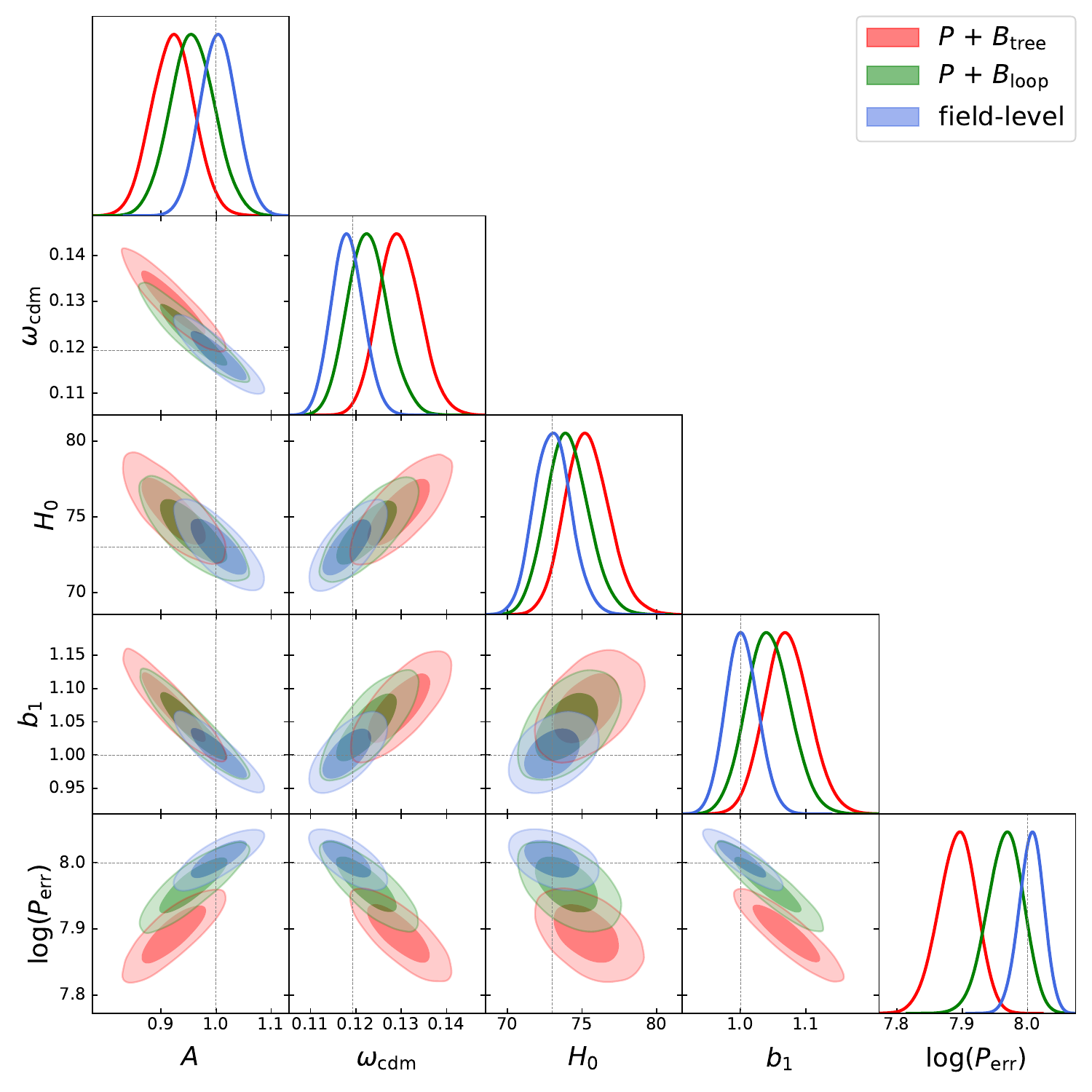}
    \caption{The 2D marginalized posterior distributions for the simple non-Gaussian example when varying the cosmological parameters, the linear bias, and the noise amplitude.
    The red contours show the constraints from the joint power spectrum and tree-level bispectrum, 
    and the green contours show the constraints when including one-loop bispectrum.
    The field-level constraints are shown in blue.}
    \label{fig:G2_b1_noise}
\end{figure}

We now turn to a more realistic but still simplified scenario in which the linear bias $b_1$ is varied alongside the cosmological parameters.  
To simplify the interpretation, we fix the noise power spectrum amplitude. 
Even though we still vary only one nuisance parameter, the degeneracy pasterns in this example are very different from the previous case with fixed~$b_1$ and varied~$P_{\rm err}$.
For instance, while previously~$A$ was constrained almost equally well by the power spectrum alone and the field-level analysis, in this example~$A$ and~$b_1$ are very strongly degenerate. This degeneracy is exact at the level of the linear power spectrum.
The only hope to break this degeneracy stems from nonlinear effects, which, in our simplified model, are captured solely by the contribution~\(P_{\cG\cG}(k)\) when relying on the power spectrum alone. 
However, as we argued in Sec.~\ref{sec:model}, the SNR for the amplitude~$A$ from the one-loop power spectrum is very small and we expect the power spectrum analysis to lead to very large error bars. 
This is indeed what we find in the left panel of Fig.~\ref{fig:G2_b1}.
\footnote{Only in this example, we adopt a broader prior for $A$ than the default, namely $A\sim {\cal U}[0.5, 2.5]$.
}

In this particular example, the~$b_1-A$ degeneracy is mainly broken by the bispectrum. Unlike in the previous case where we varied only~$P_{\rm err}$, here the inclusion of the bispectrum leads to a dramatic improvement of the error bars. 
All other features of various analyses we do are the same as before. The posterior for the joint power spectrum and bispectrum analysis agrees with the FLI posterior at the expected level, with the one-loop bispectrum fixing small biases from the analysis involving only the tree-level bispectrum. 
Also, in the same way as before, even though our simple nonlinear model generates a non-zero trispectrum along with other higher-order $n$-point functions (which depend on~$A$ and~$b_1$),
these higher-order contributions do not significantly affect the overall constraints, neither in terms of the central values nor the error bars.
 
To further illustrate this point, we do the analysis in which we combine only the linear power spectrum (without the one-loop correction) with the tree-level bispectrum. The results are  in the right panel of Fig.~\ref{fig:G2_b1}.
One can see that the error contours nearly coincide with those from the full field-level analysis (or the joint one-loop power spectrum and one-loop bispectrum analysis), although the mean values exhibit a noticeable bias.
\footnote{
This large shift in the best-fit value is again consistent with the upper bound discussed in footnote~\ref{footnote}.
In this case, the upper bound is dominated by ${\rm SNR}(P_{\rm loop})$, which is of order ${\cal O}(10)$.
}
This implies that the magnitude of the error bars in the simple model we consider is almost entirely determined by the leading nonlinear effects encoded in the tree-level bispectrum, 
whereas the higher-order terms and higher-order $n$-point functions mainly serve to shift the central values toward unbiased ones.

\subsubsection{Varying the cosmological parameters, the linear bias, and the noise amplitude}

\begin{table}[t]
    \centering
    \begin{tabular}{c|c|c|c|c|c}
        & $A$ & $\omega_{\rm cdm}$ & $H_0$ & $b_1$ & $\log(P_\text{err})$ \\ \hline
        $P+B$ & $0.957\pm0.041$ & $0.1227^{+0.0041}_{-0.0047}$ & $74.0^{+1.3}_{-1.5}$ & $1.043\pm0.034$ & $7.966^{+0.029}_{-0.026}$\\ \hline
        field-level & $1.004\pm0.034$ & $0.1182\pm0.0035$ & $73.0^{+1.1}_{-1.3}$ & $1.003\pm0.025$ & $8.006^{+0.020}_{-0.018}$\\
    \end{tabular}
    \caption{The mean values and 68\%-confidence intervals of the parameters for the simple Eulerian example.}
    \label{tab:G2_b1_noise}
\end{table}

Finally, we examine the scenario in which all relevant parameters (namely, the three cosmological parameters $\{A, \omega_{\rm cdm}, H_0\}$, the linear bias $b_1$, and the noise amplitude $P_{\rm err}$) are allowed to vary.  
In this case, breaking the degeneracies between these parameters using the power spectrum alone becomes nearly impossible.  
Therefore, we compare the field-level analysis with the joint power spectrum and bispectrum analysis.  
The results of this comparison are presented in Fig.~\ref{fig:G2_b1_noise}.

There are several interesting observations to make. First, the field-level and joint power spectrum and tree-level bispectrum posteriors are more different than before (since we vary more nuisance parameters), but these differences are still within~$\sim 20\%$ as expected in perturbation theory. 
This is true both for the central values (for example, the best-fit values of the amplitude differ by approximately~$8\%$) and the marginalized error bars. 
In the volume we consider, this difference is large enough to lead to biased estimate of the cosmological parameters, suggesting that higher-loop contributions have to be taken into account. 
Indeed, the inclusion of the one-loop bispectrum leads to unbiased results and improves the agreement with the field-level posterior. 
While the best-fit values are significantly closer, the error bars still differ by approximately~$20\%$. 
These results are summarized in Table~\ref{tab:G2_b1_noise}. 
Note that the only slightly more significant improvement from the FLI is for the amplitude of the noise, where the relative error is approximately~$30\%$ tighter. 
As we are going to see, this will remain the trend also in the more complicated setups that we will study in the next section. 

It is important to emphasize that even though the FLI leads to tighter error bars, they are not significantly smaller than in the joint power spectrum and bispectrum analysis. 
Given the error bars, one can argue that most of the information about cosmology on large scales is in the power spectrum and bispectrum. 
This is in line with the expectation from perturbation theory, where the improvement is controlled by the variance of the density field~$\Delta^2(k_{\rm max})$. 
Beyond the power spectrum and bispectrum, the remaining signal is in higher order~$n$-point functions. 
The trispectrum is expected to be the most relevant among them. 
Indeed, our results are in agreement with the typical improvements due to the inclusion of the trispectrum in the correlation function analyses~\cite{Spezzati:2025zsb}.

In conclusion, in this section we tested FLI for a simple Eulerian forward model and found that all results are in agreement with expectations. 
In particular, FLI recovers the unbiased cosmology with error bars not significantly different from the standard joint power spectrum and bispectrum analysis. 
This is in agreement with expectations in perturbation theory and the SNR estimates in Sec.~\ref{sec:model}.

\section{Comparison with the joint power spectrum and bispectrum analysis II: LPT-based model}
\label{sec:LPT}
In the previous section, we have shown that for the simple Eulerian model the field-level analysis yields constraints nearly equivalent to those obtained from the joint power spectrum and bispectrum analysis. 
While the simplicity of the model allowed for easy analytical predictions and straightforward implementation of different analyses, one may wonder if the lessons learned in this baseline setup are general enough. 
After all, the true dynamics includes large displacements not properly treated in Eulerian perturbation theory. 
Furthermore, the real galaxy density field is related to the dark matter in more complicated ways, involving more free bias parameters. It was argued in Ref.~\cite{Spezzati:2025zsb} that these additional complications cannot significantly affect the conclusions derived in Eulerian perturbation theory. 
In a nutshell, the main reasons are that on large scales (for~$k\lesssim0.1\,h/{\rm Mpc}$), for the volumes we consider in this paper, Eulerian and Lagrangian perturbation theory are practically indistinguishable and that the bias expansion, even with new contributions, is always controlled by the same small perturbation theory parameter~$\Delta^2(k_{\rm max})$. 

In this section, we test these arguments using an LPT-based model that incorporates the quadratic bias terms, described in Sec.~\ref{subsec:LPT_model}. The only possibilities that conclusions from the previous section are modified are: 
(a) the correct gravitational evolution plays an important role (the displacement field could enhance (or degrade) the constraints), and 
(b) adding higher-order bias terms significantly degrades the joint power spectrum and bispectrum analyses compared to FLI. In order to test these two hypotheses separately, we will use two different sets of LPT-based mock data, with only linear bias and all biases included. 
Furthermore, following Sec.~\ref{subsec:LPT_model}, we will do the analyses in real and redshift space. As expected, the degeneracy breaking is different in these two examples and we will explicitly check it. 

Due to the larger computational complexity associated with evaluating the one-loop bispectrum, we restrict our analysis here to the tree-level bispectrum. As we have already shown, including higher order terms can only shift the best-fit values and does not affect the error bars given that loops do not come associated with new nuisance parameters. 
In order to be conservative, we will restrict the range of scales to~$k_{\rm max} = 0.1\,h/\mathrm{Mpc}$. The same scale cuts have been used in other papers that did similar field-level analyses~\cite{Beyond-2pt:2024mqz,Nguyen:2024yth}. 
In addition, we will also provide the joint power spectrum and bispectrum Fisher forecast for each analysis we do, in order to test the accuracy of the Fisher matrix and to make comparison to previous literature when appropriate.

\begin{figure}[t]
    \centering
    {$k_\text{max} = 0.1~h/\text{Mpc}$\par\vspace{1ex}}
    \includegraphics[width=0.5\textwidth]{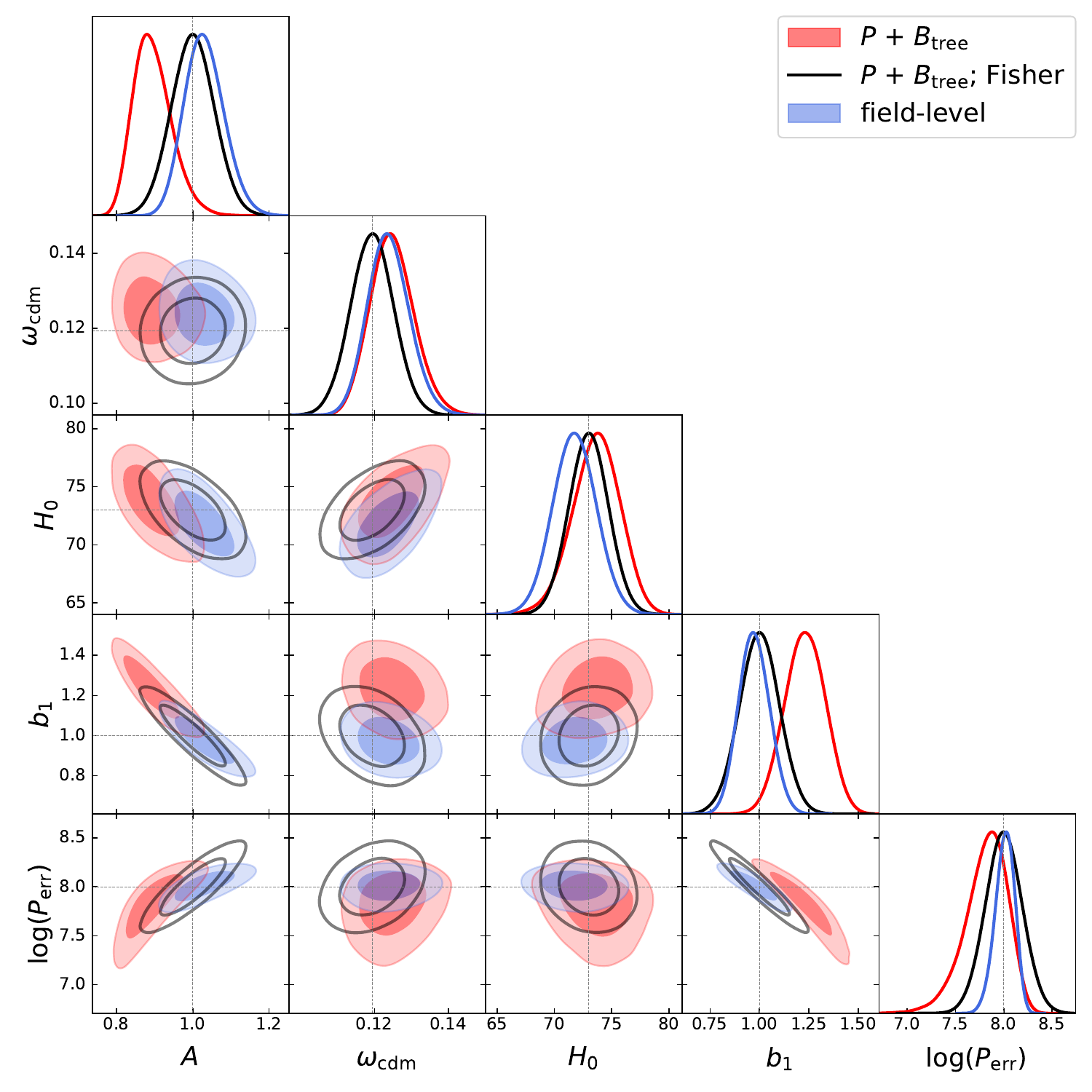}
    \caption{The 2D marginalized posterior distributions for the LPT-based model with the linear bias.
    The red contours show the constraints from the joint power spectrum and tree-level bispectrum, 
    and the blue contours show the constraints from the field-level analysis.
    The black lines indicate the inverse Fisher matrix estimate for the error contours for the joint power spectrum and the bispectrum analysis.
    Note that we also vary $c_0$ in addition to the parameters we show here.
    }
    \label{fig:LPT_real_space_lin}
\end{figure}

\subsection{Real space}
\label{subsec:LPT_real}

We begin by considering the real-space case before addressing redshift space.
First, we assess the importance of the gravitational evolution and the impact of large displacements.
To this end, we compare the field-level analysis with the joint power spectrum and bispectrum analysis for the LPT-based model in real space, employing only the linear bias.  
Here in addition to the cosmological parameters $\{A,\omega_{\rm cdm},H_0\}$, the linear bias $b_1$, and the noise power spectrum amplitude $P_{\rm err}$, we introduce a counter term coefficient (or derivative bias) $c_0$ in order to avoid a biased result
(see Sec.~\ref{sec:model} and Sec.~\ref{subsec:resolution_mismatch_signal} for further details).
Figure~\ref{fig:LPT_real_space_lin} summarizes the main result: the joint power spectrum and tree-level bispectrum analysis (displayed in red) is directly compared with the field-level analysis (shown in blue), with a focus on the cosmological parameters, linear bias, and noise amplitude.  
Additionally, the Fisher matrix estimate for the error contours of the joint power spectrum and bispectrum analysis is overlaid as black lines, and selected examples of the full 2D marginalized posterior distributions are provided in App.~\ref{app:full_posterior}.

First, one can see that the joint power spectrum and tree-level bispectrum analysis yields biased results relative to the true input values, which is similar to the behavior observed in the simple Eulerian model.  
Although a slightly lower \(k_{\rm max}\) is used in this case, the tree-level bispectrum proves insufficient to accurately capture the bispectrum at $k = 0.1\,h/\mathrm{Mpc}$ (see Fig.~\ref{fig:bispec} in App.~\ref{app:bispectrum}).
While one might expect that incorporating the one-loop bispectrum would eliminate this bias, 
the important point here is that the overall size of the error bars would remain unchanged.

Second, we find that the constraints from the field-level analysis are almost consistent with those obtained from the joint power spectrum and bispectrum analysis, particularly for the cosmological parameters.  
While the field-level approach provides a notably tighter constraint on the noise amplitude, this improvement is likely a consequence of the simplistic Gaussian noise model employed, which does not fully reflect the reality. 
Thus, it follows that neither the exact gravitational evolution (bispectrum model is based on EPT instead of LPT) nor large displacement field significantly influences the results
when comparing field-level analysis to the joint power spectrum and bispectrum analysis, 
specifically for the cosmological parameters we are primarily interested in.

\begin{figure}[t]
    \centering
    {$k_\text{max} = 0.1~h/\text{Mpc}$\par\vspace{1ex}}
    \begin{subfigure}{0.49\textwidth}
        \centering
        \includegraphics[width=\textwidth]{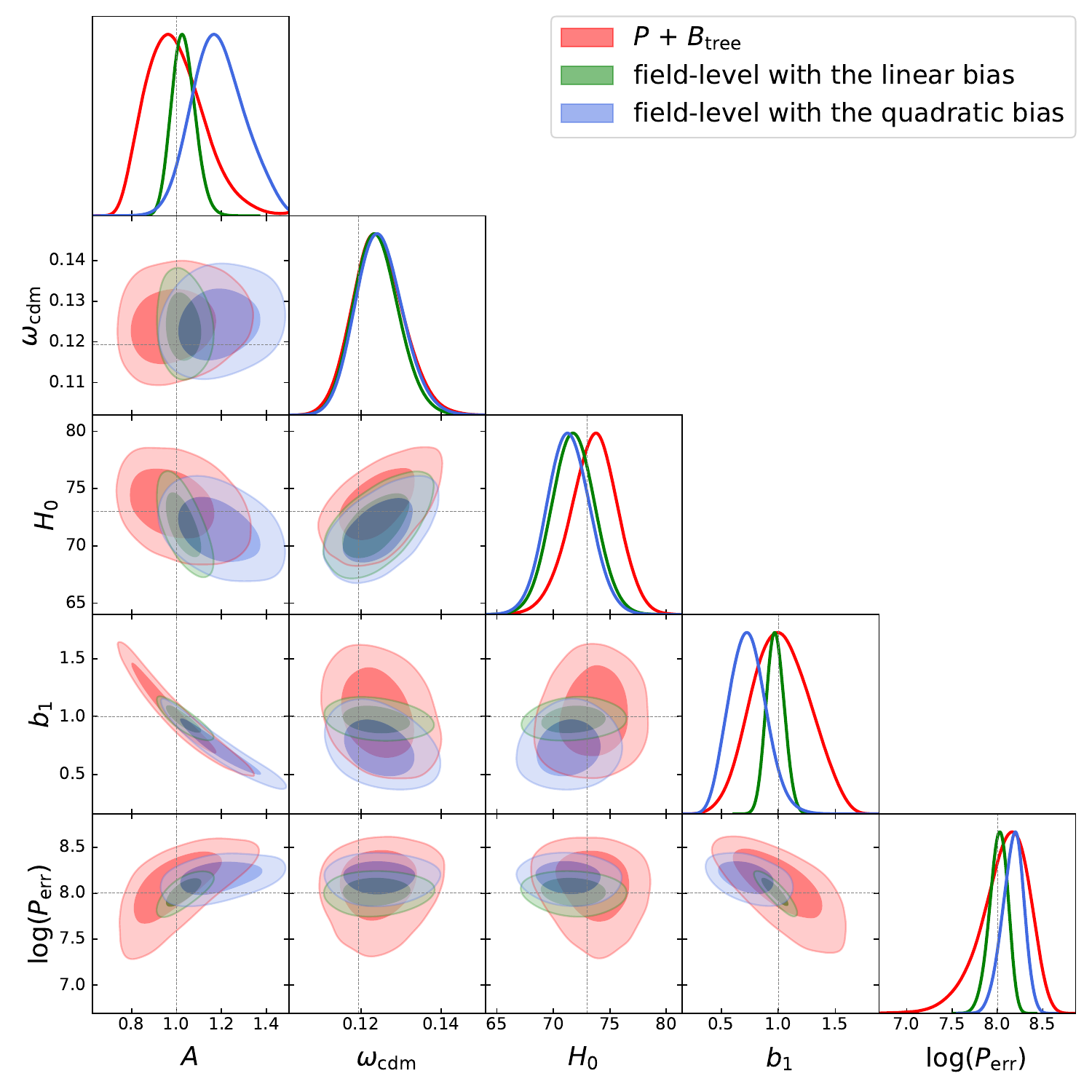}
    \end{subfigure}
    \begin{subfigure}{0.49\textwidth}
        \centering
        \includegraphics[width=\textwidth]{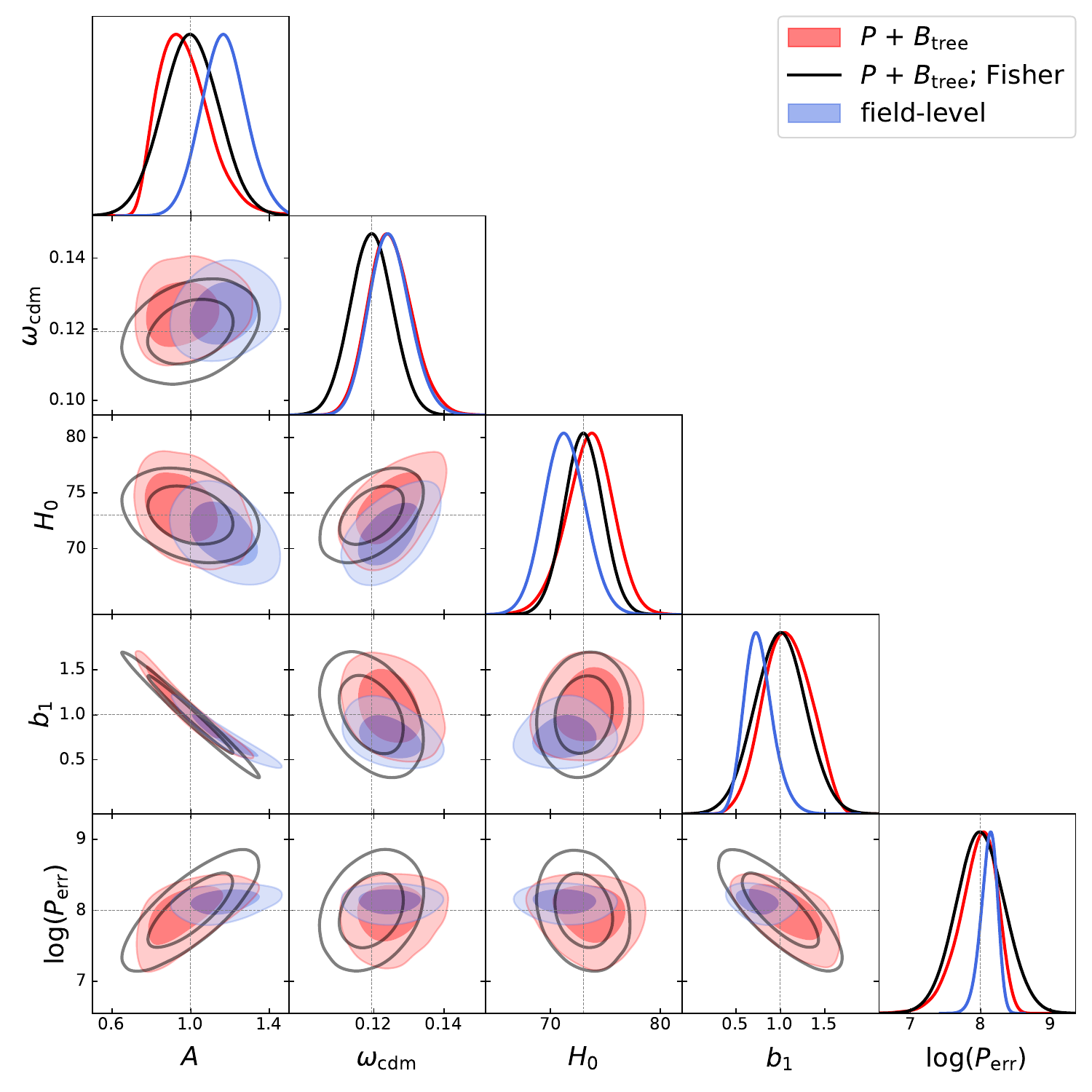}
    \end{subfigure} 
    \caption{The 2D marginalized posterior distributions for the LPT-based model with the quadratic bias in real space.
    The left panel shows the result when the mock is generated with the linear bias only, while the right panel shows the result when the mock is generated with the non-zero quadratic bias.
    All the contours show the constraints with the quadratic bias parameter included in the inference model, 
    except for the green contours in the left panel, which presents the constraints with only the linear bias included in the field-level analysis (i.e., the same posterior as the blue contours in Fig.~\ref{fig:LPT_real_space_lin}).
    The constraints from the joint power spectrum and tree-level bispectrum are shown in red, and the field-level in blue.
    The black lines indicate the inverse Fisher matrix estimate for the error contours for the joint power spectrum and the bispectrum analysis.
    }
    \label{fig:LPT_real_space_quad}
\end{figure}

We are left with the possibility that the inclusion of the higher-order bias terms could change the situation.
To investigate this, we conduct two analyses: 
(1) The inference model is augmented with quadratic bias terms, while the mock data are generated with only the linear bias (i.e., the same mock as in the previous analysis).
(2) Both the inference model and the mock data include quadratic bias terms.
In both cases we vary the cosmological parameters $\{A, \omega_{\rm cdm}, H_0\}$, the bias parameters $\{b_1, b_2, b_{\cG}, c_0\}$, and the noise power spectrum amplitude $P_\text{err}$.
The main results are shown in Fig.~\ref{fig:LPT_real_space_quad} and the full 2D marginalized posterior distributions are available in Appendix \ref{app:full_posterior}.
In the left panel of Fig.~\ref{fig:LPT_real_space_quad}, corresponding to the case (1), 
we also display the field-level posterior with only the linear bias (the same posterior shown in Fig.~\ref{fig:LPT_real_space_lin}) in green for comparison.

First, now the joint power spectrum and tree-level bispectrum posterior appears unbiased.  
However, this does not imply that the tree-level bispectrum is adequate at this scale (see Fig.~\ref{fig:bispec} in App.~\ref{app:bispectrum}).
What is happening here is that the quadratic bias terms act to absorb the residuals from the tree-level bispectrum, leading to an apparent unbiased result, albeit with larger error bars.  
In fact, the constraints on the quadratic bias parameters themselves become slightly biased (see Fig.~\ref{fig:LPT_quad_all_0.1} and Fig.~\ref{fig:LPT_quad_all_0.12}).

\begin{figure}[t]
    \centering
    \begin{subfigure}{0.49\textwidth}
        \centering
        {$k_\text{max} = 0.1~h/\text{Mpc}$\par\vspace{1ex}}
        \includegraphics[width=\textwidth]{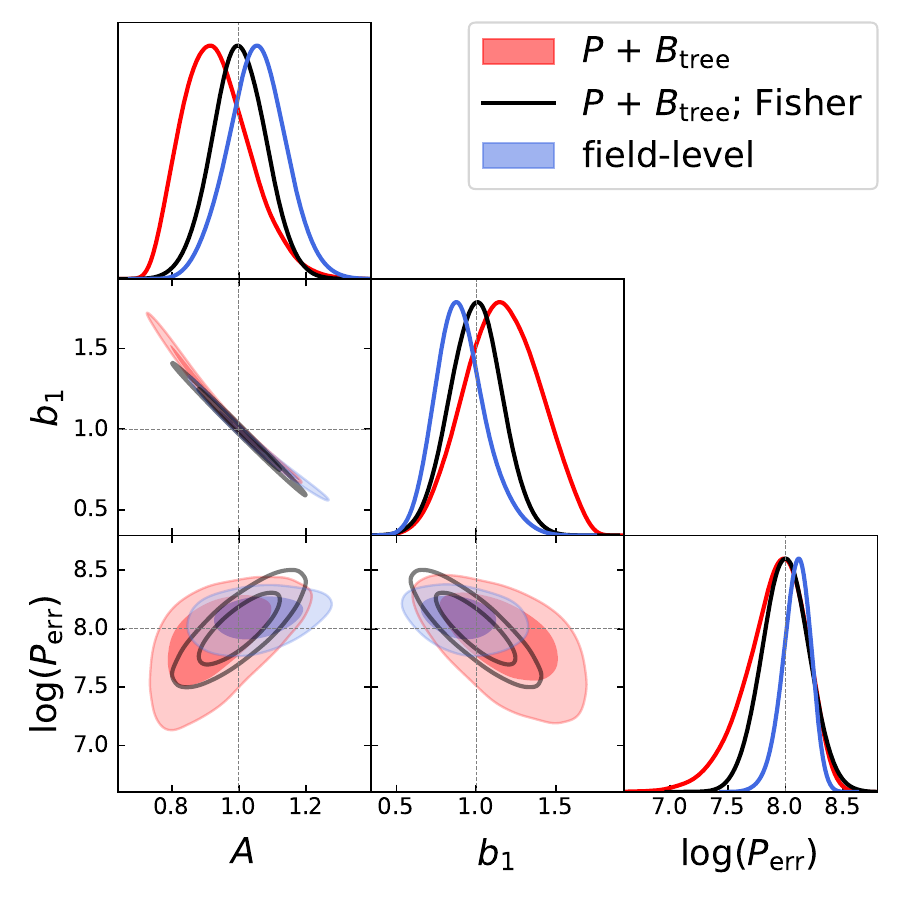}
    \end{subfigure}
    \begin{subfigure}{0.49\textwidth}
        \centering
        \includegraphics[width=\textwidth]{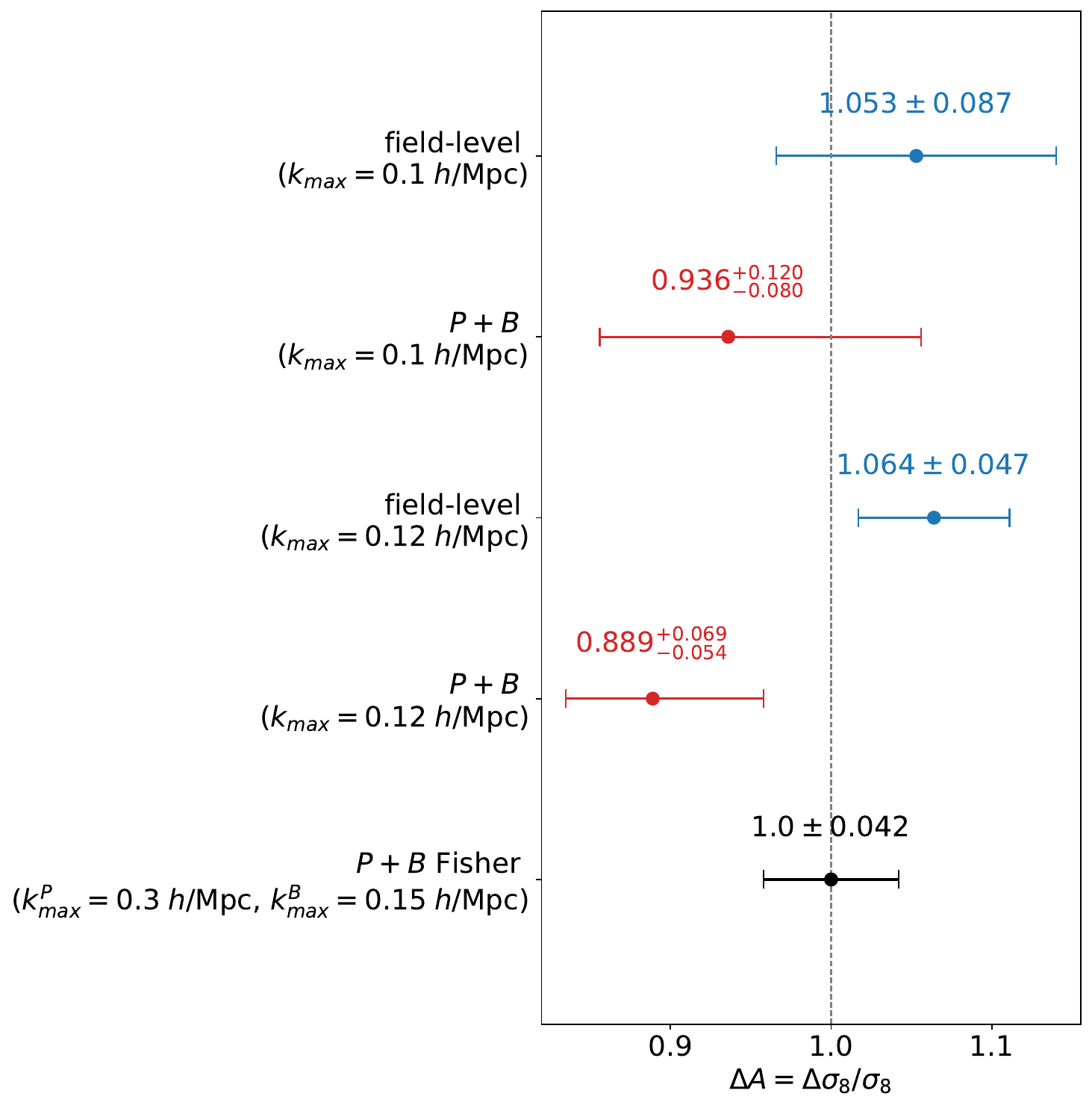}
    \end{subfigure} 
    \caption{\textit{Left panel}: The 2D marginalized posterior distributions for the LPT-based model with the quadratic bias in real space, fixing $\{\omega_{\rm cdm}, H_0\}$.
    The constraints from the joint power spectrum and tree-level bispectrum are shown in red, and the field-level in blue.
    The black lines indicate the inverse Fisher matrix estimate for the error contours for the joint power spectrum and the bispectrum analysis.
    \textit{Right panel}: A comparison of the mean values and 1-$\sigma$ error bars of the linear density field amplitude $A$ while fixing $\{\omega_{\rm cdm}, H_0\}$ (The top two results correspond to the left panel). 
    This setup matches that of Refs~\cite{Beyond-2pt:2024mqz, Nguyen:2024yth, Spezzati:2025zsb}.
    }
    \label{fig:beyond_2pt_comparison}
\end{figure}

Second, and most importantly, even with the inclusion of quadratic bias operators 
the constraints on the cosmological parameters remain consistent between the field-level analysis and the joint power spectrum and bispectrum analysis.  
As summarized in Table~\ref{tab:LPT_real_space_quad}, the constraints derived from the field-level analysis are only $\cO(10)\%$ tighter than those obtained from the joint analysis, 
which is similar to the case for the simple Eulerian model discussed in the previous section.  
Moreover, a comparison with the linear bias case (see the left panel of Fig.~\ref{fig:LPT_real_space_quad}) reveals that while the addition of quadratic bias parameters slightly degrades the constraint on $A$, 
the constraints on the other cosmological parameters ($\omega_{\rm cdm}$ and $H_0$) remain essentially unaffected.
We expect this to happen, since the information on \(\omega_{\rm cdm}\) is largely extracted from the shape of the power spectrum, 
while the information on $H_0$ is extracted from the overall physical scale of the spectra (i.e., $h/\mathrm{Mpc}$ and $(\mathrm{Mpc}/h)^3$).
These properties are less sensitive to the higher-order bias terms.

\begin{table}[t]
    \centering
    \begin{tabular}{c|c|c|c|c|c}
        & $A$ & $\omega_{\rm cdm}$ & $H_0$ & $b_1$ & $\log(P_\text{err})$ \\ \hline
        $P+B_\text{tree}$ & $0.971^{+0.094}_{-0.15}$ & $0.1248^{+0.0056}_{-0.0066}$ & $73.6^{+2.2}_{-2.0}$ & $1.08\pm 0.26$ & $7.95^{+0.33}_{-0.20}$\\ \hline
        field-level & $1.17\pm {0.11}$ & $0.1246^{+0.0054}_{-0.0062}$ & $71.3 \pm 1.9$ & $0.76^{+0.13}_{-0.18}$ & $8.12^{+0.13}_{-0.097}$\\
    \end{tabular}
    \caption{The mean values and 68\%-confidence intervals of the parameters for the LPT-based model with the quadratic bias parameters, 
    corresponding to the right panel of Fig.~\ref{fig:LPT_real_space_quad}.
    }
    \label{tab:LPT_real_space_quad}
\end{table}

Finally let us comment on a comparison with the previous studies~\cite{Beyond-2pt:2024mqz, Nguyen:2024yth}.
The left panel of Fig.~\ref{fig:beyond_2pt_comparison} follows the same analysis as the right panel of Fig.~\ref{fig:LPT_real_space_quad}, except that only the linear amplitude $A$ is varied, while $\omega_{\rm cdm}$ and $H_0$ are held fixed, in accordance with Refs.~\cite{Beyond-2pt:2024mqz, Nguyen:2024yth}.
The right panel shows the mean values and 1-$\sigma$ errors of $A$ for several choices of $k_{\rm max}$, which can be directly compared with Fig.~22 in Ref.~\cite{Beyond-2pt:2024mqz} and Fig.~11 and Tab.~IV in Ref.~\cite{Spezzati:2025zsb}.
Note that our $P+B$ errors correspond to the ``$P+B$, restricted'' case in Ref.~\cite{Beyond-2pt:2024mqz}, since here we do not include the third-order bias as well as the non-Gaussian stochasticity in the bispectrum.
Similarly, the ``EFT FBI, $k_{\rm max}=0.1~h/{\rm Mpc}$'' error bar in Ref.~\cite{Beyond-2pt:2024mqz} should be compared with our field-level result with $k_{\rm max} = 0.12~h/{\rm Mpc}$ 
since the formar case uses the cubic cutoff in Fourier space.

It is clear that the errors on $A$ agree between the field-level and the joint power spectrum and bispecrum analyses when compared with the same $k_{\rm max}$, 
consistent with our discussion so far.
Furthermore, the results shown in the right panel of Fig.~\ref{fig:beyond_2pt_comparison} are also in good agreement with those in Refs.~\cite{Beyond-2pt:2024mqz, Spezzati:2025zsb},
although the detailed comparison is not straightforward due to the different choices of the redshift and the fiducial parameters.
\footnote{
The absolute size of the field-level error bars is larger than that reported in Refs~.\cite{Beyond-2pt:2024mqz, Nguyen:2024yth}, most likely due to differences in the fiducial parameters.
The point here is that the difference between the field-level and $P+B$ error bars is similar to what is found in Refs.~\cite{Beyond-2pt:2024mqz, Spezzati:2025zsb}.
}
The only noticeable exception lies in the ``SBI $P+B$'' results in Ref.~\cite{Nguyen:2024yth}, where the ``SBI $P+B$'' errors on $A$ are larger than the field-level errors by factor of a few,
which seems to be contradictory to the results with the standard $P+B$ analysis in this paper.
A full investigation on the cause of this discrepancy is beyond the scope of this paper, but see the discussion in Ref.~\cite{Spezzati:2025zsb}. 

In summary, these findings lead us to the key conclusion that, even for the more realistic model incorporating both the large displacement field and the higher-order bias terms, 
the field-level analysis yields constraints that are nearly equivalent to those obtained from the joint power spectrum and bispectrum analysis.
We also explicitly show that including the 2LPT displacement does not change the result in App.~\ref{app:2LPT}.

\subsection{Redshift space}
\label{subsec:LPT_RSD}

\begin{figure}[t]
    \centering
    {$k_\text{max} = 0.1~h/\text{Mpc}$\par\vspace{1ex}}
    \begin{subfigure}{0.49\textwidth}
        \centering
        \includegraphics[width=\textwidth]{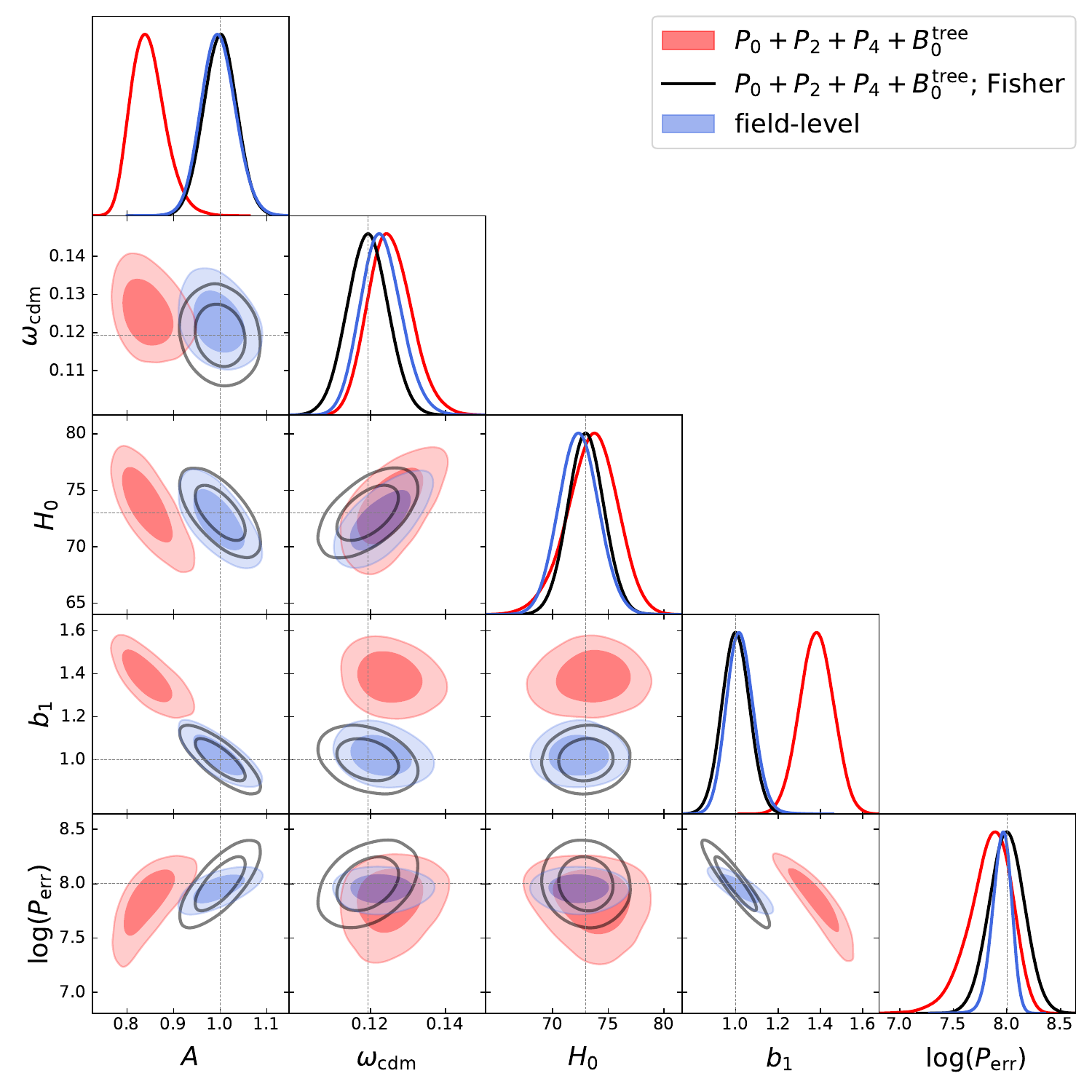}
    \end{subfigure}
    \begin{subfigure}{0.49\textwidth}
        \centering
        \includegraphics[width=\textwidth]{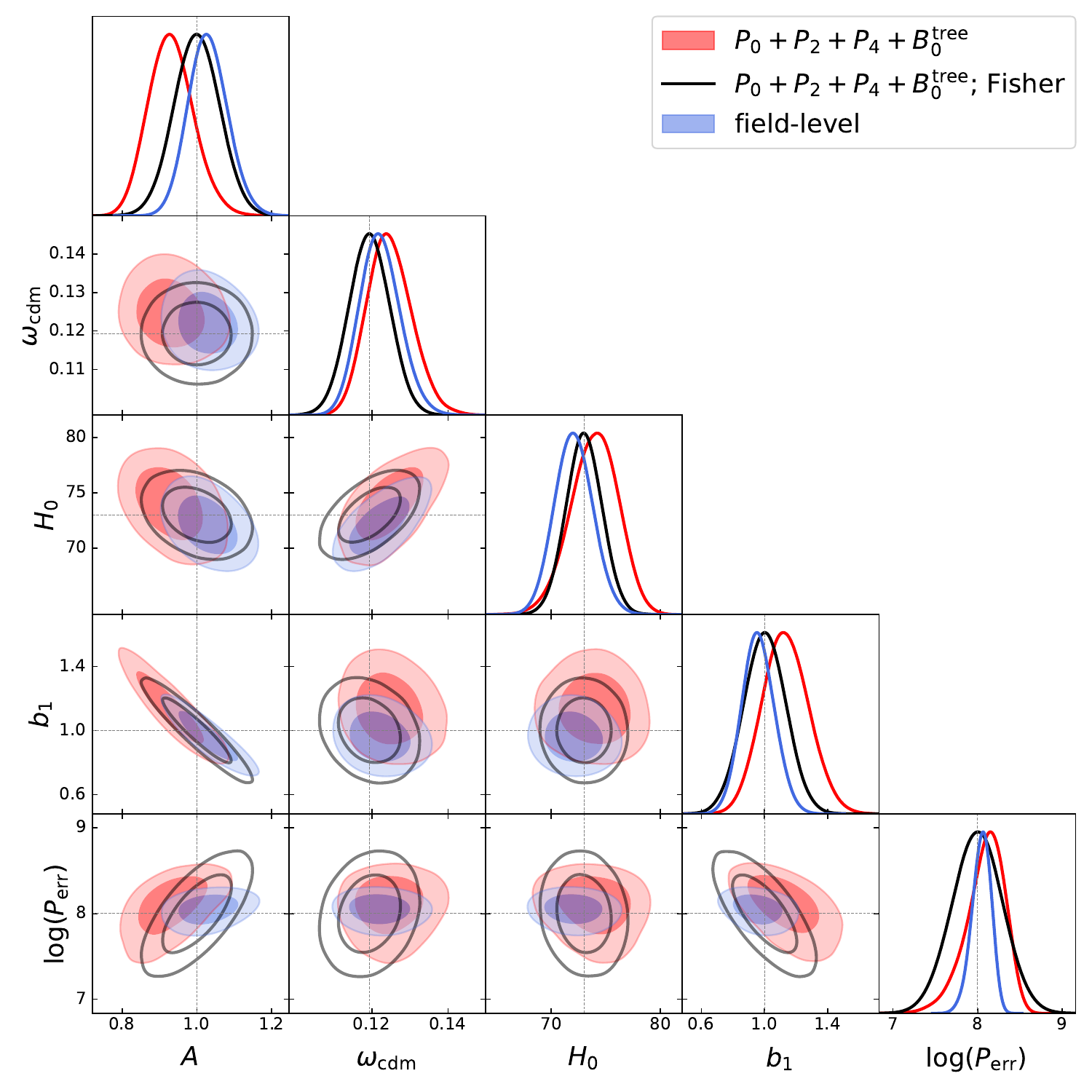}
    \end{subfigure} 
    \caption{The left and the right panels show the 2D marginalized posterior distributions for the LPT-based model with the linear bias and the quadratic bias in redshift space, respectively.
    The constraints from the joint power spectrum and tree-level bispectrum are shown in red, and the field-level in blue.
    The black lines indicate the inverse Fisher matrix estimate for the error contours for the joint power spectrum and the bispectrum analysis.
    }
    \label{fig:LPT_redshift_space_P+B}
\end{figure}

We now extend our analysis to include redshift-space distortions (RSD).
A key difference from the real-space case is the ability of RSD to break the degeneracy between $A$ and $b_1$ at the level of the linear power spectrum, 
owing to its sensitivity to the velocity field, which is expected to be unbiased on large scales.  
In redshift space, the power spectrum is characterized by its multipoles $P_0$, $P_2$, and $P_4$, each of which carries distinct information about $A$ and $b_1$.  
Therefore, it is of particular interest to compare the constraining power of various approaches, including the full field-level analysis, 
the power spectrum analysis based on $\{P_0,\,P_2,\,P_4\}$, and joint analyses that incorporate the bispectrum (e.g., $\{P_0,\,B_0\}$ or $\{P_0,\,P_2,\,P_4,\,B_0\}$).

Let us start with a comparison of the field-level analysis to the joint power spectrum and bispectrum analysis based on the data set $\{P_0, P_2, P_4, B_0\}$.
In Fig.~\ref{fig:LPT_redshift_space_P+B}, we show the 2D marginalized posterior distributions for the LPT-based model in redshift space,
with the left panel corresponding to the case with the linear bias alone and the right panel to that with the quadratic bias.
Tab.~\ref{tab:LPT_rsd_space_quad} provides a summary of the mean values and 68\%-confidence intervals of the parameters for the model that includes quadratic bias terms.  
Once again, we find that the error bars derived from the field-level analysis are comparable to those obtained from the joint analysis, 
reinforcing the conclusion that, in redshift space, the combination of the power spectrum and bispectrum capture almost as much information as the full field-level approach.

\begin{figure}[t]
    \centering
    \begin{subfigure}{0.49\textwidth}
        \centering
        {$k_\text{max} = 0.1~h/\text{Mpc}$\par\vspace{1ex}}
        \includegraphics[width=\textwidth]{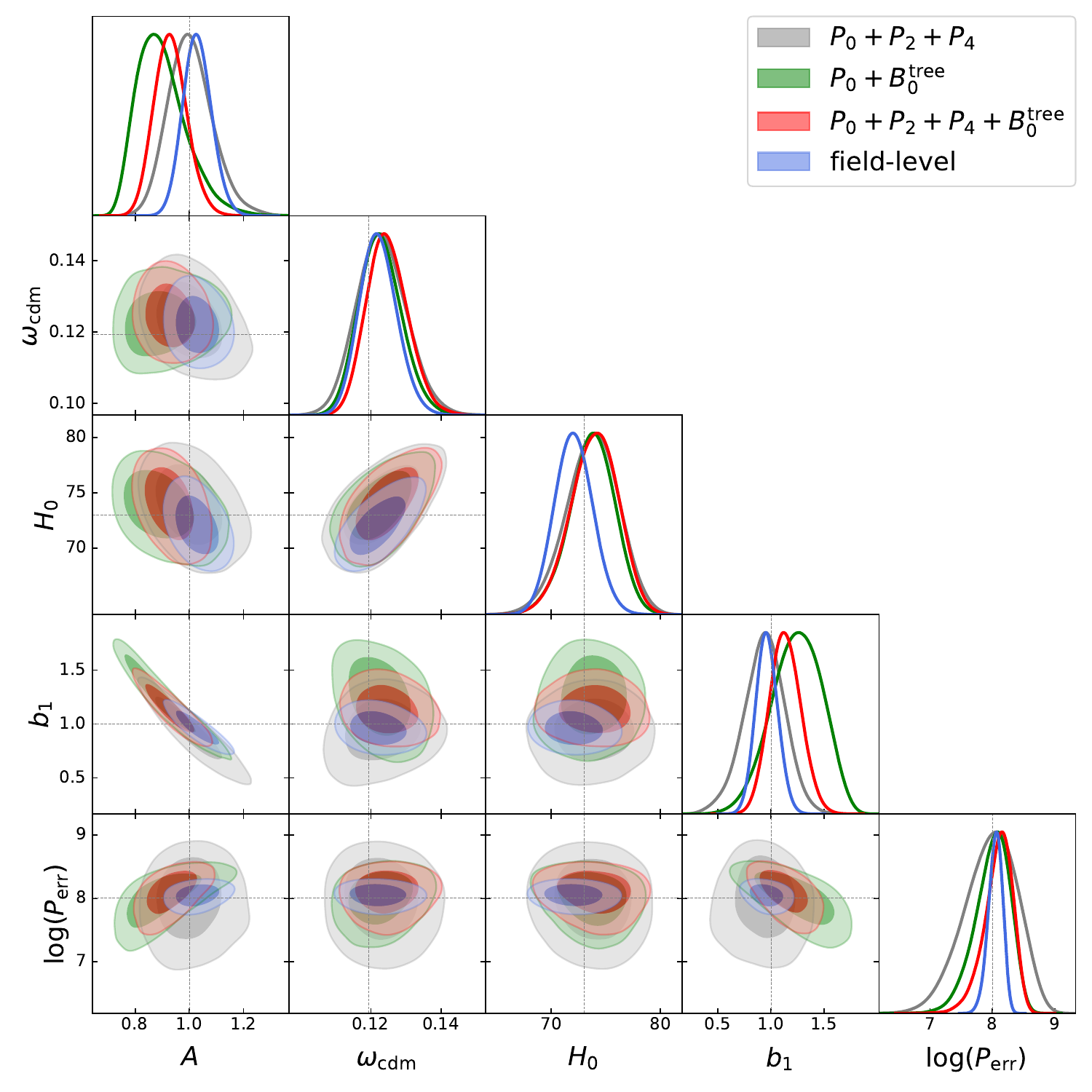}
    \end{subfigure}
    \begin{subfigure}{0.49\textwidth}
        \centering
        \includegraphics[width=\textwidth]{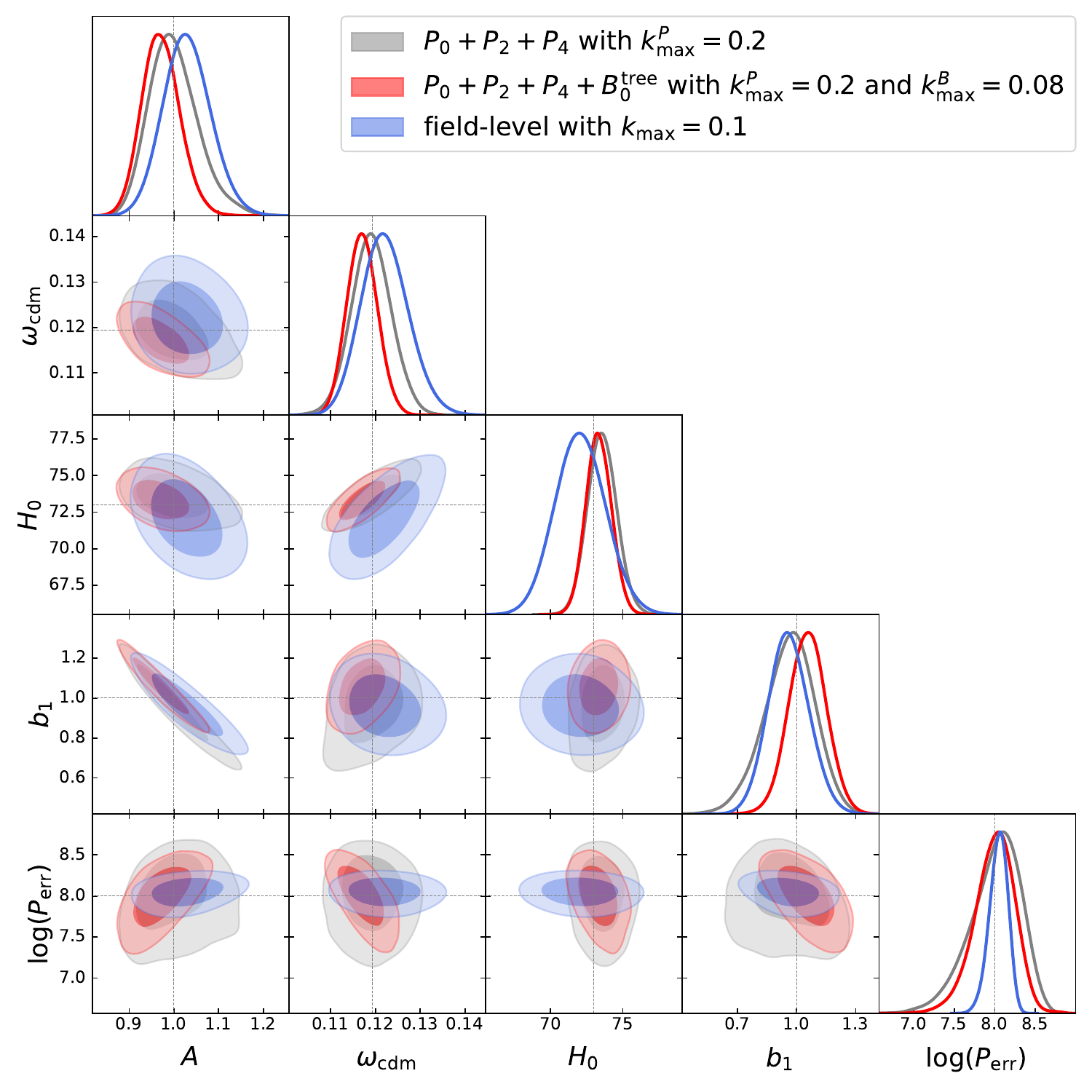}
    \end{subfigure} 
    \caption{The 2D marginalized posterior distributions for the LPT-based model with the quadratic bias in redshift space 
    at $k_\text{max} = 0.1~h/\text{Mpc}$ (left) and various choices of $k_\text{max}$ (right), where the unit of $k_\text{max}$ is $h/\text{Mpc}$.
    The gray contours show the results of the power spectrum multipoles alone,
    the green ones show the power spectrum monopole and the bispectrum monopole,
    the red ones show the power spectrum multipoles plus the bispectrum monopole,
    and the blue ones show the field-level.
    }
    \label{fig:LPT_redshift_space_quad_P_P+B}
\end{figure}

\begin{table}[t]
    \centering
    \begin{tabular}{c|c|c|c|c|c}
        & $A$ & $\omega_{\rm cdm}$ & $H_0$ & $b_1$ & $\log(P_\text{err})$ \\ \hline
        $P_0 + P_2 + P_4$ & $1.006^{+0.071}_{-0.092}$ & $0.1231^{+0.0066}_{-0.0075}$ & $73.8^{+2.5}_{-2.2}$ & $0.93^{+0.20}_{-0.18}$ & $7.97^{+0.48}_{-0.36}$\\ \hline
        $P_0 + B_0^\text{tree}$ & $0.901^{+0.063}_{-0.11}$ & $0.1227^{+0.0055}_{-0.0064}$ & $73.7^{+2.2}_{-1.9}$ & $1.24^{+0.26}_{-0.22}$ & $7.99^{+0.33}_{-0.22}$\\ \hline
        $P_0 + P_2 + P_4 + B_0^\text{tree}$ & $0.931^{+0.055}_{-0.065}$ & $0.1247^{+0.0054}_{-0.0063}$ & $73.9^{+2.3}_{-2.0}$ & $1.13\pm 0.15$ & $7.99^{+0.33}_{-0.22}$\\ \hline
        field-level & $1.029\pm 0.053$ & $0.1221^{+0.0049}_{-0.0056}$ & $72.1 \pm 1.7$ & $0.96 \pm 0.1$ & $8.05^{+0.13}_{-0.10}$\\
    \end{tabular}
    \caption{The mean values and 68\%-confidence intervals of the parameters for the LPT-based model with the quadratic bias parameters in redshift space
    at $k_\text{max} = 0.1~h/\text{Mpc}$,
    corresponding to the right panel of Fig.~\ref{fig:LPT_redshift_space_P+B} and the left panel of Fig.~\ref{fig:LPT_redshift_space_quad_P_P+B}.
    }
    \label{tab:LPT_rsd_space_quad}
\end{table}

The left panel of Fig.~\ref{fig:LPT_redshift_space_quad_P_P+B} presents a comparison among analyses based on $\{P_0,P_2,P_4\}$, $\{P_0,B_0\}$, $\{P_0,P_2,P_4,B_0\}$, and the full field-level approach, all performed with $k_{\rm max}=0.1\,h/\mathrm{Mpc}$.  
More detailed numerical values can be found in Table~\ref{tab:LPT_rsd_space_quad}.  
Notably, the analysis based on $\{P_0,B_0\}$ yields constraints similar to, and slightly stronger than, those from $\{P_0,P_2,P_4\}$, 
which is consistent with the simple SNR estimate in Eq.~\eqref{eq:SNR_RSD}.
This trend becomes even more pronounced, as $k_{\rm max}$ increases (see Fig.~\ref{fig:LPT_redshift_space_quad_P_P+B_0.12} for $k_{\rm max}=0.12\,h/\mathrm{Mpc}$), 
implying that beyond this scale the bispectrum monopole is more informative than the power spectrum quadrupole.  
In practice, however, achieving unbiased results near this scale necessitates evaluating the one-loop correction for the bispectrum; 
hence, in typical applications one would opt for a smaller $k_{\rm max}$ for the bispectrum (where the tree-level prediction remains valid) 
while employing a larger $k_{\rm max}$ for the power spectrum (with the one-loop correction included).  
For instance, we illustrate in the right panel of Figs.~\ref{fig:LPT_redshift_space_quad_P_P+B} and~\ref{fig:LPT_redshift_space_quad_P_P+B_0.12} a case where $k_{\rm max}=0.2\,h/\mathrm{Mpc}$ is used for the power spectrum and $k_{\rm max}=0.08\,h/\mathrm{Mpc}$ for the bispectrum, 
alongside field-level results obtained with $k_{\rm max}=0.1$ and $0.12\,h/\mathrm{Mpc}$, respectively.  
Although this comparison is not entirely fair in terms of mode counting due to the different $k_{\rm max}$ values, 
this exercise nonetheless offers a practical reference for real analyses, particularly given the challenges associated with achieving accurate field-level modeling at higher $k_{\rm max}$.  
We find that the power spectrum analysis employing $k_{\rm max}=0.2\,h/\mathrm{Mpc}$ yields constraints that are comparable to, or for $\omega_{\rm cdm}$ and $H_0$ even tighter than, 
those obtained from the field-level analysis at $k_{\rm max}=0.1$ or $0.12\,h/\mathrm{Mpc}$.
In this case, adding the bispectrum with $k_{\rm max}=0.08\,h/\mathrm{Mpc}$ does not lead to a significant improvement in the constraints, 
which is consistent with the findings of Refs.~\cite{Ivanov:2021kcd,Philcox:2022frc,DAmico:2022osl}.  
In contrast, adopting a larger $k_{\rm max}$ is especially advantageous for constraining parameters that affect 
the shape of the linear power spectrum.

\section{The consequences of the resolution mismatch}
\label{sec:resolution_mismatch}

\begin{figure}[t]
    \centering
    \begin{subfigure}{0.49\textwidth}
        \centering
        {$k_\text{max} = 0.12~h/\text{Mpc}$\par\vspace{1ex}}
        \includegraphics[width=\textwidth]{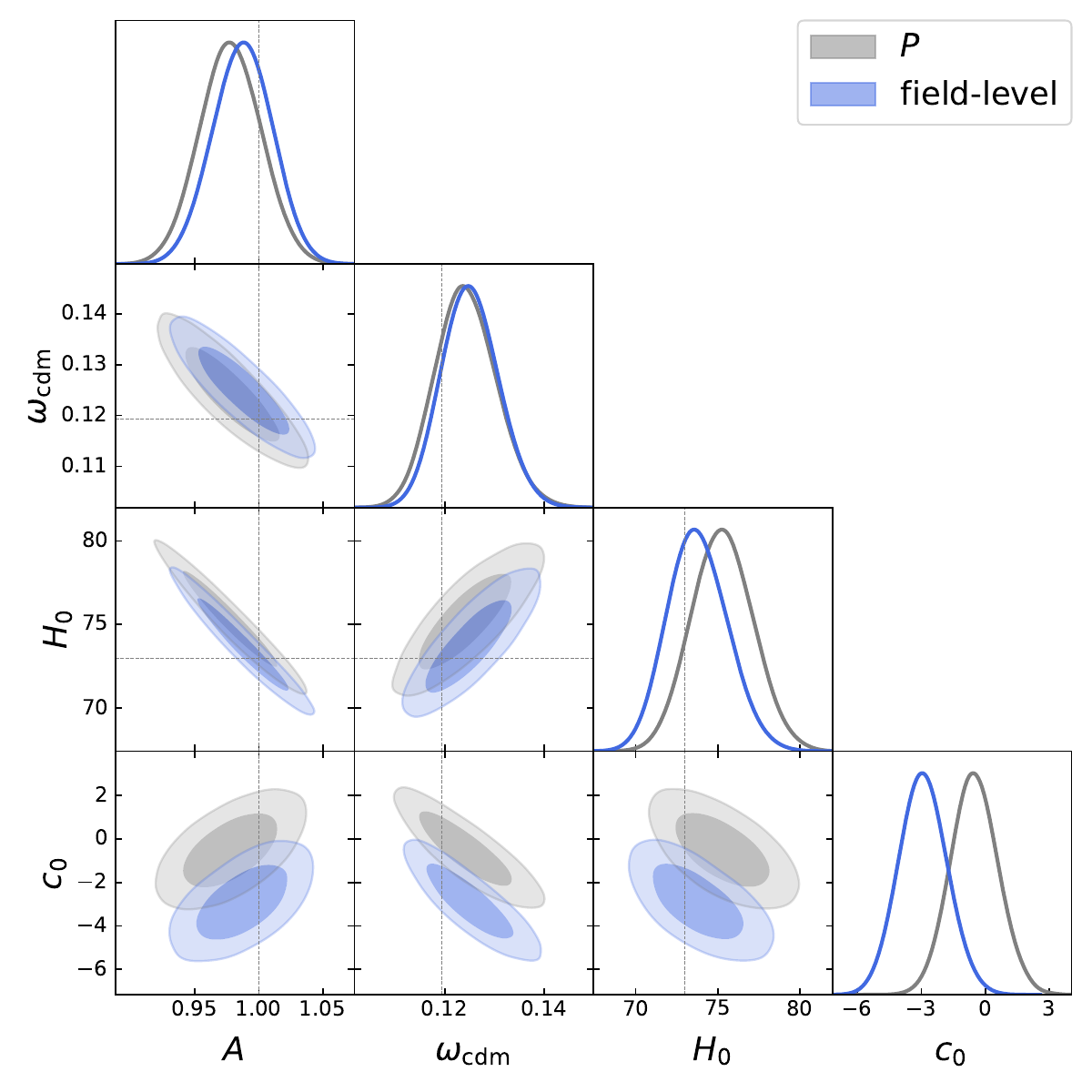}
    \end{subfigure}
    \begin{subfigure}{0.49\textwidth}
        \centering
        \includegraphics[width=0.9\textwidth]{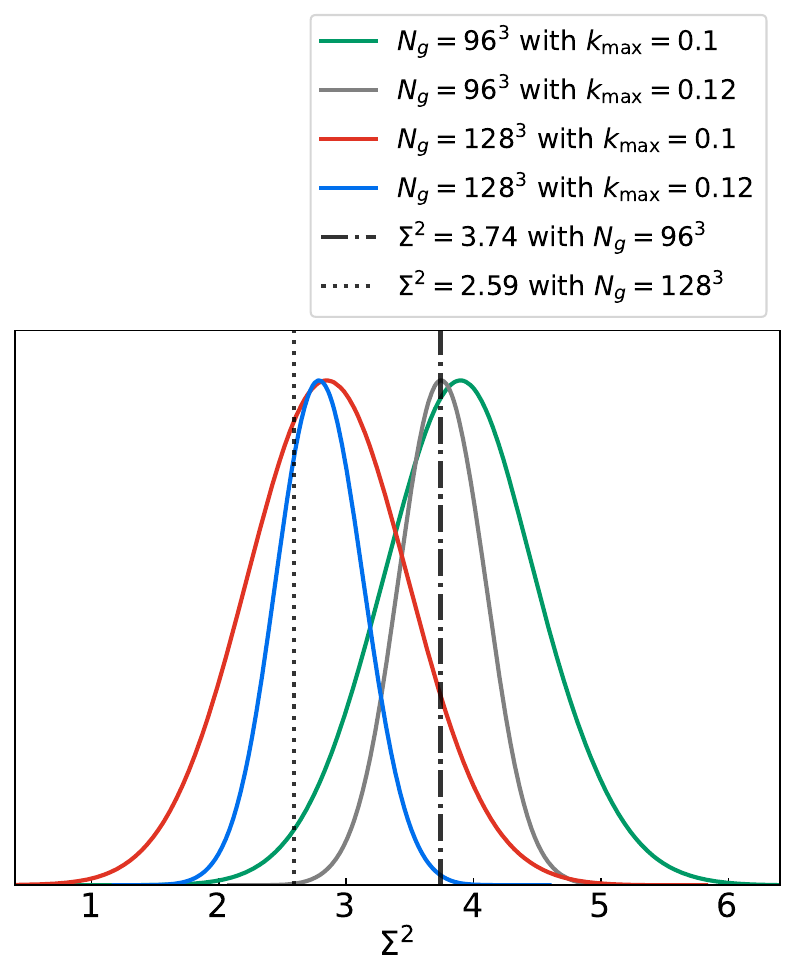}
    \end{subfigure} 
    \caption{\textit{Left panel}: The 2D marginalized posterior distributions for Zel'dovich matter model in real space, while fixing the noise power spectrum.
    The gray and blue contours correspond to the result of the power spectrum and field-level analysis, respectively.
    \textit{Right panel}: The posteriors for $\Sigma^2$ (see Eq.~\eqref{eq:Zeldovich_matter_smooth}), while fixing all the other parameters.
    Results are shown for two grid resolutions: $N_g = 96^3$ (green and grey lines) and $N_g = 128^3$ (red and blue lines), 
    each evaluated at two values of $k_{\rm max}$: $0.1~h/{\rm Mpc}$ and $0.12~h/{\rm Mpc}$.
    The dash-dotted and dotted vertical lines indicate the corresponding theoretical predictions for each grid (Eq.~\eqref{eq:Sigma2_S}).
    }
    \label{fig:LPT_matter}
\end{figure}

So far we have focused on the comparison between the field-level analysis and the joint power spectrum and bispectrum analysis 
and concluded that they yield quite similar results.
However, important differences also exist between the two approaches.  
The purpose of this section, as well as the next, is to discuss these differences and examine their implications.

\subsection{Impact on the signal}
\label{subsec:resolution_mismatch_signal}

Let us begin with what the resolution mismatch causes in the field-level analysis.
As a concrete example, we first consider the Zel'dovich matter case in real space, with only the cosmological parameters and $c_0$ varied.
The left panel of Fig~\ref{fig:LPT_matter} shows the posterior distributions for these parameters, derived from the power spectrum analysis (gray) and the field-level analysis (blue), with $k_{\rm max} = 0.12~h/\text{Mpc}$.
As we have discussed in Sec.~\ref{subsec:G2_case}, the two analyses give rise to similar constraints on the cosmological parameters.
However, one can notice that the constraints on $c_0$ from the field-level analysis implies a non-zero value, while the power spectrum analysis yields a consistent result with zero.
Indeed, we expect that the resolution mismatch in the field-level analysis leads to a non-zero value of $c_0$, as we explain below.

The field-level expression for the Zel'dovich matter is given by
\begin{align}
    \delta_g({\bk}) = \int \dd^3{\bf q}~e^{-i \bk \cdot (\bq+ \bPsi(\bq))}.
    \label{eq:zeldovich_matter}
\end{align}
In order to discuss the effect of the resolution mismatch, we decompose the linear displacement field $\bPsi$ into the long- and short-modes as
\footnote{This decomposition is exact only for the linear displacement field.}
\begin{align}
    \bPsi(\bq) = \bPsi_{\rm L}(\bq) + \bPsi_{\rm S}(\bq),
\end{align}
where
\begin{align}
    \bPsi_{\rm L}(\bq) = \int^{\Lambda'} \frac{\dd^3\bk}{(2\pi)^3} ~ \frac{i \bk}{k^2}\delta_1(\bk),
    \qquad
    \bPsi_{\rm S}(\bq) = \int_{\Lambda'} \frac{\dd^3\bk}{(2\pi)^3} ~ \frac{i \bk}{k^2}\delta_1(\bk).
\end{align}
In other words, the mock data contains all the modes, while the inference model contains only the long modes, 
implying that the short modes present in the mock data act as the stochasticity in the inference model.
Averaging Eq.~\eqref{eq:zeldovich_matter} over the short modes we obtain
\begin{align}
    \langle \delta_g({\bk}) \rangle_\text{S}
    = &
    \int \dd^3{\bf q}~e^{-i \bk \cdot (\bq+ \bPsi_{\rm L}(\bq))} \langle e^{-i \bk \cdot \bPsi_{\rm S}(\bq)} \rangle_\text{S}
    \nonumber \\
    = &
    e^{-\frac12 k^2 \Sigma^2_{\rm S}}\int \dd^3{\bf q}~e^{-i \bk \cdot (\bq+ \bPsi_{\rm L}(\bq))},
    \label{eq:Zeldovich_matter_smooth}
\end{align}
where we have used the cummulant expansion theorem in the second equality and $\Sigma^2_{\rm S}$ is the small-scale variance of the displacement,
\begin{align}
    \Sigma^2_{\rm S} = \frac13 \int_{\Lambda'} \frac{\dd^3\bk}{(2\pi)^3} ~ \frac{P_\text{lin}(k)}{k^2} .
    \label{eq:Sigma2_S}
\end{align}
Eq.~\eqref{eq:Zeldovich_matter_smooth} means that the mock data should look a Gaussian smoothing of the Zel'dovich matter field in the inference model.
This is the origin of the quadratic counterterm, $c_0$, and its amplitude is given by $c_0 = -\frac12 \Sigma^2_{\rm S}$.
Substituting $\Lambda' = \Lambda$ (the cubic cutoff in the initial conditions, corresponding to $N^3_g=128^3$) yields $c_0\simeq -1.3~({\rm Mpc}/h)^2$, 
which is consistent with the resultant posterior of $c_0$ in the left panel of Fig~\ref{fig:LPT_matter}.
On the other hand, the power spectrum analysis does not suffer from this effect, 
since the integral in the theoretical prediction does not have such a cutoff.
The same Gaussian damping can be derived for the biased tracer as well, which is discussed in App.~\ref{app:sigma2_bias}.

As we discuss in App.~\ref{app:reconstructed_IC}, we cannot reconstruct the true initial conditions beyond $k_\text{max}$.
However, this does \textit{not} imply that the relevant cutoff here is set by $k_\text{max}$ used in the likelihood.
While it is true that the phases of Fourier modes beyond $k_\text{max}$ are incorrect, their squared amplitudes are still correct on average, as imposed by the prior.
To demonstrate this point, we present the posterior for $\Sigma^2_{\rm S}$ in the right panel of Fig.~\ref{fig:LPT_matter}, 
where we use the exact form of Eq.~\eqref{eq:Zeldovich_matter_smooth} as the forward model, vary only $\Sigma^2_{\rm S}$ and fix all the other parameters.
We run the field-level inference for four different combinations of initial condition resolution and $k_\text{max}$: 
$(N^3_g, k_\text{max}) = (96^3,~0.1~h/\text{Mpc}),~(96^3,~0.12~h/\text{Mpc}),~(128^3,~0.1~h/\text{Mpc}),~(128^3,~0.12~h/\text{Mpc})$.
As clearly shown, the posterior mean is determined by $N_g$ (or $\Lambda$) whereas $k_\text{max}$ just narrows the posterior width.
The inferred values are in agreement with the theoretical predictions of Eq.~\eqref{eq:Sigma2_S} with $\Lambda'=\Lambda$.

\subsection{Impact on the noise}
\label{subsec:resolution_mismatch_noise}

\begin{figure}[t]
    \centering
    \begin{subfigure}{0.49\textwidth}
        \centering
        \includegraphics[width=\textwidth]{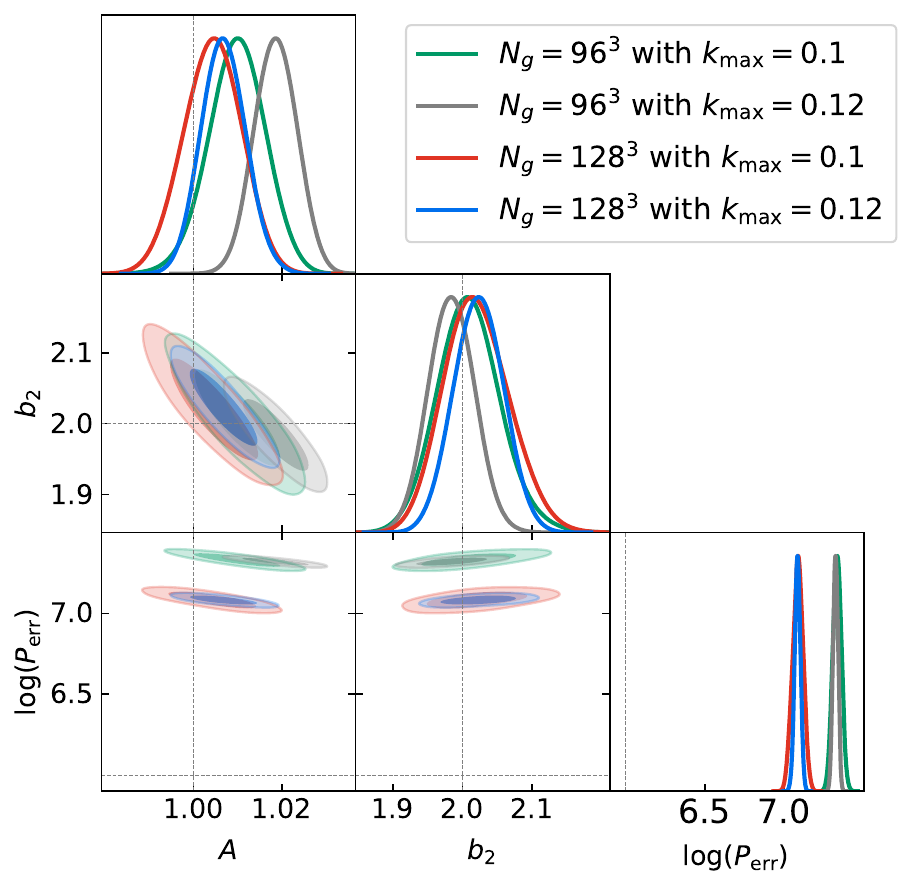}
    \end{subfigure}
    \begin{subfigure}{0.49\textwidth}
        \centering
        \includegraphics[width=0.9\textwidth]{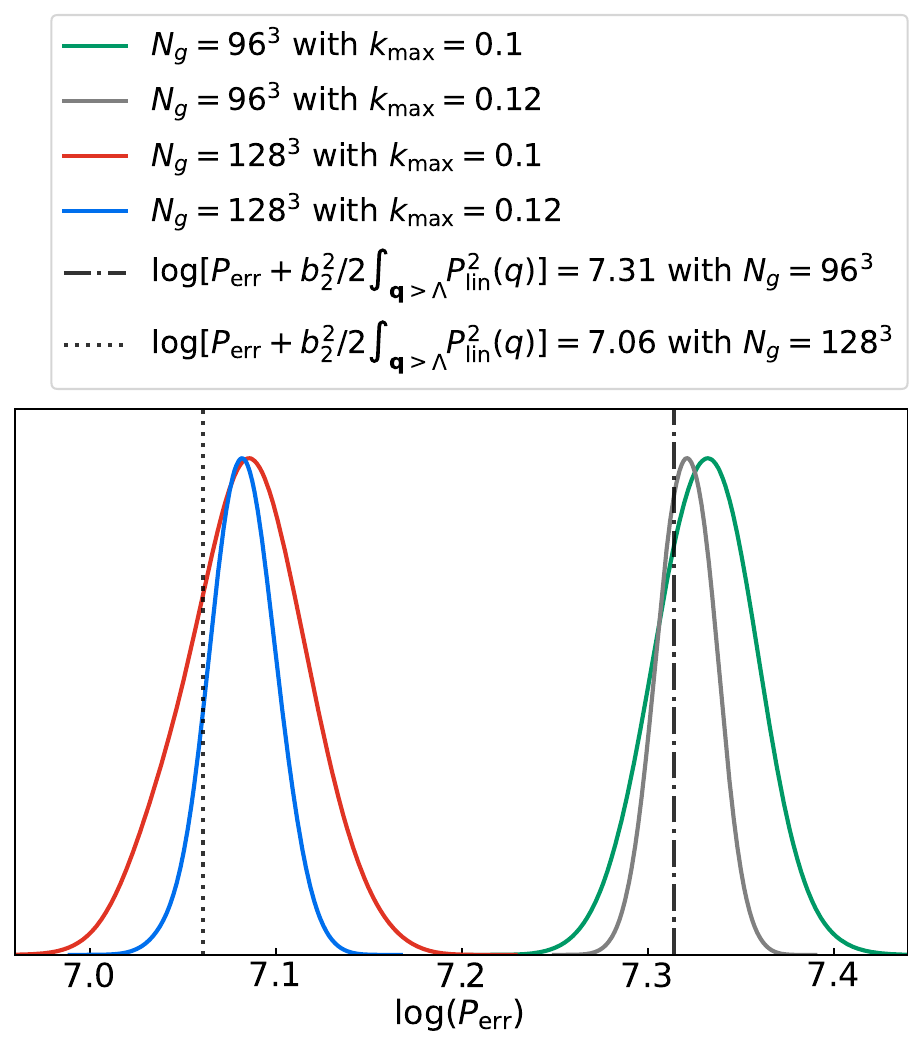}
    \end{subfigure} 
    \caption{\textit{Left panel}: The 2D marginalized posterior distributions for the toy model (Eq.~\eqref{eq:b2_model}), varying only $A$, $b_2$ and $P_{\rm err}$.
    Results are shown for two grid resolutions: $N_g = 96^3$ (green and grey lines) and $N_g = 128^3$ (red and blue lines), 
    each evaluated at two values of $k_{\rm max}$: $0.1~h/{\rm Mpc}$ and $0.12~h/{\rm Mpc}$.
    \textit{Right panel}: The 1D posterior distributions of $P_{\rm err}$ (the zoom-in version of the bottom right panel).
    The dash-dotted and dotted vertical lines indicate the corresponding theoretical predictions for each grid (Eq.~\eqref{eq:Perr_S}).
    }
    \label{fig:b2_noise}
\end{figure}

Not only does this issue affect the forward model (i.e., the signal), but it also impacts the field-level noise.
Here we focus on the dominant effect from the $b_2$ term.
It is well known that the term proportional to $\delta_1^2$ at the field level generates the white-noise-like term to the power spectrum:
\begin{align}
    \frac{b_2^2}{4} \langle \delta_1^2 \delta_1^2 \rangle^\prime(k) = \frac{b_2^2}{2} \int\frac{{\rm d}^3{\bf q}}{(2\pi)^3} P_{\rm lin}(|{\bf k}-{\bf q}|) P_{\rm lin}(q)
     \to  \frac{b_2^2}{2} \int\frac{{\rm d}^3{\bf q}}{(2\pi)^3} P^2_{\rm lin}(q)~(k\to 0).
\end{align}
This contribution is typically absorbed into the (residual) shot noise in the standard power spectrum analysis, since this is fully degenerate with the constant shot noise.
\footnote{The same term gives rise to a white-noise-like contribution in the bispectrum as well.}
While this term can be treated as part of the signal in field-level inference, the finite resolution of the initial conditions causes it to behave as an effective shot noise as well~\cite{Schmittfull:2018yuk,Obuljen:2022cjo,Kostic:2022vok,Foreman:2024kzw,Kokron:2025yma}.\footnote{The impact of this effect on the difference between power spectrum analysis and field-level inference will be explored in a forthcoming paper~\cite{Akitsu_2025}.}
To illustrate this effect, we consider the following toy model:
\begin{align}
    \hat{\delta}_g = \delta_1 + \frac{b_2}{2} \left(\delta_1^2 - \sigma^2 \right) + \epsilon.
    \label{eq:b2_model}
\end{align}
Splitting the linear density field into the long- and short-modes as $\delta_1 = \delta_{1,{\rm L}} + \delta_{1,{\rm S}}$ as before, the effective noise in this example becomes
\begin{align}
    \tilde{\epsilon} = \frac{b_2}{2} \delta^2_{1,\rm S} + \epsilon,
\end{align}
whose power spectrum in the $k\to0$ limit is given by
\begin{align}
    P_\text{err} = \frac{b_2^2}{2} \int_{\Lambda'}\frac{{\rm d}^3{\bf q}}{(2\pi)^3} P^2_{\rm lin}(q) + P_\epsilon.
    \label{eq:Perr_S}
\end{align}
This means that the inferred value of $P_{\rm err}$ is expected to shift depending on the values of $b_2$, $\Lambda'$ and $P_{\epsilon}$.
We again stress that the cutoff is set by the resolution of the initial conditions, i.e., $\Lambda'=\Lambda$.
For our default choice of $b_2 = -0.5$, $N^3_g = 128^3$ and $P_{\epsilon} = e^8~({\rm Mpc}/h)^3$, this shift is difficult to observe.
To make the effect more visible, we create a new mock based on the toy model (Eq.~\eqref{eq:b2_model}) where we set $b_2= 2.0$ and $P_\epsilon=e^6$.
We then perform the field-level inference by varying only $A$, $b_2$ and $\log(P_{\rm err})$, using four different combinations of initial condition resolution and $k_\text{max}$ as in the previous Zel'dovich example.

The results are shown in Fig.~\ref{fig:b2_noise}.
First, in fact the posteriors of $\log(P_{\rm err})$ significantly deviate from the value corresponding to the power spectrum of the injected Gaussian noise field.
Second, these deviations are well accounted for once the contribution from the $b_2$ term is included, with the cutoff set to $\Lambda'=\Lambda$ (Eq.~\eqref{eq:Perr_S}).
Third, one can notice that the posteriors in the $A-b_2$ plane are also biased, in particular when $N^3_g=96^3$ and $k_\text{max}=0.12~h/{\rm Mpc}$.
This could be ascribed to the fact that the effective noise is no longer Gaussian, since it originates from $\delta_1^2$, which is clearly a non-Gaussian field.
We now investigate the impact of such non-Gaussian field-level noise.

\section{The non-Gaussian noise at the field level}
\label{sec:non-Gaussian_noise}

So far, we have imposed a Gaussian form on the field-level likelihood by manually injecting the Gaussian noise at the field level.
For a realistic biased tracer, however, the residual noise exhibits non-Gaussian features. 
Here we investigate the potential impact of non-Gaussian noise on the field-level inference.
To this end, we consider the following two non-Gaussian noise examples: (a) Poissonian noise and (b) density-dependent noise.

\subsection{Poissonian noise}
\label{subsec:Poisson_noise}

\begin{figure}[t]
    \centering
    {$k_\text{max} = 0.12~h/\text{Mpc}$\par\vspace{1ex}}
    \begin{subfigure}{0.49\textwidth}
        \centering
        {Gaussian linear forward model}
        \includegraphics[width=\textwidth]{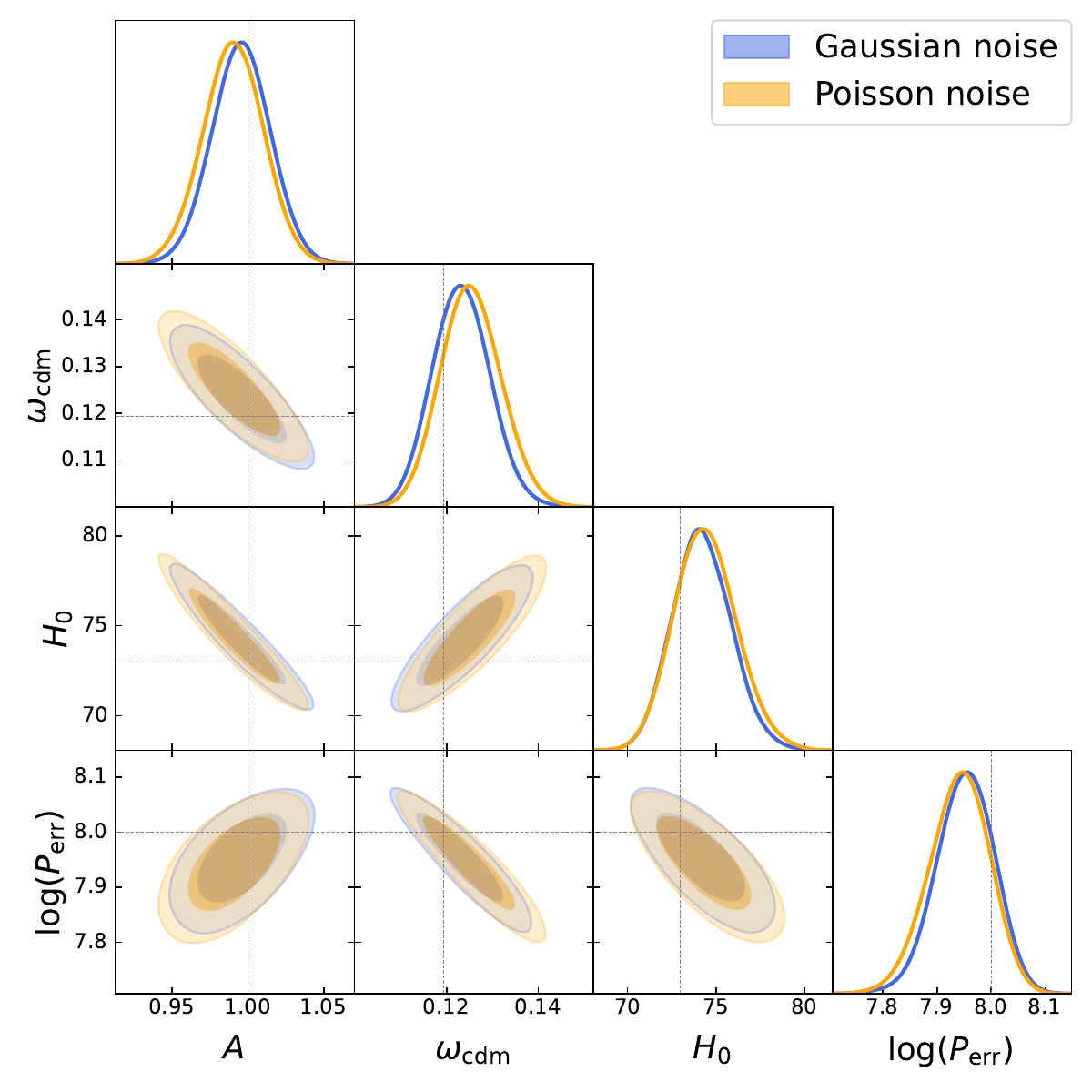}
    \end{subfigure}
    \begin{subfigure}{0.49\textwidth}
        \centering
        {Eulerian nonlinear forward model}
        \includegraphics[width=\textwidth]{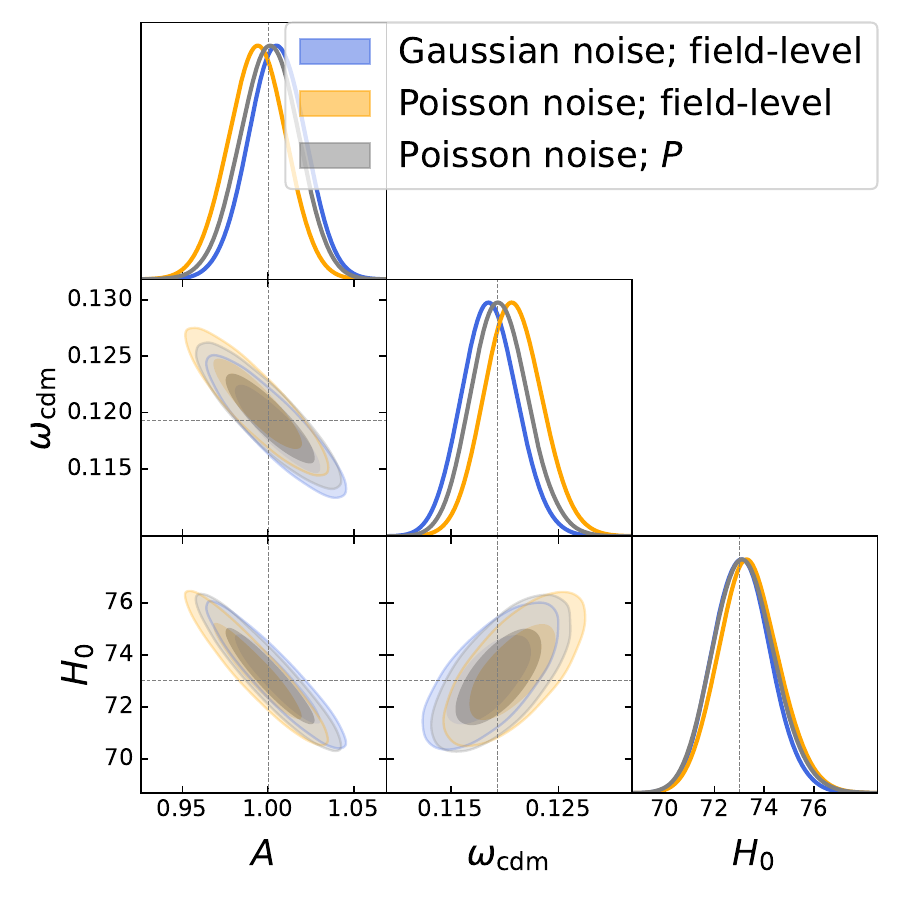}
    \end{subfigure} 
    \caption{The 2D marginalized posterior distributions from the field-level inference,
    shown for the linear forward model (left) and the simple Eularian forward model (right).
    In both panels, the blue contours represent the cases with injected Gaussian noise, while the orange contours represent the cases with injected Poissonian noise.
    In the right panel, the grey contours show the power spectrum analysis for the mock with the Poissonian noise.
    }
    \label{fig:Poisson_noise_cosmo}
\end{figure}

\begin{figure}[t]
    \centering
    {$k_\text{max} = 0.12~h/\text{Mpc}$\par\vspace{1ex}}
    \begin{subfigure}{0.49\textwidth}
        \centering
        \includegraphics[width=\textwidth]{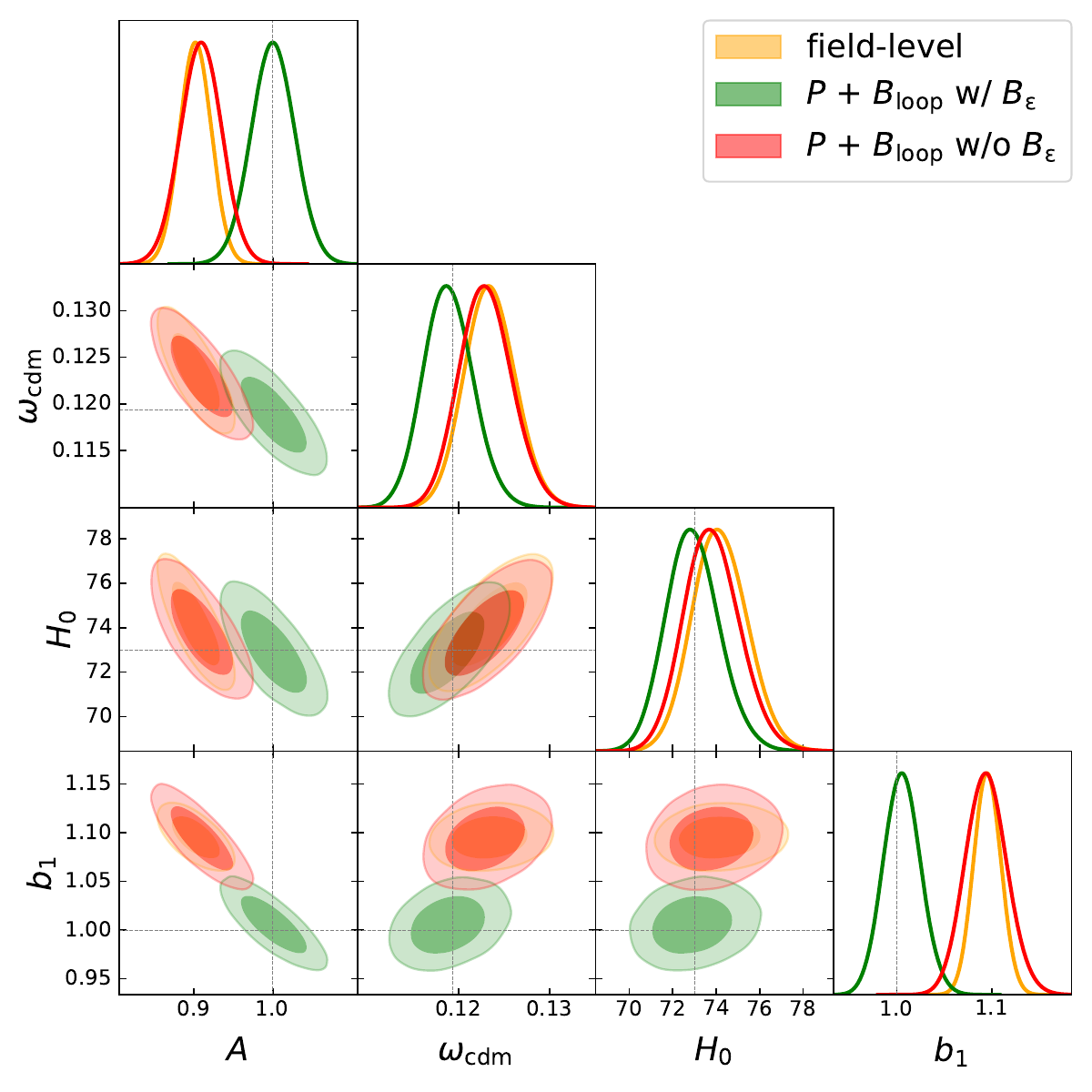}
    \end{subfigure}
    \begin{subfigure}{0.49\textwidth}
        \centering
        \includegraphics[width=\textwidth]{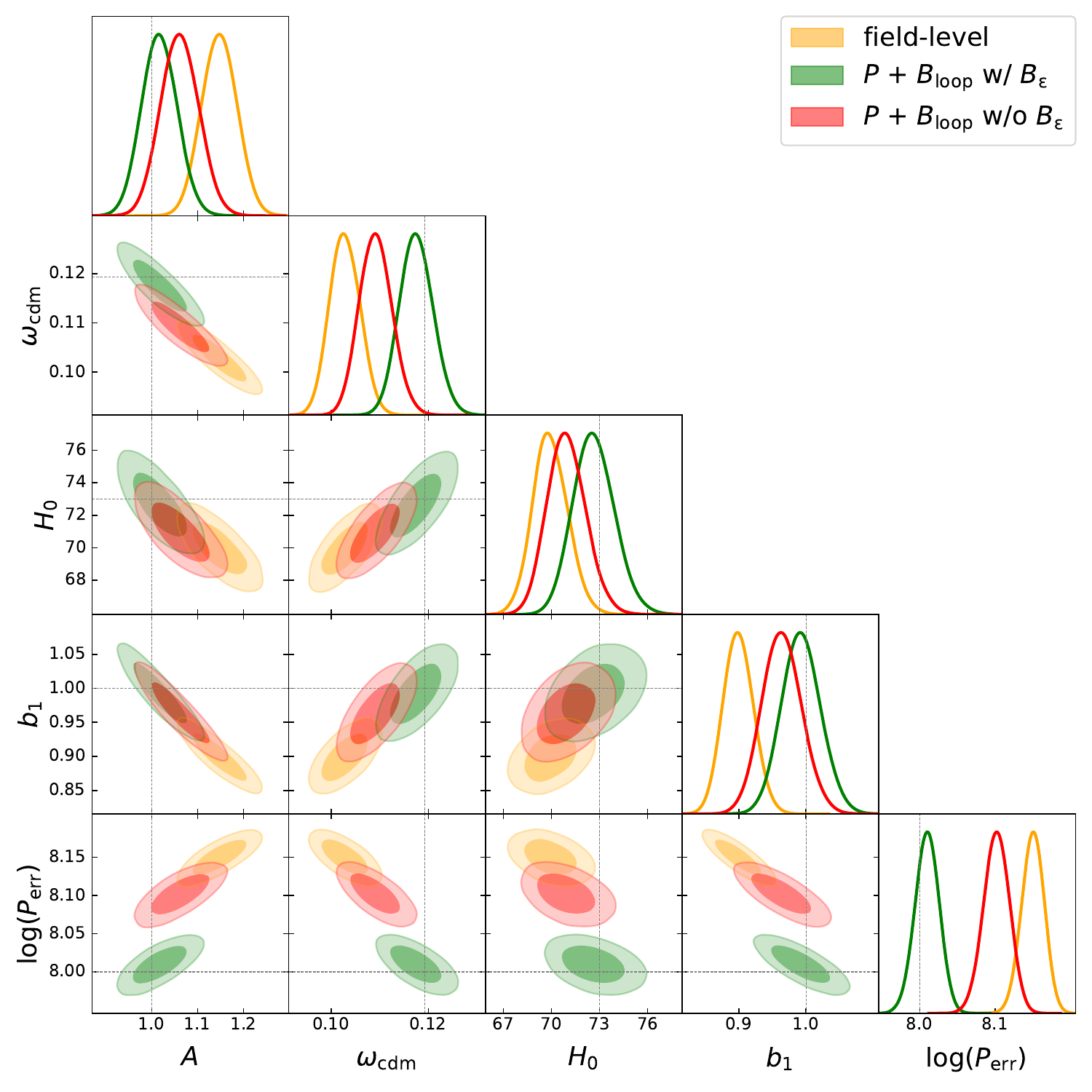}
    \end{subfigure} 
    \caption{The 2D marginalized posterior distributions for the simple Eulerian mock with the Poissonian noise.
    In the left panel, $\{A, \omega_{\rm cdm}, H_0, b_1\}$ are varied while in the right panel, $\{A, \omega_{\rm cdm}, H_0, b_1, \log(P_\text{err})\}$ are varied.
    In both panels, the orange contours represent field-level inference, the green contours represent the joint power spectrum and bispectrum analysis with the full modeling, 
    and the red contours represent the joint power spectrum and bispectrum analysis without $B_\epsilon$.
    }
    \label{fig:Poisson_noise_b1_noise}
\end{figure}

In many galaxy clustering studies, the residual noise in the galaxy density field is often approximated by a Poisson distribution.
In this subsection, we repeat similar analyses to Sec.~\ref{sec:EPT} but with Poissonian noise to quantify the impact of the non-Gaussian field-level noise.
To simulate Poisson shot noise, we scatter particles by sampling their positions from a uniform distribution over the box and then construct the noise field, $\epsilon_\text{p}$, by measuring the density field from these particles.
Specifically, we randomly place $N = 139^3$ particles so that the noise power spectrum is given by $P_\text{err} = \bar{n}^{-1} \simeq e^8~({\rm Mpc}/h)^3$.

As a first test, we perform field-level inference using the Gaussian field-level likelihood on the mock data defined by:
\begin{align}
    \hat{\delta}_g = \delta_1 + \epsilon_\text{p},
\end{align}
where $\delta_1$ is the linear Gaussian field.
The results are shown in the left panel of Fig.~\ref{fig:Poisson_noise_cosmo}, alongside the Gaussian noise case (identical to the left panel of Fig.~\ref{fig:Gaussian}).
Despite the mismatch between the true noise distribution and the assumed Gaussian likelihood,
the resulting posterior remains unbiased and agrees with the Gaussian noise case.
This can be understood as follows:
since the signal consists solely of the Gaussian random field, the inference relies entirely on power spectrum information.
As shown in Sec.~\ref{subsec:Gaussian_case} the power spectrum analysis yields the same posterior.
The point here is that in this case the inference does not use the bispectrum (or any other non-Gaussian) information.

Next, we repeat a similar comparison with the following non-Gaussian mock data:
\begin{align}
    \hat{\delta}_g = b_1 (\delta_1 + {\cal G}_2) + \epsilon_\text{p}.
    \label{eq:G2_Poisson}
\end{align}
Note that the corresponding power spectrum and bispectrum are the same as Eq.~\eqref{eq:P_B_G2}, except that now the bispectrum has the contribution from the non-Gaussian stochasticity:
\begin{align}
    B_g(k_1, k_2, k_3) &=  b_1^3 \big( B_{\rm tree}(k_1, k_2, k_3) + B_{{\cal G}_2{\cal G}_2{\cal G}_2}(k_1, k_2, k_3) \big) + B_{\epsilon_{\rm p}}\;, \\
    B_{\epsilon_{\rm p}} &= \langle \epsilon_{\rm p} \epsilon_{\rm p} \epsilon_{\rm p} \rangle' = \frac{1}{\bar{n}^2} = P^2_{\epsilon_{\rm p}}.
\end{align}
In the joint power spectrum and bispectrum analyses presented below, we include the noise bispectrum $B_{\epsilon_{\rm p}}$ in the bispectrum model, 
with $B_{\epsilon_{\rm p}}=P^2_{\epsilon_{\rm p}}$ assumed from the Poisson statistics.
\footnote{We note that the covariance for the power spectrum and bispectrum is estimated from the ensemble of mocks generated according to Eq.~\eqref{eq:G2_Poisson}, which includes the Poisson noise.}
This ensures that the number of parameters varied in the field-level inference and in the joint power spectrum and bispectrum analysis is consistent.

As in Sec.~\ref{subsec:G2_case}, we first focus on the case in which only the cosmological parameters $\{A, \omega_{\rm cdm}, H_0\}$ are varied while $b_1$ and $P_{\rm err}$ are fixed.
The results are shown in the right panel of Fig.~\ref{fig:Poisson_noise_cosmo}.
Similar to the previous linear Gaussian case, the field-level result (the orange contour) remains unbiased and consistent with the power spectrum analysis (the grey contour).
This aligns with our expectation, since in this case the bispectrum (or generally any non-Gaussian signal) does not help to break the degeneracy and has much smaller SNR than the power spectrum, as argued in Sec.~\ref{subsec:G2_case}.
In other words, as long as the non-Gaussian information in the signal is not effective, non-Gaussian corrections to the noise can be safely neglected.

However, when we allow the bias parameter $b_1$ to vary, the situation becomes different.
In the left panel of Fig.~\ref{fig:Poisson_noise_b1_noise}, one can see that the field-level inference for the Poissonian noise mock (the orange contours) fails to recover the input values.
This result meets our expectation. Now the inference becomes sensitive not only to the power spectrum but also to higher-order statistics such as the bispectrum to break the $b_1-A$ degeneracy.
However, the assumption of the Gaussian likelihood would treat the noise as Gaussian,
causing the noise bispectrum to be misinterpreted as part of the signal bispectrum. 
This results in biased posteriors, akin to using an incorrect model for the power spectrum or bispectrum (see, e.g., the red and orange contours in Fig.~\ref{fig:G2_b1}).\footnote{In this toy model, the noise bispectrum has a comparable amplitude to the one-loop bispectrum, i.e., $B_{\epsilon_{\rm p}}\sim |B_{{\cal G}_2{\cal G}_2{\cal G}_2}|$.}

On the other hand, and crucially, the standard power spectrum and bispectrum analysis does remain unbiased (see the green countours), as long as the noise bispectrum ($B_{\epsilon_{\rm p}}$) is included in the theoretical template,
while using the Gaussian likelihood for the power spectrum and bispectrum.
If, however, the noise bispectrum is not taken into account, the standard power spectrum and bispectrum also becomes biased (the red contours).
The overlap of the red and orange contours confirms our account of the origin of the bias in the field-level inference here.

This trend is more prominent in the right panel of Fig.~\ref{fig:Poisson_noise_b1_noise}, where we allow the noise power spectrum to vary as well, i.e., we vary $\{A, \omega_{\rm cdm}, H_0, b_1, \log(P_\text{err})\}$.
As discussed in Sec.~\ref{subsubsec:G2_all}, in this case the field-level inference is sensitive to the trispectrum as well.
Consequently, the field-level inference becomes more biased since the Gaussian field-level likelihood is an inappropriate approximation that neglects the noise contributions to the bispectrum as well as the trispectrum.
This is confirmed by the fact that now the joint power spectrum and bispectrum without the noise bispectrum no longer overlaps with the field-level result.
In contrast, we again emphasize that the joint power spectrum and bispectrum analysis with the noise bispectrum is not biased.
This indicates that one must use a correct field-level likelihood whenever the non-Gaussian signal is informative for the parameter constraints, which is the case for many practical applications.
In summary, it is essential to account for non-Gaussian corrections to the noise at the same order as the signal, 
to avoid systematic biases in the inference.
\footnote{
We also note that, in this Poisson noise example, the error on $\log(P_{\rm err})$ at the field level is now comparable to that from the joint power spectrum and bispectrum, in contrast to the case shown in Fig.~\ref{fig:G2_b1_noise}, although a more detailed study is required to clarify the underlying cause.
}

\subsection{The density-dependent noise}
\label{subsec:density_dep_noise}
\begin{figure}[t]
    \centering
    {$N_\text{max} = 64$\par\vspace{1ex}}
    \includegraphics[width=0.8\textwidth]{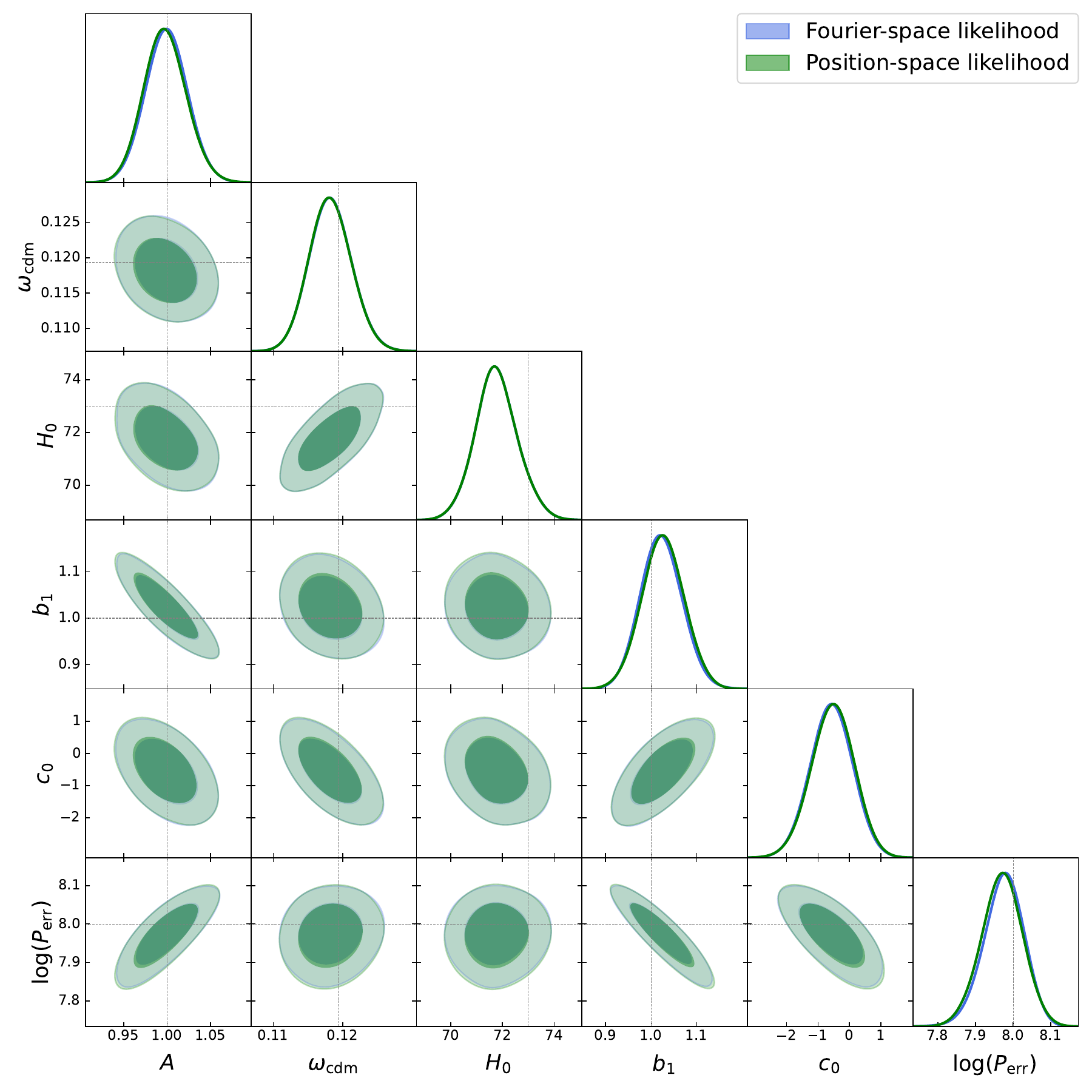}
    \caption{The 2D marginalized posterior distributions for the LPT-based model with the linear bias.
    The blue contours show the results with the Fourier-space likelihood, while the green contours show the results with the Position-space likelihood.
    }
    \label{fig:position_space_like}
\end{figure}

\begin{figure}[t]
    \centering
    {$N_\text{max} = 64$\par\vspace{1ex}}
    \includegraphics[width=0.9\textwidth]{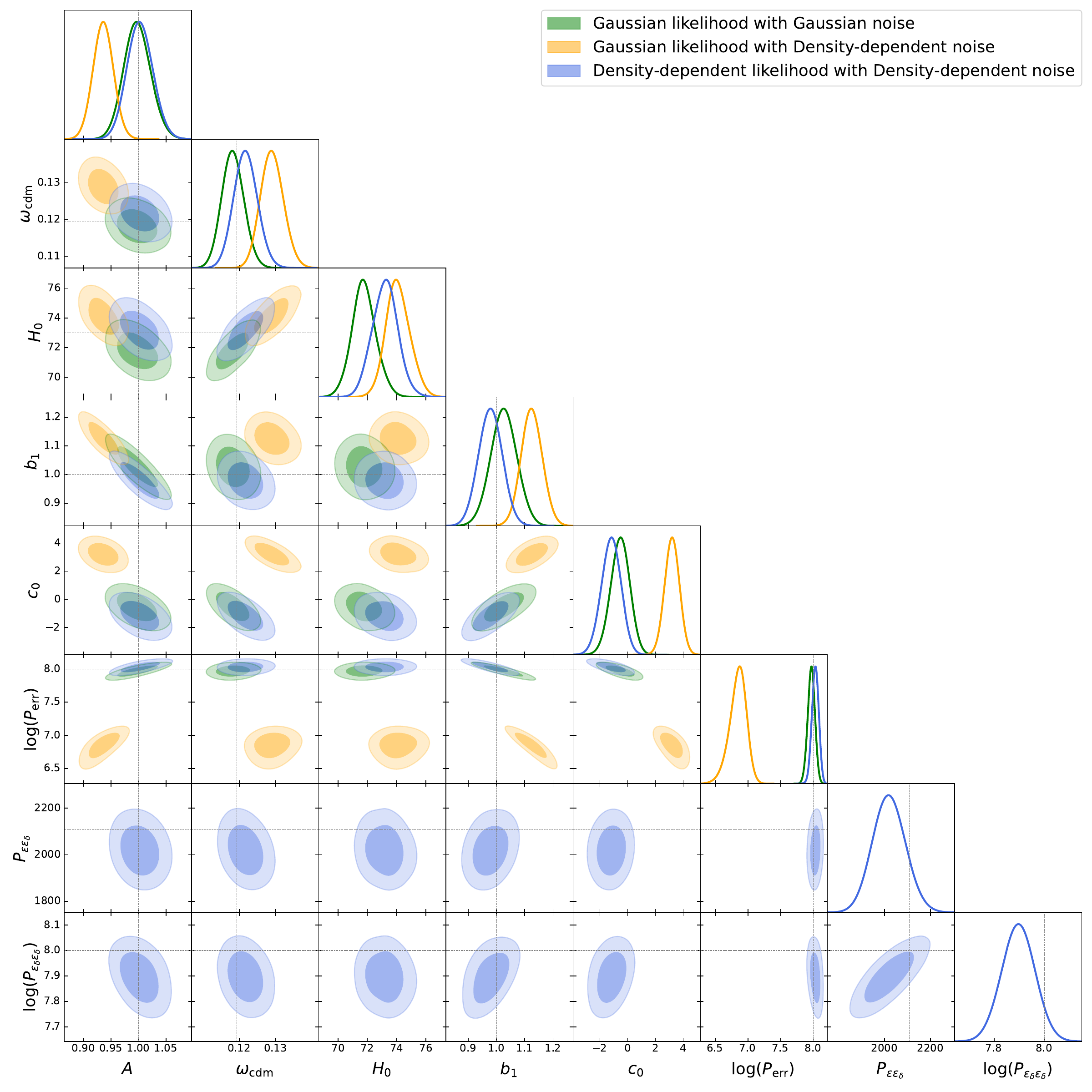}
    \caption{
    The 2D marginalized posterior distributions when the mock contains the density-dependent noise.
    The forward model employed is the LPT-based model with linear bias.
    The orange contours represent the results without density-dependent variance (Eq.~\eqref{eq:likelihood_position}), 
    while the while blue contours include the density-dependent variance (Eq.~\eqref{eq:likelihood_position_pos_dep}).
    For comparison, the results with Gaussian noise (identical to those in Fig.~\ref{fig:position_space_like}) are also shown.
    Note that we also vary $c_0$ in addition to the parameters we show here.
    }
    \label{fig:density_dep_noise}
\end{figure}

Another known source of non-Gaussian noise is the density-dependent noise, arising from the correlation between long- and short-wavelength modes.
One way to implement the density-dependent noise in the field-level analysis is to work in position space~\cite{Cabass:2020nwf}.
Here we first demonstrate that the Fourier space likelihood is equivalent to the position space likelihood under some assumptions.
Mathematically this follows from Parseval's identity. 
Specifically, if one chooses a cubic cutoff in the likelihood, i.e., picking up those Fourier modes with $|k_i| < k_\text{max}$ for each $i=x,y,z$ (as is done for the prior term) 
and consider the constant noise spectrum, $P_\text{err}(k) = P_\text{err}$, 
then Eq.~\eqref{eq:likelihood} can be rewritten as
\begin{align}
    \cL [\hat{\delta}_g|\btheta,\delta({\bf k})]
    =
    \left( \prod_{{\bf x}^i}\sqrt{\frac{V}{2\pi N_\text{max}^3 P_{\rm err}}} \right)
    \exp \left(
    -\frac{V}{2 N_\text{max}^3} \sum_{{\bf x}^i} \frac{|\hat{\delta}_g({\bf x}^i) - \delta_g({\bf x}^i;\btheta, \delta)|^2}{P_{\rm err}}
    \right),
    \label{eq:likelihood_position}
\end{align}
where the discrete sum is written explicitly with ${\bf x}^i$ denoting $i$-th voxel in position space 
and $N^3_\text{max} = (L k_\text{max}/\pi)^3$ the total number of independent modes in the likelihood.
We explicitly confirm the equivalence between the Fourier-space likelihood and the position-space likelihood, which is presented in Fig.~\ref{fig:position_space_like}. 
Here we employ the LPT-based model with the linear bias as a forward model with $N^3_{\rm max} = 64^3$.

We next consider the case where the observed data contains the following densitiy-dependent noise:
\begin{align}
    \hat{\delta}_g({\bf x}) = \delta_g({\bf x}) + \epsilon({\bf x}) + \epsilon_\delta({\bf x})\delta_1({\bf x}).
    \label{eq:density_noise_mock}
\end{align}
We assume the noise fields obey the Gaussian distribution with the covariance given by
\begin{align}
    {\rm C}_\epsilon = 
    \begin{pmatrix}
        \langle \epsilon \epsilon \rangle^\prime& \langle \epsilon \epsilon_\delta \rangle^\prime \\
        \langle \epsilon \epsilon_\delta \rangle^\prime & \langle \epsilon_\delta \epsilon_\delta \rangle^\prime
    \end{pmatrix}
    = 
    \begin{pmatrix}
        P_{\epsilon\epsilon} & P_{\epsilon\epsilon_\delta} \\
        P_{\epsilon\epsilon_\delta} & P_{\epsilon_\delta\epsilon_\delta}
    \end{pmatrix}.
\end{align}
Neglecting the scale-dependence in the covariance, one can write down the corresponding field-level likelihood in position space, which incorporates the density-dependent noise~\cite{Cabass:2020nwf}:
\begin{align}
    \cL [\hat{\delta}_g|\btheta,\delta({\bf k})]
    =
    \left( \prod_{{\bf x}^i}\sqrt{\frac{V}{2\pi N_\text{max}^3 P_{\rm err}[\delta({\bf x}^i)]}} \right)
    \exp \left(
    -\frac{V}{2 N_\text{max}^3} \sum_{{\bf x}^i} \frac{|\hat{\delta}_g({\bf x}^i) - \delta_g({\bf x}^i;\btheta, \delta)|^2}{P_{\rm err}[\delta({\bf x}^i)]}
    \right).
    \label{eq:likelihood_position_pos_dep}
\end{align}
The difference between Eq.~\eqref{eq:likelihood_position} and Eq.~\eqref{eq:likelihood_position_pos_dep} is that the noise power spectrum is now density-dependent:
\begin{align}
    P_{\rm err}[\delta({\bf x})] = P_{\epsilon\epsilon} + 2 P_{\epsilon\epsilon_\delta}\delta_1({\bf x}) + P_{\epsilon_\delta\epsilon_\delta}\delta_1^2({\bf x}).
\end{align}
We implement this likelihood to our field-level inference pipeline as well.

To test the impact of the density-dependent noise, we generate a mock according to Eq.~\eqref{eq:density_noise_mock} using the LPT-based model with linear bias as the forward model. 
We set the noise power spectra to $P_{\epsilon\epsilon} = P_{\epsilon_\delta\epsilon_\delta} = \sqrt{2} P_{\epsilon\epsilon_\delta} = e^8$
\footnote{The correlated Gaussian fields, $\epsilon$ and $\epsilon_\delta$, are generated by $\epsilon_\delta=\epsilon_0$ and $\epsilon=(\epsilon_0 + \epsilon_1)/\sqrt{2}$, 
where $\epsilon_0$ and $\epsilon_1$ are independent Gaussian random fields whose spectra are $P_{\epsilon_0\epsilon_0} = P_{\epsilon_1\epsilon_1} = e^8$.}
and for other parameters we use their fiducial values (Tab.~\ref{tab:params}).
We then perform field-level inference using the position-space likelihood, both with and without the density-dependent variance (Eq.~\eqref{eq:likelihood_position} and Eq.~\eqref{eq:likelihood_position_pos_dep}).
The results are presented in Fig.~\ref{fig:density_dep_noise}.
As a reference, we also plot the purely Gaussian case in which the mock contains only Gaussian noise and inference is performed using a Gaussian likelihood without density-dependent variance (identical to the case shown in Fig.~\ref{fig:position_space_like}).

We confirm that the field-level inference is successful even for the mock with the density-dependent noise, when using the likelihood with the density-dependent variance, as indicated by the blue contours.
On the other hand, similar to the Poissonian noise case, we again find that using \textit{incorrect} field-level likelihood leads to the biased result, as indicated by the orange contours.
In particular, the inferred $P_{\rm err}$ deviates significantly from its true value.
As a result, the uncertainties on the cosmological parameters are slightly underestimated compared to those obtained using the \textit{correct} likelihood.

\subsection{Summary}
As we have shown, using the \textit{correct} field-level likelihood is crucial in field-level inference, in order to get the unbiased result.
Although it is often assumed that the field-level likelihood takes the Gaussian form in the literature, there is no guarantee that this assumption holds.
From a theoretical perspective, the residual noise is expected to be non-Gaussian (e.g. the Poisson noise and the density-dependent noise discussed above).
Ref.~\cite{Schmittfull:2018yuk} explicitly demonstrated that the distribution of the residuals is indeed non-Gaussian by comparing a perturbative forward model to $N$-body halos.
We would like to note here that when we apply our field-level pipeline with the Gaussian likelihood to $N$-body halos, the MCMC fails to converge. 
In particular, the inferred noise amplitude keeps decreasing without reaching a stable value.

On the other hand, the power spectrum (and the bispectrum) analysis can perturbatively take into account these non-Gaussian field-level noise in their theoretical templates, 
while still relying on a Gaussian likelihood.
In other words, even if the underlying field-level noise is non-Gaussian, the likelihood for the power spectrum or bispectrum remains Gaussian thanks to the central limit theorem,
since many independent Fourier modes are averaged in each bin.
This argument does not apply to field-level inference, where individual Fourier modes are compared directly.
In this sense, the power spectrum (and the bispectrum) analysis is more robust against non-Gaussian field-level noise.

\section{Summary and Conclusions}
\label{sec:conclusion}

In this work, we have explored the potential of field-level inference (FLI) with perturbative forward modeling for constraining cosmological parameters. 
Our main findings are summarized below.

\begin{itemize}
    \item The power spectrum analysis offers comparable constraining power to FLI, even for non-Gaussian fields, when varying only cosmological parameters with no free nuisance parameters.
    This is because the additional higher-order non-Gaussian information does not help to break the degeneracies in this case.
    In the perturbative regime, all higher-order effects, including higher-order loop corrections and higher-order correlators, are suppressed by the small density variance, $\Delta^2(k_{\rm max})\ll 1$.
    \item When nuisance parameters are varied, non-Gaussian information does help to break degeneracies between cosmological and nuisance parameters.
    In the perturbative regime, however, adding the bispectrum, which is the leading-order non-Gaussian statistic, is already sufficient to recover the results from the FLI.
    This implies that the trispectrum and higher-order correlators carry little additional information.
    This is theoretically expected when the inverse perturbative model is accurate on large scales and in the zero-noise limit,
    and we explicitly confirm that it still holds in the presence of finite noise.
    \item We also verify that the FLI is almost equivalent to the joint power spectrum and bispectrum analysis in real space, even when the large displacement and the higher-order galaxy bias (up to the quadratic order) are taken into account.
    The FLI posterior for cosmological parameters is just ${\cal O}(10\%)$ tighter than the joint power spectrum and bispectrum analysis at $k_{\rm max}\lesssim 0.12~h/{\rm Mpc}$.
    \item In redshift space, joint constraints from the power spectrum multipoles ($P_0+P_2+P_4$) and the bispectrum monopole ($B_0$) 
    reproduce the FLI constraints to within $\lesssim 10\%$ for cosmological parameters at $k_{\rm max} \lesssim 0.12~h/{\rm Mpc}$. 
    While $P_0+P_2+P_4$ and $P_0 + B_0$ give comparable errors at $k_{\rm max}\lesssim 0.1~h/{\rm Mpc}$,
    the bispectrum monopole becomes more informative than the power spectrum quadrupole at higher $k_{\rm max}$, as expected from SNR considerations.
    In reality, however, extending the bispectrum analysis beyond $0.1~h/{\rm Mpc}$ would require one-loop terms with the cubic and quartic bias parameters.
    A more realistic comparison is therefore to extend the power spectrum analysis to higher $k_{\max}$ while keeping the bispectrum analysis with lower $k_{\max}$.
    We find that pushing the power spectrum to $k_{\rm max}=0.2~h/{\rm Mpc}$ while keeping the bispectrum at $k_{\rm max}=0.08~h/{\rm Mpc}$ yields constraints that are as tight as, 
    or for shape-sensitive parameters even tighter than, the FLI results obtained with $k_{\rm max} \leq 0.12~h/{\rm Mpc}$. 
    \item A finite Nyquist cutoff $\Lambda$ in the sampled initial conditions affects both the forward-model signal and the noise at the field level.
    While its impact on the forward-model signal can be absorbed by the standard EFT counter terms,
    it can also alter the noise amplitude and even modify the field-level noise properties.
    We also confirm that its impact scales with the grid resolution $\Lambda$ rather than the likelihood cutoff $k_{\rm max}$.
    In other words, in order to minimize these effects it is better to choose larger $\Lambda$.
    \item While the Gaussian form of the field-level likelihood is commonly assumed in the literature, 
    we explicitly show that this assumption becomes incompatible with the practical goal of extracting non-Gaussian information from the data.
    In contrast, non-Gaussian field-level noise can be systematically accounted for within perturbative power spectrum and bispectrum analyses.
\end{itemize}

Our conclusions apply to the perturbative regime where non-Gaussianity is suppressed by the small parameter $\Delta^2(k_{\rm max})\ll 1$.
It is plausible to hope that field-level inference overtakes standard power spectrum and bispectrum analyses once we enter the non-perturbative regime,
yet this hope depends critically on the field-level likelihood.
The results we have obtained on non-Gaussian stochasticity have an important ramification there:
the growing non-Gaussianity that could offer additional information also renders the Gaussian field-level likelihood unreliable and converts any neglected stochasticity into a significant source of bias in the inference.
Thus, the construction of a correct field-level likelihood that properly accounts for the non-Gaussian stochasticity remains the key requirement, especially beyond the perturbative regime.
Addressing this issue appears to be the critical step toward practical applications of field-level inference.\footnote{This might not be the case for map-level inference of the weak lensing, where each 2D Fourier mode effectively averages over many line-of-sight modes, potentially recovering the Gaussian likelihood~\cite{Jeong:2011rh}.
A dedicated study is still required to confirm this.}

\section*{Acknowledgements}
KA is supported by Fostering Joint International Research (B) under Contract No.21KK0050 and the Japan Society for the Promotion of Science (JSPS) KAKENHI
Grant No.JP24K17056. SC acknowledges support from the National Science Foundation at the IAS through NSF/PHY 2207583. Support for this work was provided by NASA through the NASA Hubble Fellowship grant
\#HST-HF2-51572.001 awarded by the Space Telescope Science Institute, which is operated by the
Association of Universities for Research in Astronomy, Inc., for NASA, under contract
NAS5-26555.
Numerical computations were carried out on Typhon cluster and Apollo GPU nodes at the Institute for Advanced Study and 
on Cray XD2000 system at Center for Computational Astrophysics, National Astronomical Observatory of Japan.

\appendix

\section{Expressions for the bispectrum}
\label{app:bispectrum}

\begin{figure}[t]
    \centering
    \includegraphics[width=\textwidth]{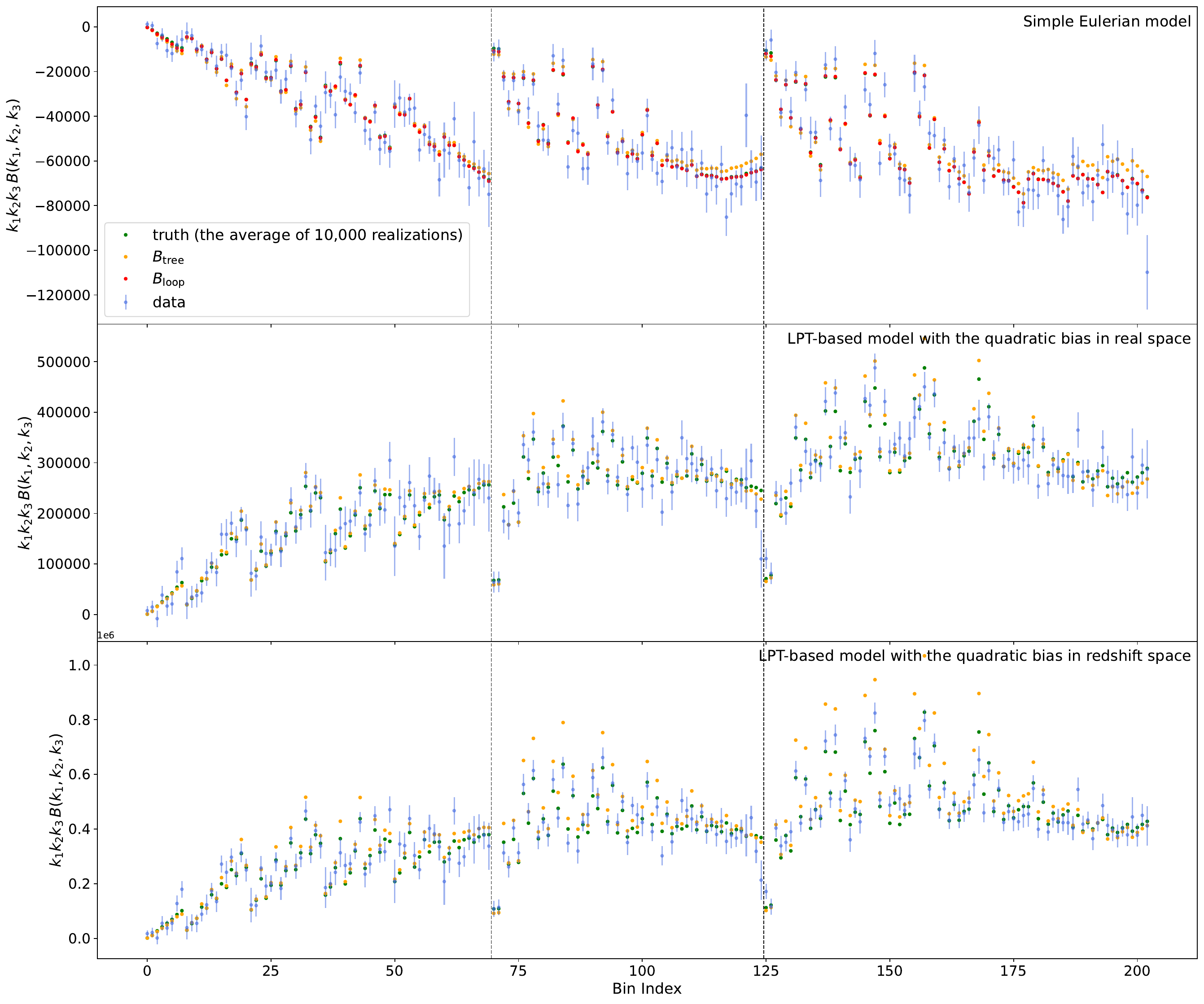}
    \caption{
    \textit{From top to bottom:} the bispectra for the simple Eulerian model, and the LPT-based model with the quadratic bias in real space and redshift space.
    The green points represent the mean bispectra over 10,000 realizations (``truth''),  
    the orange points the tree-level prediction,  
    the red points the one-loop prediction (shown only for the simple Eulerian model),  
    and the blue points with error bars a single mock realization.  
    Dashed vertical lines indicate where the maximum wavenumber $k_3$ exceeds $0.08~h/{\rm Mpc}$ (grey) and $0.1~h/{\rm Mpc}$ (black), 
    under $k_1 \leq k_2 \leq k_3$.
    }
    \label{fig:bispec}
\end{figure}

Following Ref.~\cite{Scoccimarro:1999ed},
we can choose the following parametrization for angles 
\begin{align}
\mu_1 = \hat{\bk}_1 \cdot \hat{\boldsymbol{n}}\;, \qquad \mu_2 = \mu_1 \cos \theta - \sqrt{1-\mu_1^2} \sin \theta \cos \phi \;, \qquad \mu_3 = - \frac{k_1}{k_3} \mu_1 - \frac{k_2}{k_3} \mu_2 \;,
\end{align}
where $\cos\theta\equiv x= \hat{\bk}_1\cdot \hat{\bk}_2$.
The bispectrum monopole can be computed integrating over~$\mu_1$ and~$\phi$
\begin{align}
    B_{g,{\rm tree}}^0 (k_1,k_2,k_3) = \frac{1}{4\pi} \int_0^{2\pi} {\rm d}\phi \int_{-1}^1 {\rm d} \mu_1 \; B_{g,{\rm tree}}(\bk_1,\bk_2,\bk_3,\hat{\boldsymbol{n}}) \;.
\end{align}
The integrals can be done analytically at this order in perturbation theory. Focusing on a single permutation that contributes to the tree-level result, we can define a new monopole kernel as follows   
\begin{align}
Z_2^0 (\bk_1,\bk_2) & \equiv \frac{1}{4\pi} \int_0^{2\pi} {\rm d}\phi \int_{-1}^1 {\rm d} \mu_1 \; Z_1(\bk_1)Z_1(\bk_2)Z_2(\bk_1,\bk_2)   \nonumber \\
& = \left[ \frac{b_2}{2} + \left( b_{\mathcal G_2} - \frac{b_1}{2} \right) (x^2 -1) \right] Q_1 (x) + (b_1+1) \left[ \frac12 + \frac 12 x \left( \frac{k_1}{k_2} + \frac{k_2}{k_1} \right) + \frac12 x^2 \right] Q_1(x) \nonumber \\
& +f  \left[ \frac 12 x \left( \frac{k_1}{k_2} + \frac{k_2}{k_1} \right) + x^2 \right] Q_2(k_1,k_2,x) + f Q_3(k_1,k_2,x) \;. 
\end{align}
Here we have defined  
\begin{align}
Q_1(x) \equiv \frac{15 b_{\rm E}^2 + 10 b_{\rm E} f +f^2 + 2 f^2 x^2}{15} \;,
\end{align}
\begin{align}
Q_2(k_1,k_2,x) \equiv 2f(x^2-1) \frac{k_1^2+k_2^2}{k_3^2} \frac{7b_{\rm E} + 3f}{105} + \frac{35 b_{\rm E}^2 + 42 b_{\rm E} f + 9 f^2 + 6 f^2 x^2}{105} \;,
\end{align}
\begin{align}
Q_3(k_1,k_2,x) & \equiv x \frac{k_3^2}{k_1 k_2} \frac{105 b_{\rm E}^3 + 189 b_{\rm E}^2 f + 99 b_{\rm E} f^2 + 15 f^3 + 36 b_{\rm E} f^2 x^2 + 20 f^3 x^2}{630} \nonumber \\
& \qquad - (x^2-1) \frac{35 b_{\rm E}^3 + 35 b_{\rm E}^2 f + 9 b_{\rm E} f^2 + f^3 + 12 b_{\rm E} f^2 x^2 + 4 f^3 x^2}{105} \;.
\end{align}
Again, the real-space bispectrum can be obtained by setting $f=0$ in the equation above.

Fig.~\ref{fig:bispec} shows the comparisons between the theoretical predictions and the measurements of the bispectra
for the simple Eulerian model (the top panel) and the LPT-based model with the quadratic bias in real space (the middle panel) and redshift space (the bottom panel).
Note that the bottom panel shows the bispectrum monopole.

\begin{figure}[t]
    \centering
    {$k_\text{max} = 0.1~h/\text{Mpc}$\par\vspace{1ex}}
    \includegraphics[width=0.9\textwidth]{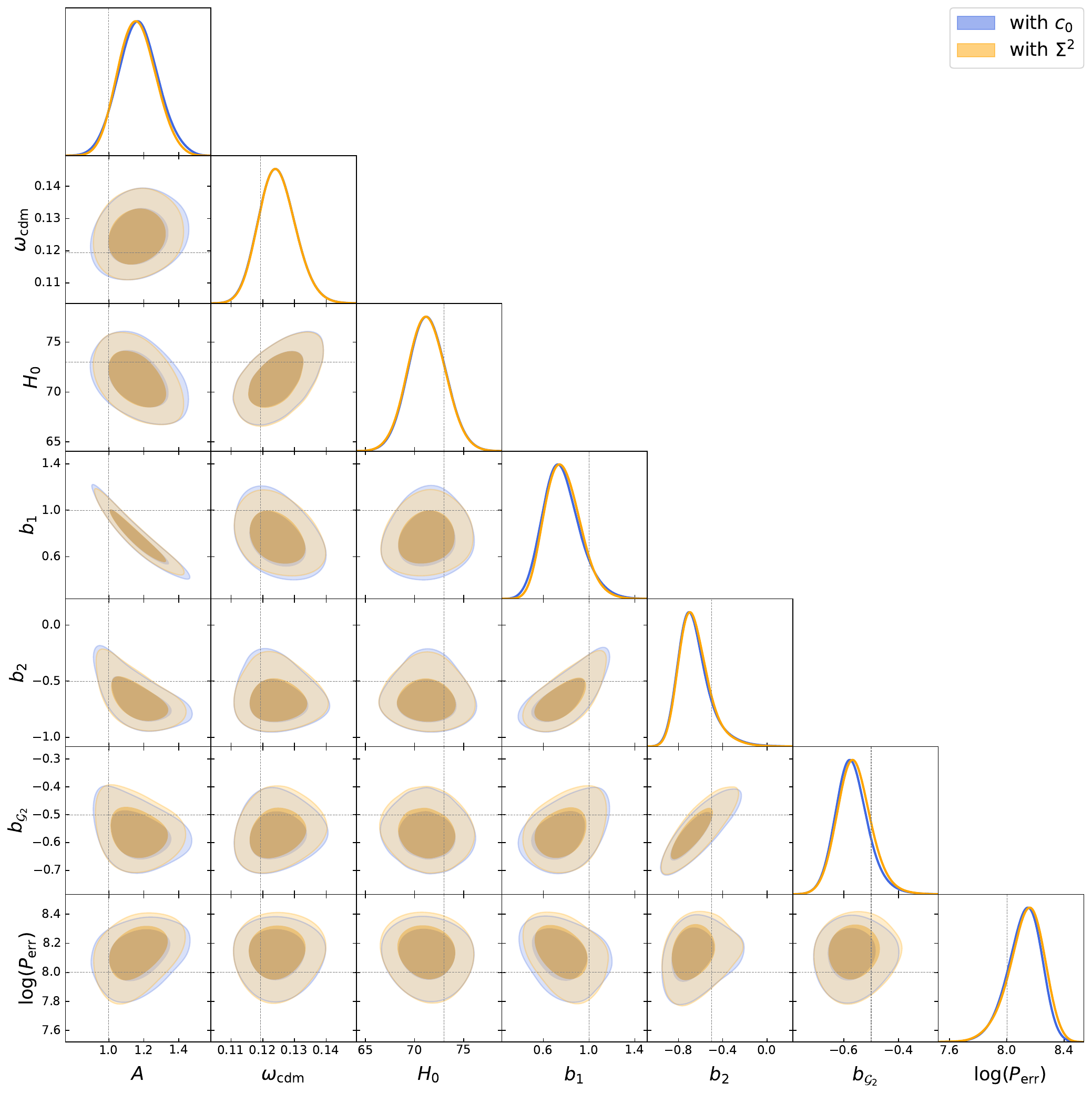}
    \caption{The 2D marginalized posterior distributions for the LPT-based model with the quadratic bias.
    The blue contours show the results with the $c_0$ counter term, while the orange contours show the results with the exact damping with $\Sigma^2$.
    }
    \label{fig:Sigma2_biased_tracer}
\end{figure}

\section{Zel'dovich damping in the biased tracer}
\label{app:sigma2_bias}

Starting from Eq.~\eqref{eq:LPT_model_field},
now we can split all linear fields into long and short. Then we get
\begin{align}
\delta_g({\bk}) & = \int \dd^3{\bf q}~\left( b_1 \delta_{1,{\rm L}}(\bq)+ \frac {b_2^2}2 \delta_{2,{\rm L}}(\bq) + b_{\mathcal G_2} \mathcal G_{2,{\rm L}}(\bq) + \mathcal O(\bq) \right)e^{-i \bk \cdot (\bq+ \bPsi_{\rm L}(\bq)+ \bPsi_{\rm S}(\bq))} ,
\end{align}
where 
\begin{align}
    \mathcal O(\bq) = \mathcal O_{\rm S}(\bq) + \mathcal O_{\rm LS}(\bq) = \left(  b_1 \delta_{1,{\rm S}}(\bq)+ \frac {b_2^2}2 \delta_{2,{\rm S}}(\bq) + b_{\mathcal G_2} \mathcal G_{2,{\rm S}}(\bq) \right) + \left( b_2^2 \delta_{1,{\rm L}}(\bq) \delta_{1,{\rm S}}(\bq)  + b_{\mathcal G_2} \mathcal G_{2,{\rm L}{\rm S}}(\bq) \right) \;.
\end{align}
We can see that 
\begin{align}
    \langle \mathcal O_{\rm S}(\bq) e^{-i \bk \cdot \bPsi_{\rm S}(\bq)} \rangle_{\rm S} = 0 ,
\end{align}
because this has exactly the form of short-wavelength $\delta_g$ whose mean is zero by construction. One can see that also 
\begin{align}
    \langle \mathcal O_{\rm LS}(\bq) e^{-i \bk \cdot \bPsi_{\rm S}(\bq)} \rangle_{\rm S} = 0 ,
\end{align}
for similar reasons and the fact that the displacement field does not correlate with density in the same point. 
These considerations lead to
\begin{align}
    \langle\delta_g({\bk}) \rangle_{\rm S} & = e^{-\frac 12 k^2 \Sigma_{\rm S}^2} \int \dd^3{\bf q}~\left( b_1 \delta_{1,{\rm L}}(\bq)+ \frac {b_2^2}2 \delta_{2,{\rm L}}(\bq) + b_{\mathcal G_2} \mathcal G_{2,{\rm L}}(\bq) \right)e^{-i \bk \cdot (\bq+ \bPsi_{\rm L}(\bq) )}.
\end{align}
Thus, the same damping factor appears for the biased tracer as in Sec.~\ref{subsec:resolution_mismatch_signal}.

Fig.~\ref{fig:Sigma2_biased_tracer} compares the posteriors of the LPT-based model with the quadratic bias using the standard counterterm (Eq.~\eqref{eq:counterterm}, the blue contours) and the exact form derived above (the orange contours).
The two are essentially indistinguishable, given that the exponent is very small, $k^2\Sigma^2_{\rm S} \lesssim 0.01$, for $k\lesssim 0.1~h/{\rm Mpc}$.
These results explicitly justify the use of the standard counterterm in the main text.

\section{Impact of the second-order displacement}
\label{app:2LPT}

\begin{figure}[t]
    \centering
    {$k_\text{max} = 0.1~h/\text{Mpc}$\par\vspace{1ex}}
    \includegraphics[width=0.9\textwidth]{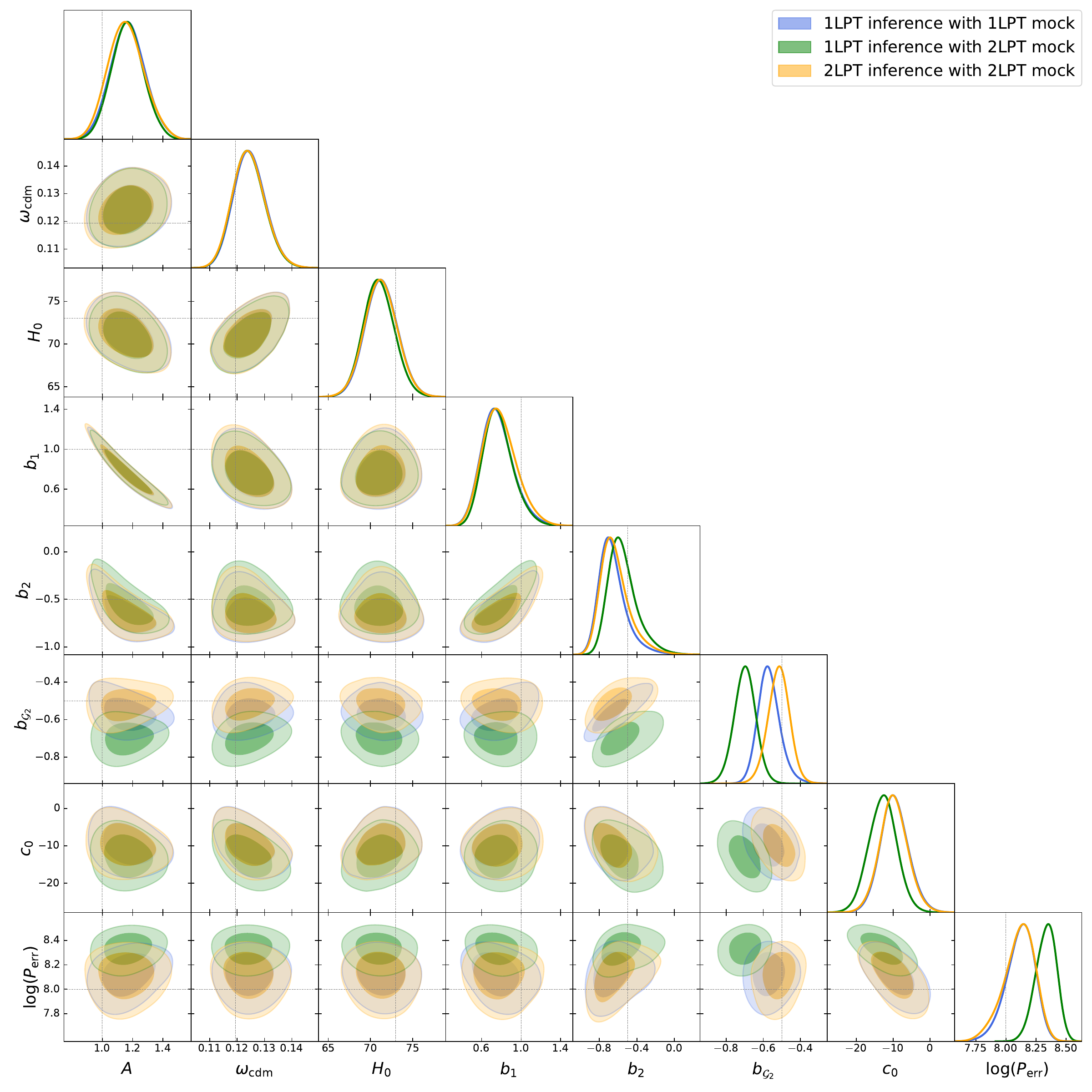}
    \caption{
    The 2D marginalized posterior distributions for the LPT-based model with the quadratic bias.
    The blue contours represent the results using the Zel'dovich (1LPT) displacement in both the inference model and the mock data (identical to the blue contours in the right panel of Fig.~\ref{fig:LPT_real_space_quad}).
    The green and orange contours correspond to the cases where the mock is generated with 2LPT displacements, while the inference uses 1LPT and 2LPT displacements, respectively.    }
    \label{fig:2LPT}
\end{figure}

Our main results in Sec.~\ref{sec:LPT} are all based on the model with the linear (1LPT) displacement field.
One might wonder if the higher-order displacement plays a important role.
In this appendix, we demonstrate that their impact is negligible.
To assess the impact of the higher-order displacement, we perform the field-level inference using a mock that includes the 2LPT displacement.
Specifically, we run the inference using forward models both with and without the 2LPT correction on this 2LPT mock.
The forward model used here is the LPT-based model with the quadratic bias, and we set $k_{\rm max}=0.1~h/{\rm Mpc}$.
Fig.~\ref{fig:2LPT} shows the comparison:
the orange corresponds to the 2LPT model with the 2LPT mock,
the green to the 1LPT model with the 2LPT mock,
and the blue to the 1LPT model with the 1LPT mock, identical to the result in the right panel of Fig.~\ref{fig:LPT_real_space_quad}.

It is evident that the 2LPT correction has no significant impact on the constraints of cosmological parameters. The only notable difference between the orange and blue contours appears in $b_{{\cal G}_2}$, which is expected since ${\cal G}_2$ represents the 2LPT potential.
This serves as an explicit example that the nonlinear term of the shifted operators correctly captures the effects of higher-order displacements.
In contrast, the green contours show a shift in the noise power spectrum as well. 
This can likely be attributed to the term $\sim \frac{b_2^2}{2}\delta^2_1({\bf q})e^{-i{\bf k}\cdot {\boldsymbol{\Psi}_{\rm 2LPT}({\bf q})}}$, which is absent in the 1LPT model and introduces a constant shot-noise-like contribution on large scales, as discussed in Sec.~\ref{sec:resolution_mismatch}.
These results demonstrate that higher-order displacements do not affect the conclusions presented in the main text.

\section{Results of the LPT-based model with $k_{\rm max} = 0.12~h/{\rm Mpc}$}
\label{app:lpt_0.12}

\begin{figure}[t]
    \centering
    {$k_\text{max} = 0.12~h/\text{Mpc}$\par\vspace{1ex}}
    \begin{subfigure}{0.49\textwidth}
        \centering
        \includegraphics[width=\textwidth]{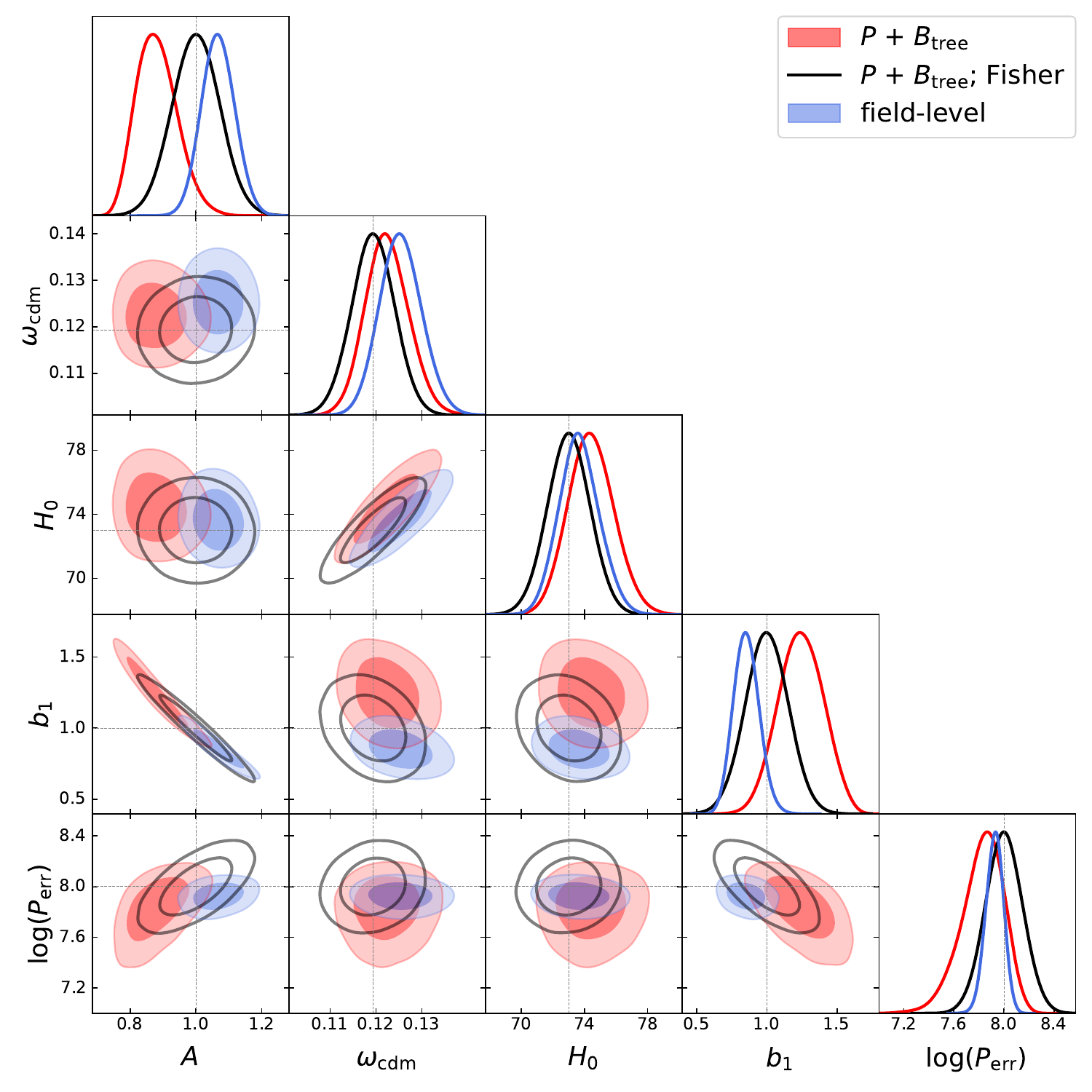}
    \end{subfigure}
    \begin{subfigure}{0.49\textwidth}
        \centering
        \includegraphics[width=\textwidth]{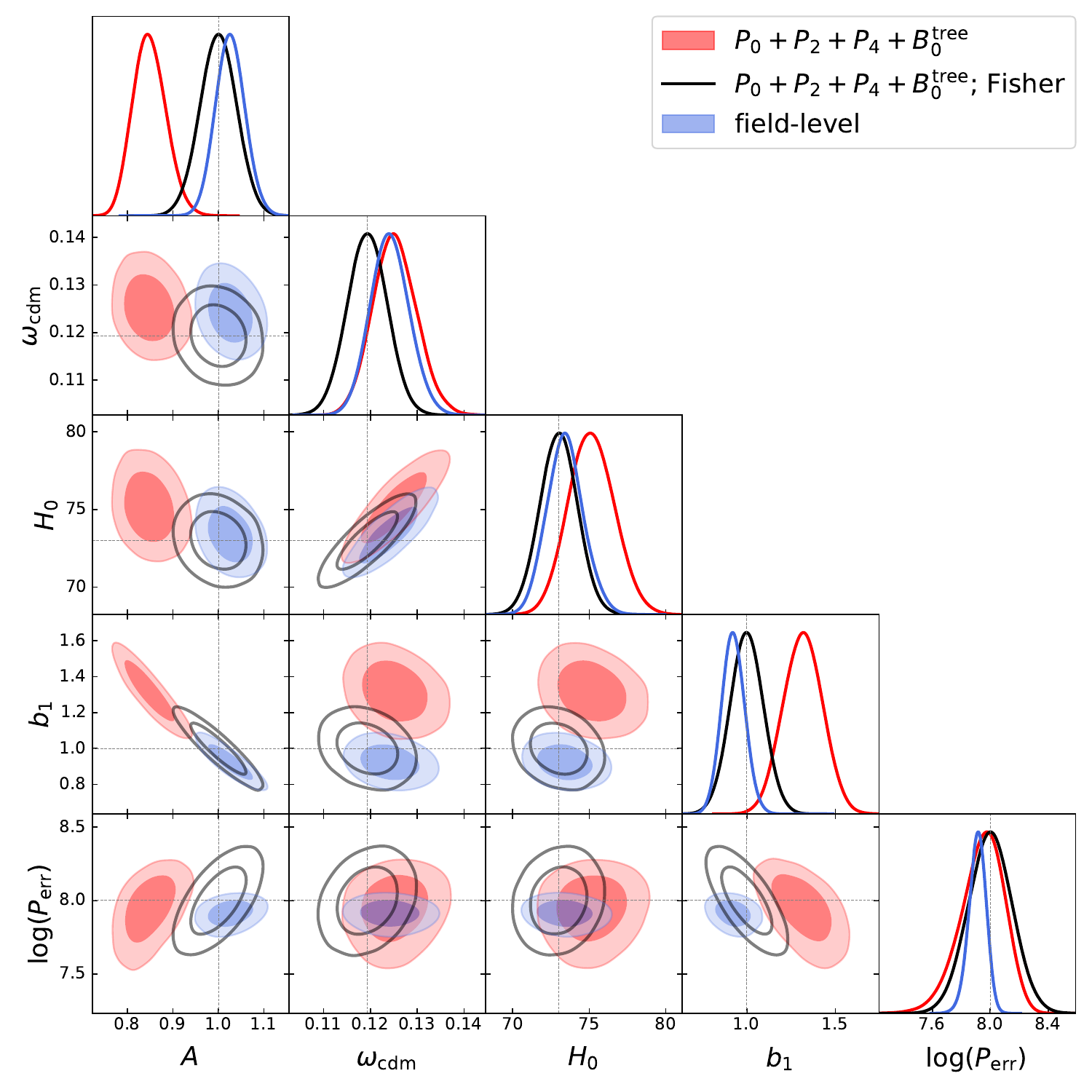}
    \end{subfigure} 
    \caption{The left and the right panels show the 2D marginalized posterior distributions for the LPT-based model with the quadratic bias in real and redshift space, respectively.
    The constraints from the joint power spectrum and tree-level bispectrum are shown in red, and the field-level in blue.
    The black lines indicate the inverse Fisher matrix estimate for the error contours for the joint power spectrum and the bispectrum analysis.
    }
    \label{fig:LPT_quad_0.12}
\end{figure}

\begin{table}[t]
    \centering
    \begin{tabular}{c|c|c|c|c|c}
        & $A$ & $\omega_{\rm cdm}$ & $H_0$ & $b_1$ & $\log(P_\text{err})$ \\ \hline
        $P+B_\text{tree}$ & $0.882^{+0.054}_{-0.070}$ & $0.1225^{+0.0045}_{-0.0050}$ & $74.5\pm 1.5$ & $1.24\pm 0.16$ & $7.83^{+0.19}_{-0.14}$\\ \hline
        field-level & $1.068\pm {0.050}$ & $0.1254\pm 0.0046$ & $73.7 \pm 1.3$ & $0.854\pm 0.091$ & $7.925\pm 0.071$\\
    \end{tabular}
    \caption{The mean values and 68\%-confidence intervals of the parameters for the LPT-based model with the quadratic bias parameters at $k_{\rm max} = 0.12~h/{\rm Mpc}$, 
    corresponding to the left panel of Fig.~\ref{fig:LPT_quad_0.12}.
    }
    \label{tab:LPT_real_space_quad_0.12}
\end{table}

\begin{figure}[t]
    \centering
    \begin{subfigure}{0.49\textwidth}
        \centering
        {$k_\text{max} = 0.12~h/\text{Mpc}$\par\vspace{1ex}}
        \includegraphics[width=\textwidth]{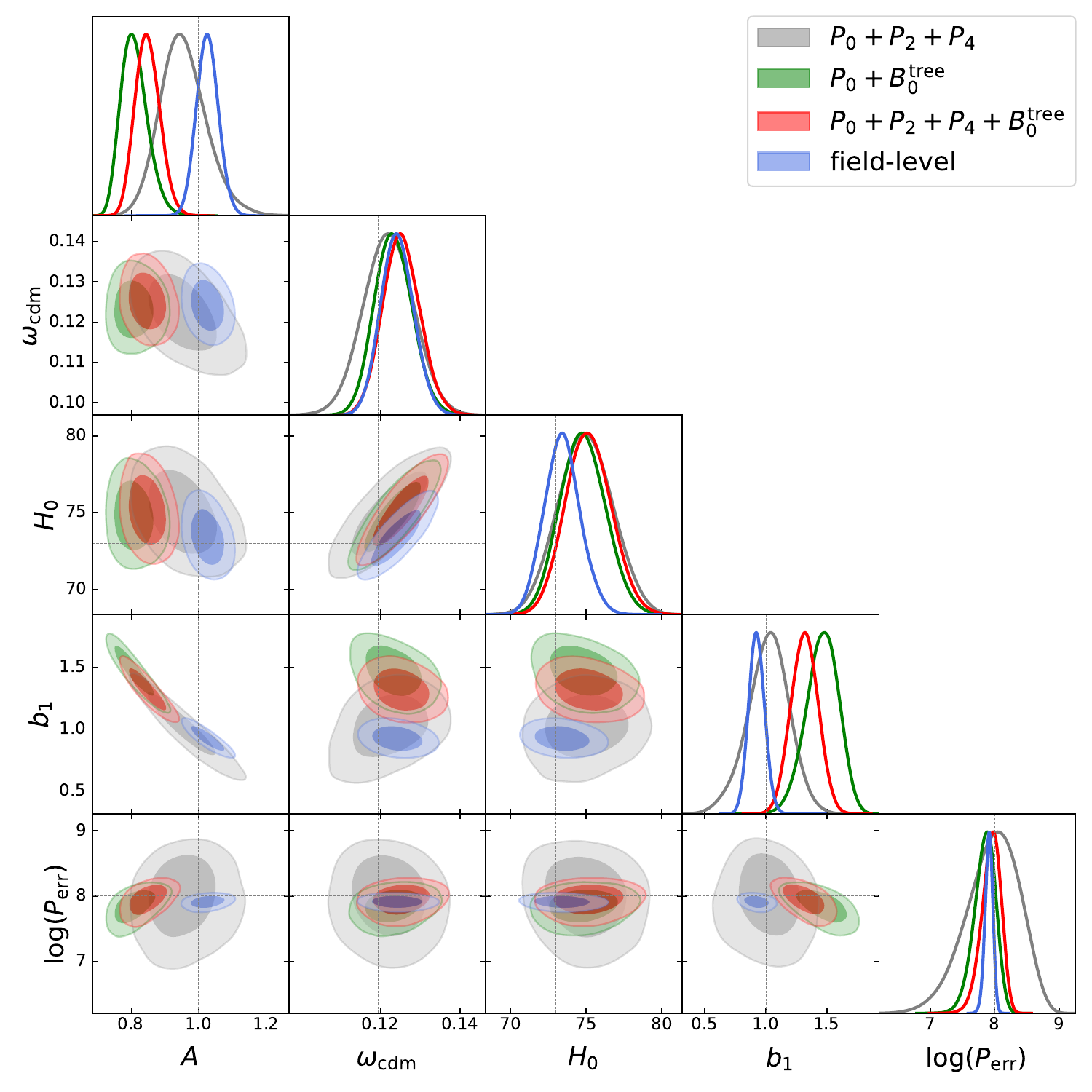}
    \end{subfigure}
    \begin{subfigure}{0.49\textwidth}
        \centering
        \includegraphics[width=\textwidth]{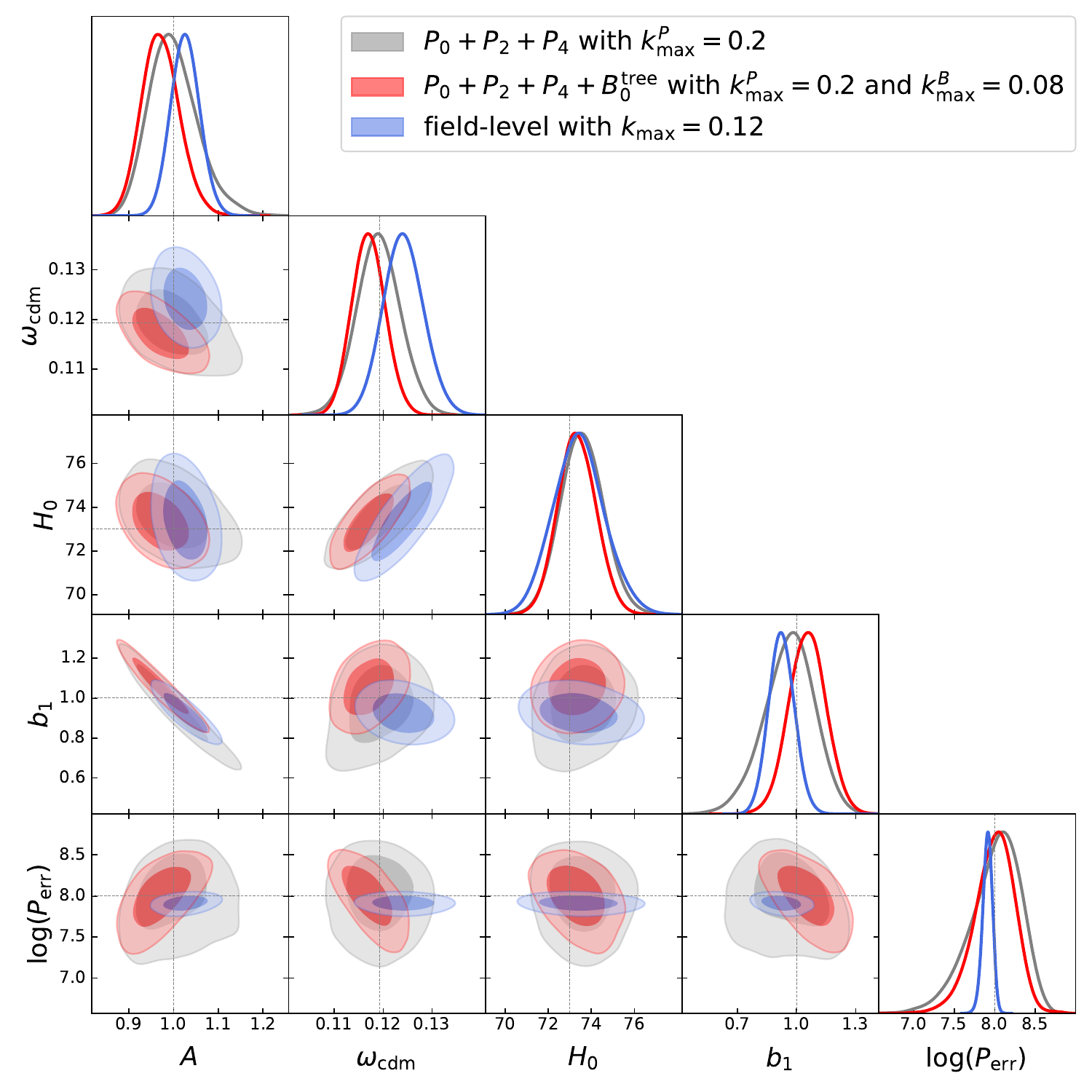}
    \end{subfigure} 
    \caption{The 2D marginalized posterior distributions for the LPT-based model with the quadratic bias in redshift space 
    at $k_\text{max} = 0.12~h/\text{Mpc}$ (left) and various choices of $k_\text{max}$ (right).
    The contours indicate 68\% and 95\% credible intervals.
    The gray contours show the results of the power spectrum multipoles alone,
    the green ones show the power spectrum monopole and the bispectrum monopole,
    the red ones show the power spectrum multipoles plus the bispectrum monopole,
    and the blue ones show the field-level.
    }
    \label{fig:LPT_redshift_space_quad_P_P+B_0.12}
\end{figure}

\begin{table}[t]
    \centering
    \begin{tabular}{c|c|c|c|c|c}
        & $A$ & $\omega_{\rm cdm}$ & $H_0$ & $b_1$ & $\log(P_\text{err})$ \\ \hline
        $P_0 + P_2 + P_4$ & $0.956^{+0.059}_{-0.076}$ & $0.1218\pm 0.0063$ & $74.9\pm 1.8$ & $1.02^{+0.18}_{-0.16}$ & $7.97^{+0.47}_{-0.34}$\\ \hline
        $P_0 + B_0^\text{tree}$ & $0.810^{+0.032}_{-0.045}$ & $0.1232\pm 0.0047$ & $74.8\pm 1.5$ & $1.46^{+0.15}_{-0.12}$ & $7.84^{+0.19}_{-0.14}$\\ \hline
        $P_0 + P_2 + P_4 + B_0^\text{tree}$ & $0.849^{+0.034}_{-0.039}$ & $0.1252\pm 0.0047$ & $75.2\pm 1.5$ & $1.32\pm 0.11$ & $7.94^{+0.17}_{-0.13}$\\ \hline
        field-level & $1.026\pm 0.032$ & $0.1242\pm 0.0041$ & $73.4 \pm 1.2$ & $0.924 \pm 0.065$ & $7.911^{+0.063}_{-0.055}$\\
    \end{tabular}
    \caption{The mean values and 68\%-confidence intervals of the parameters for the LPT-based model with the quadratic bias parameters in redshift space
    at $k_\text{max} = 0.12~h/\text{Mpc}$,
    corresponding to the right panel of Fig.~\ref{fig:LPT_quad_0.12} and the left panel of Fig.~\ref{fig:LPT_redshift_space_quad_P_P+B_0.12}.
    }
    \label{tab:LPT_rsd_space_quad_0.12}
\end{table}

In the main text, we show the results with $k_{\rm max} = 0.1~h/{\rm Mpc}$ for the LPT-based model, 
given that the tree-level bispectrum is not capable of capturing the non-Gaussian features on larger scales.
Fig.~\ref{fig:LPT_quad_0.12} and Fig.~\ref{fig:LPT_redshift_space_quad_P_P+B_0.12} extend the comparison in the main text to $k_{\rm max}=0.12~h/{\rm Mpc}$.
In both real and redshift space the field-level posterior (blue) and the joint power spectrum and bispectrum posterior (red) remains consistent in terms of the error bars,
although the latter posterior gets more biased.
Tab.~\ref{tab:LPT_real_space_quad_0.12} and Tab.~\ref{tab:LPT_rsd_space_quad_0.12} show the mean values and $68\%$ confidence intervales of the parameters for read and redshift space, respectively.
As noted in Sec~\ref{subsec:LPT_RSD}, $P_0 + B_0$ becomes more informative than $P_0+P_2+P_4$ at this scale.

\section{Impact of the non-Gaussian covariance in the joint power spectrum and bispectrum analysis}
\label{app:cov_NG}

\begin{figure}[t]
    \centering
    {$k_\text{max} = 0.12~h/\text{Mpc}$\par\vspace{1ex}}
    \begin{subfigure}{0.49\textwidth}
        \centering
        \includegraphics[width=\textwidth]{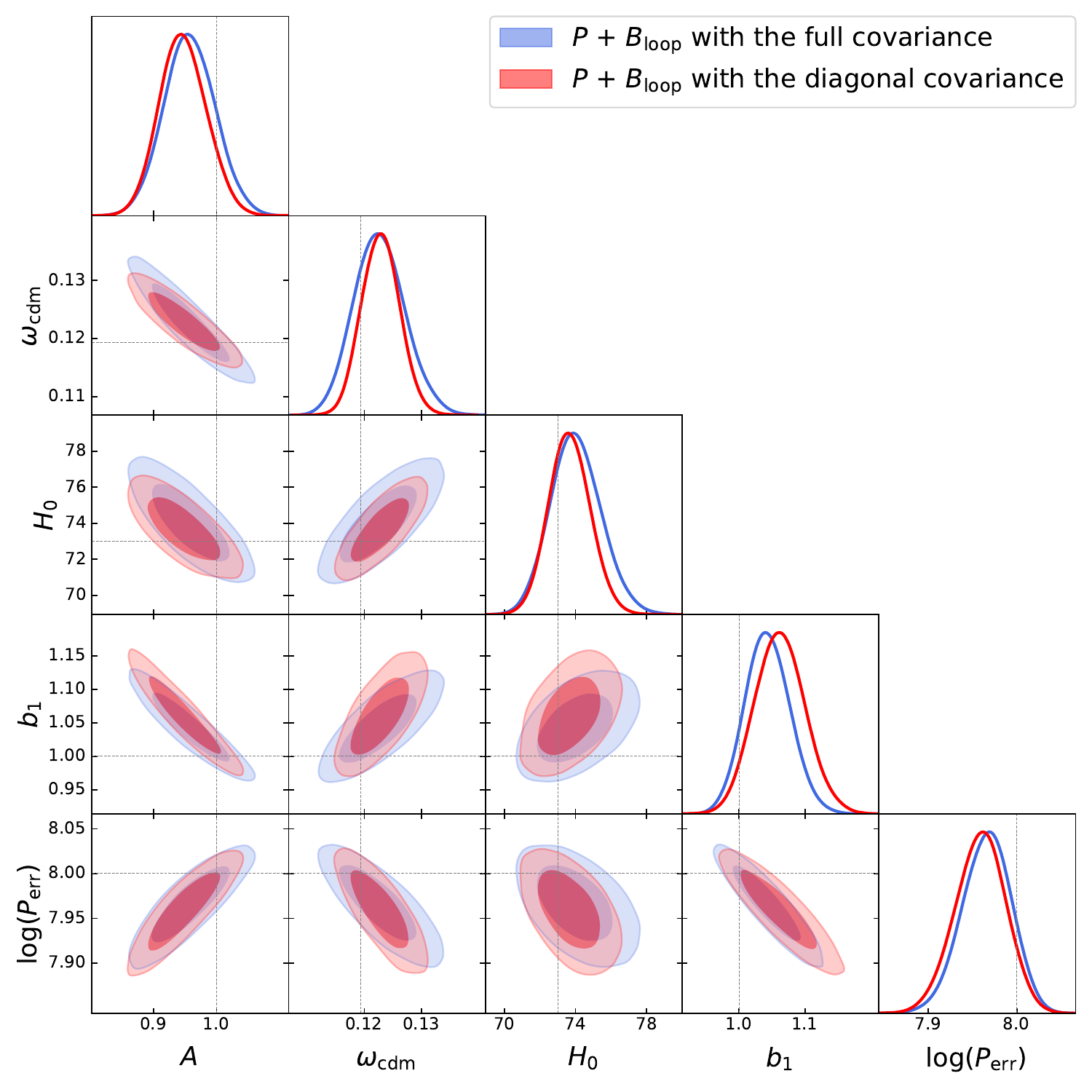}
    \end{subfigure}
    \begin{subfigure}{0.49\textwidth}
        \centering
        \includegraphics[width=\textwidth]{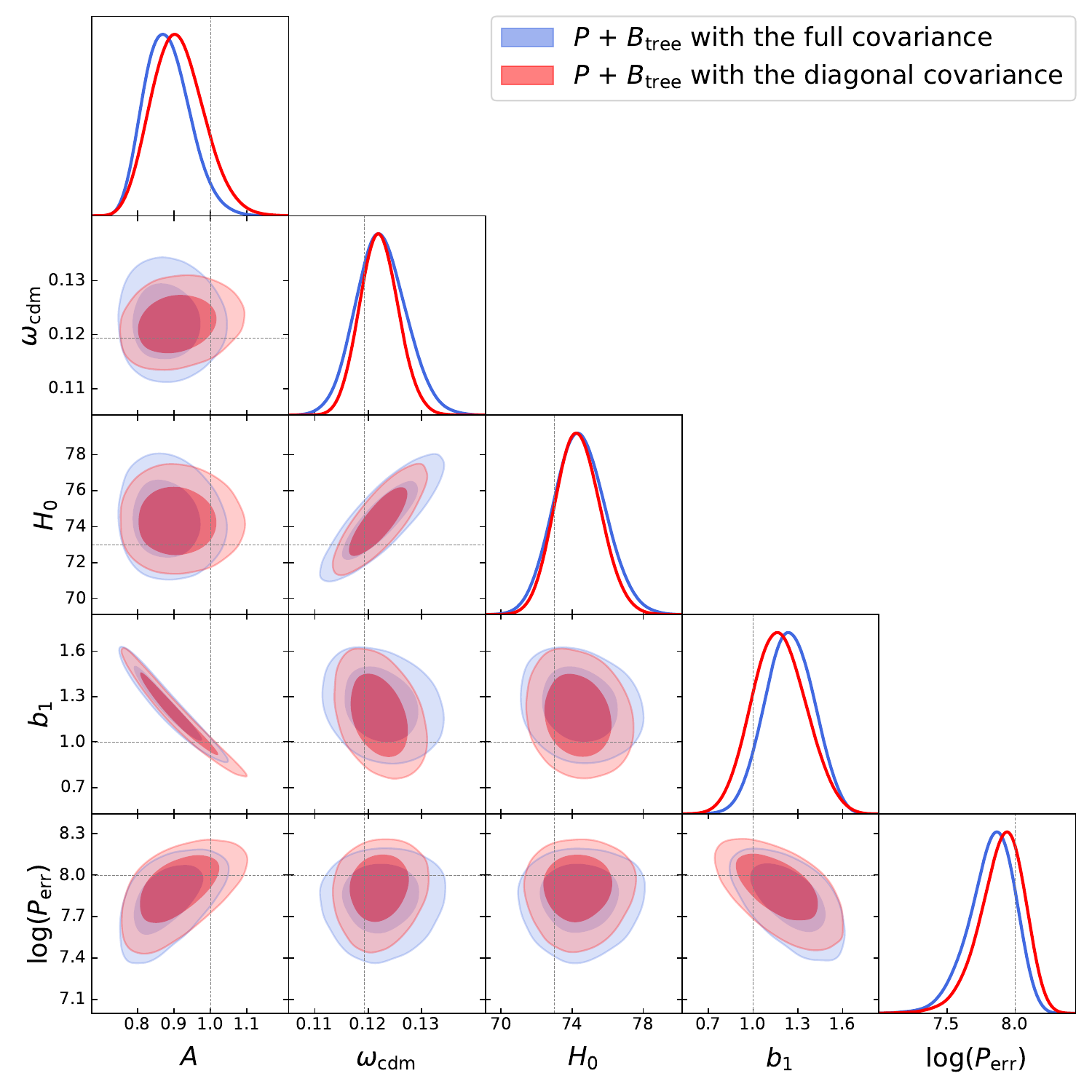}
    \end{subfigure}
    \caption{The left panel shows 2D marginalized posterior for the simple Eulerian model with the full covariance (the blue) and the diagonal covariance (the red).
    The blue contours are identical to the green contours in Fig.~\ref{fig:G2_b1_noise}.
    The right panel shows the same comparison, but for the LPT-based model with the quadratic bias.
    The blue contours are identical to the red contours in the left panel of Fig.~\ref{fig:LPT_quad_0.12}.
    }
    \label{fig:cov_NG}
\end{figure}

In order to test whether the non-Gaussian components of the covariance matters for the joint power spectrum and bispectrum analysis at the scales considered, 
we compare the results obtained using the full, mock-estimated covariance matrix with those from a diagonal covariance in Fig.~\ref{fig:cov_NG}.
The left panel shows the comparison for the simple Eulerian model and the right panel shows the comparison for the LPT-based model with the quadratic bias.

Although the two posteriors are indeed different, their difference is small and of order $\Delta^2(k_{\rm max}) \lesssim 0.1$, as expected.
We also note that the difference between the two becomes slightly smaller for the LPT-based model with the quadratic bias,
likely becauase parameter degeneracies in larger parameter space partially compensate the impact of the non-Gaussian covariance.

\section{Reconstructed initial conditions}
\label{app:reconstructed_IC}

\begin{figure}[htbp]
    \centering
    \begin{subfigure}[b]{\linewidth}
      \centering
      \hspace*{-0.5em}%
      \includegraphics[width=\linewidth]{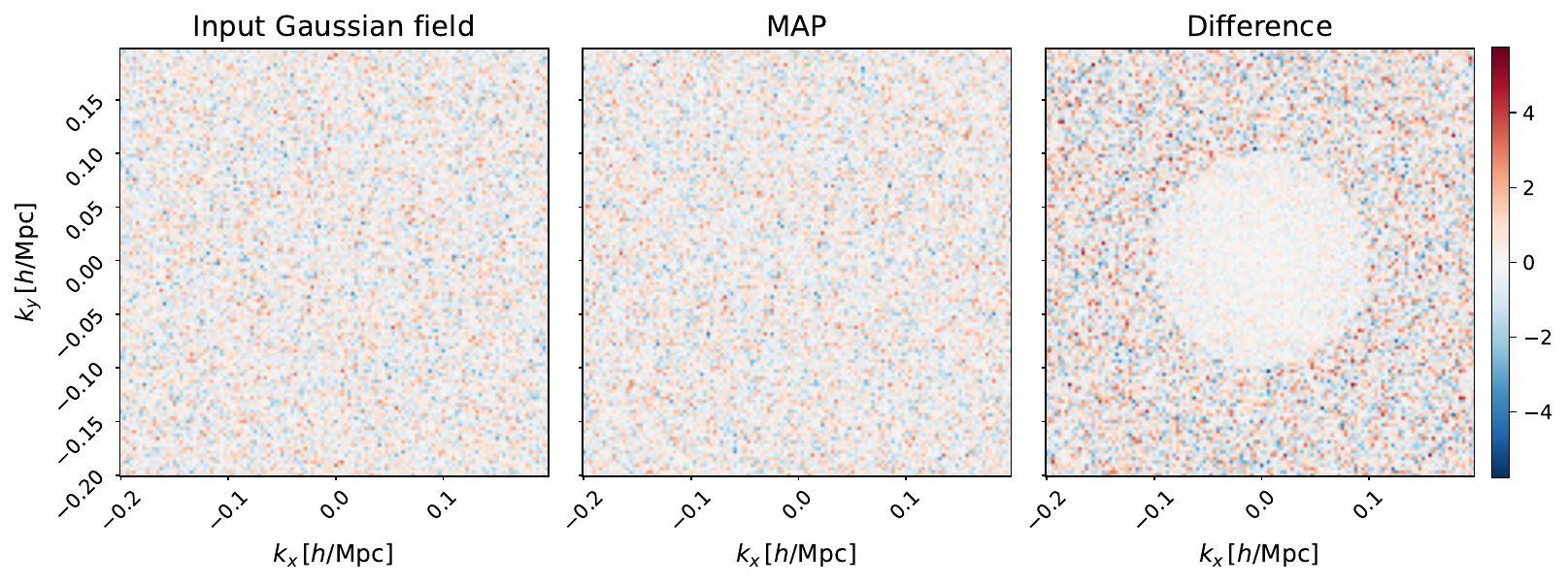}
    \end{subfigure}
  
    \begin{subfigure}[b]{\linewidth}
      \centering
      \includegraphics[width=\linewidth]{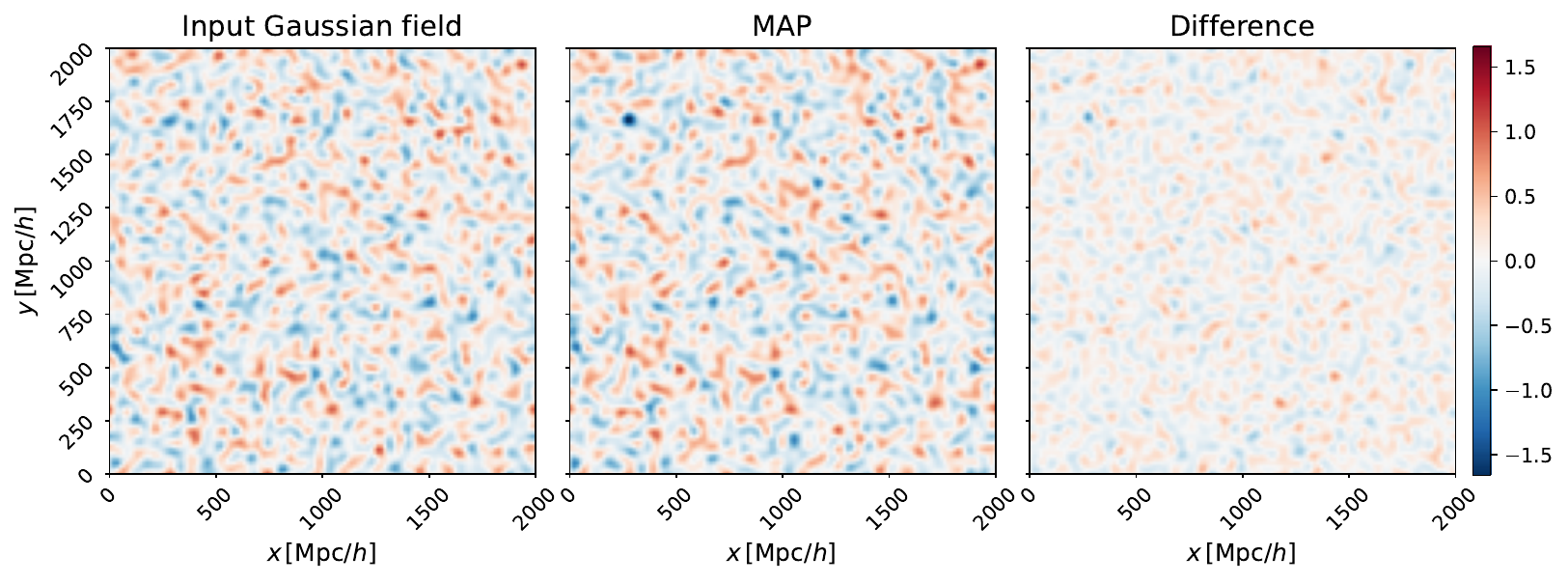}
    \end{subfigure}
  
    \begin{subfigure}[b]{\linewidth}
      \centering
      \includegraphics[width=\linewidth]{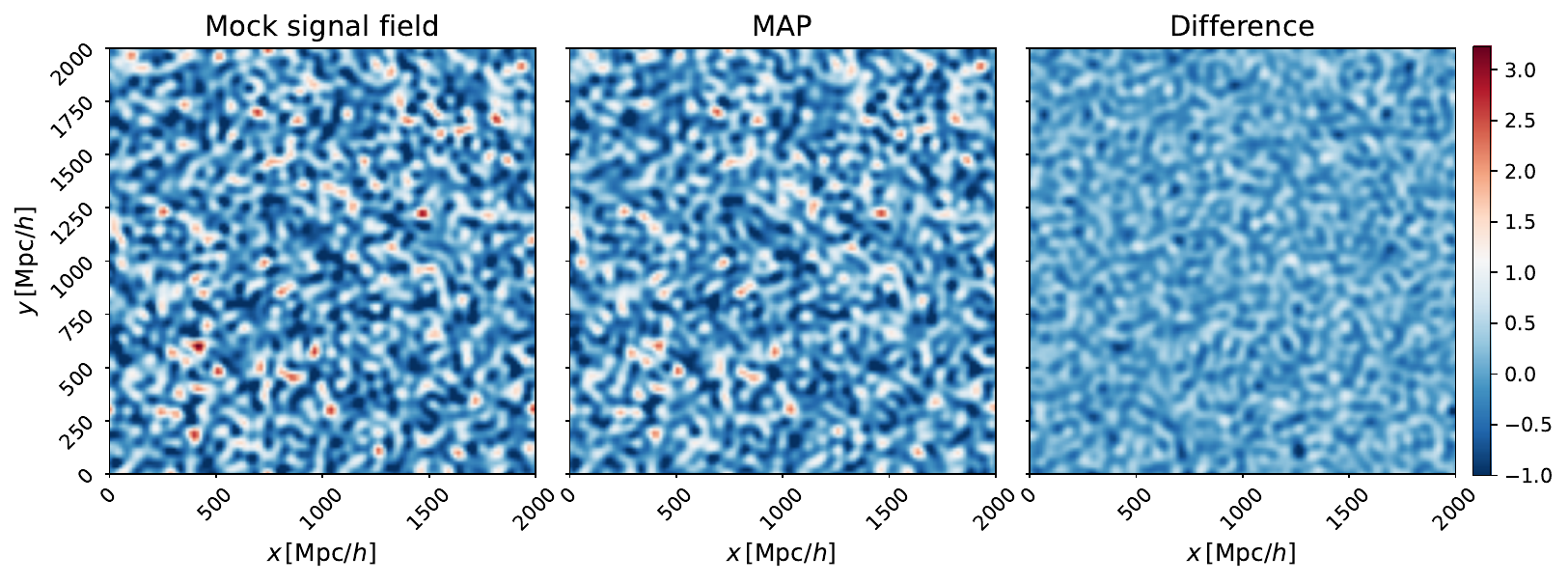}
    \end{subfigure}
  
    \caption{2D slices of the input (left), the MAP sample (center), and their difference (right) for the LPT-based model with the quadratic bias.
    The top panel shows the map for the initial Gaussian field in Fourier space at the $k_z=0$ plane, 
    the middle panel shows the map for the initial density field in position space, 
    and the bottom panel shows the map for the signal in position space.
    }
    \label{fig:recon_maps}
\end{figure}

\begin{figure}[t]
    \centering
    \begin{subfigure}{0.49\textwidth}
        \centering
        \includegraphics[width=\textwidth]{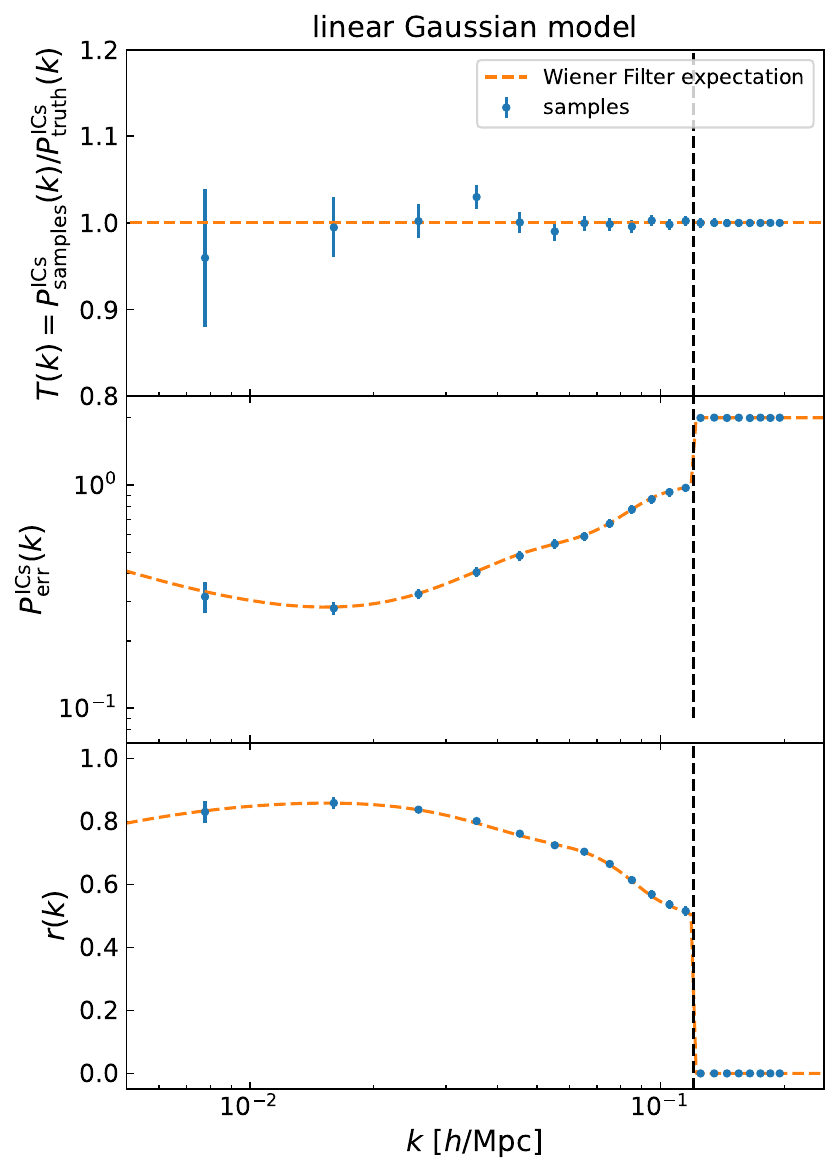}
    \end{subfigure}
    \begin{subfigure}{0.49\textwidth}
        \centering
        \includegraphics[width=\textwidth]{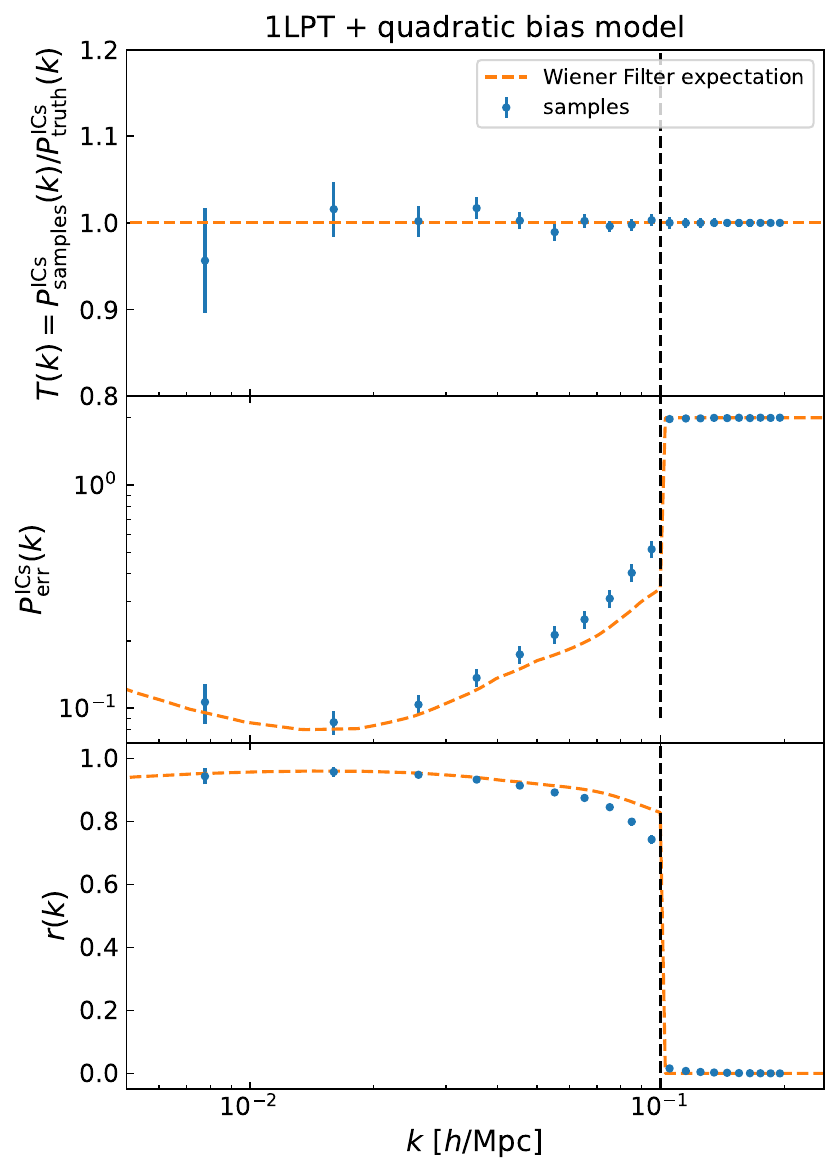}
    \end{subfigure} 
    \caption{
    \textit{From top to bottom:} The transfer function of the initial power spectrum $T(k)$, the power spectrum of the residuals $P_{\rm err}^{\rm ICs}(k)$, the cross-correlation coefficient of the initial conditions $r(k)$. 
    The left and right panels show the results for the linear Gaussian mock and the LPT-based model with the quadratic bias, respectively.
    The dashed vertical lines indicate $k_\text{max}$ used in each analysis.
    }
    \label{fig:ICs_statistics}
\end{figure}

While our focus is the posterior distribution of the cosmological parameters, 
the field-level inference enables us to reconstruct the initial conditions as a byproduct,
since what we sample is the joint posterior distribution of the cosmological parameters and the initial conditions, $\cP[\btheta, \delta|\hat{\delta}_g]$ (see Eq.~\eqref{eq:log_posterior}).
Given the huge dimensionality of the initial conditions, it is impractical to keep the full samples of the initial conditions (e.g. for our case the total size is $128^3 \times \text{\# of samples} \gtrsim 10^{11}$).
Hence we only keep track the MAP (maximum a posteriori) sample and the mean of the samples for the initial conditions, although our implementation allows to keep the full samples for the initial conditions as well.

As an example, we show the 2D slices of the reconstructed initial conditions for the LPT-based model with the quadratic bias (discussed in Sec.~\ref{subsec:LPT_real}) in Fig.~\ref{fig:recon_maps}.
The top and middle panels display 2D slices of the Fourier-space and the position-space maps of the initial conditions, respectively.
Specifically, the top panels show the initial Gaussian random field, ${\rm Re}~g({\bf k})\sim{\cal N}(0,1)$, which is related to the initial density field via
\begin{align}
    \delta_1({\bf k}) =  \sqrt{\frac{P_{\rm lin}(k)}{2}} g({\bf k}).
\end{align}
It is clear from the Fourier-space map that we are able to reconstruct the initial modes up to the scale of $k_\text{max}$ in the likelihood, 
regardless of the cutoff scale in the prior ($\Lambda$ in Eq.~\eqref{eq:log_posterior}).
We therefore apply low-pass filter to the initial modes in Fourier space to remove the high-$k$ modes above $k_\text{max}$ when making the figues in position space (the middle and bottom panels).
The bottom panel shows the 2D slices of the final density field in position space.
The reconstructed map is of course not perfect, but qualitatively captures the large-scale structure of the initial conditions as well as the final density field.

We can examine the statistical properties of the reconstructed initial conditions more quantitatively.
To this end, we compute the following spectra of $g({\bf k})$, instead of $\delta_1({\bf k})$, in order to isolate the statistical properties of the initial Gaussian field:
\begin{itemize}
    \item 
the transfer function, which is the ratio of the power spectrum of the sampled $g({\bf k})$ to the one of the true $g({\bf k})$, defined as 
\begin{align}
    T(k) = \frac{\langle| g_{\rm sample}({\bf k})|^2  \rangle}{\langle |g_{\rm truth}({\bf k}) |^2  \rangle} = \frac{P_{\rm sample}^{\rm ICs}(k)}{P_{\rm truth}^{\rm ICs}(k)}.
\end{align}
    \item 
the error power spectrum of the sampled $g({\bf k})$ to the one of the true $g({\bf k})$, defined as 
\begin{align}
    P^{\rm ICs}_{\rm err} = \langle |g_{\rm sample}({\bf k}) - g_{\rm truth}({\bf k}) |^2  \rangle.
\end{align}
    \item 
the cross-correlation coefficient between the sampled $g({\bf k})$ and the true $g({\bf k})$, defined as 
\begin{align}
    r(k) = \frac{\langle g_{\rm sample}({\bf k}) g^*_{\rm truth}({\bf k}) \rangle}{\sqrt{ P_{\rm sample}^{\rm ICs}(k) P_{\rm truth}^{\rm ICs}(k) }}.
\end{align}
\end{itemize}
Note that these quantities are related to each other. In particular, the error power spectrum can be written as 
\begin{align}
    P_{\rm err}^{\rm ICs} (k) = P_{\rm truth}^{\rm ICs}(k) \left[1 + T(k) - 2r(k)\sqrt{T(k)} \right].
\end{align}
These spectra are presented in Fig.~\ref{fig:ICs_statistics} from top to bottom.
As a reference we show the results of the linear problem (discussed in Sec.~\ref{subsec:Gaussian_case}) in the left, while we show the results of the LPT-based model with the quadratic bias in the right.

For the linear problem one can construct the optimal estimator for $g({\bf k})$, known as the Wiener-filter solution:
\begin{align}
    g_{\rm WF}({\bf k}) = \frac{1}{\sqrt{P_{\rm lin}(k) + P_{\rm err}}}  \hat{\delta}_g({\bf k}),
    \label{eq:WF_solution}
\end{align}
where $P_{\rm err} = \langle |\epsilon({\bf k})|^2 \rangle$.
The Wiener-filter solution can be used to obtain theoretical expectations for above statistics.
With Eq.~\eqref{eq:WF_solution} one can show that
\begin{align}
    T_{\rm WF}(k) &= 1
    \\
    P_{\rm err; WF}^{\rm ICs}(k) &= \frac{2P_{\rm err}(k)}{P_{\rm lin}(k) + P_{\rm err}(k)}
    \\
    r_{\rm WF}(k) &= \frac{P_{\rm lin}(k)}{P_{\rm lin}(k) + P_{\rm err}(k)},
\end{align}
which are shown as the orange dashed lines in Fig.~\ref{fig:ICs_statistics}.
Although this calculation is only meaningful for the linear problem, 
we also show the Wiener-filter expectations for the LPT-based model by replacing $P_{\rm lin}(k)$ with $P_{\rm signal}(k)$.

First, in the linear case shown in the left panels of Fig.~\ref{fig:ICs_statistics},
the statistics of the initial conditions are in excellent agreement with the Wiener filter prediction,
confirming that our field-level inference pipeline correctly samples the initial conditions.
In contrast, for the LPT-based model shown in the right panels,
the agreement is not as strong as in the linear case.
However, the Wiener filter prediction still captures the overall trend and matches the measurements on large scales, where nonlinear corrections are negligible.
Second, the transfer function remains unity even beyond $k_{\rm max}$,
indicating that the sampled initial conditions have, on average, the correct squared amplitude.
This is expected, as this condition is explicitly imposed by the prior on the initial conditions (Eq.~\eqref{eq:prior}).
Third, the phases of the initial conditions can be accurately reconstructed up to $k_{\rm max}$.
Beyond this scale, however, the phases of the sampled initial conditions become completely uncorrelated with the true values,
as this information is determined by the constraints imposed by the likelihood.

\section{Convergence of the MCMC chains}
\label{app:mcmc}

\begin{figure}[t]
    \centering
    \includegraphics[width=0.9\textwidth]{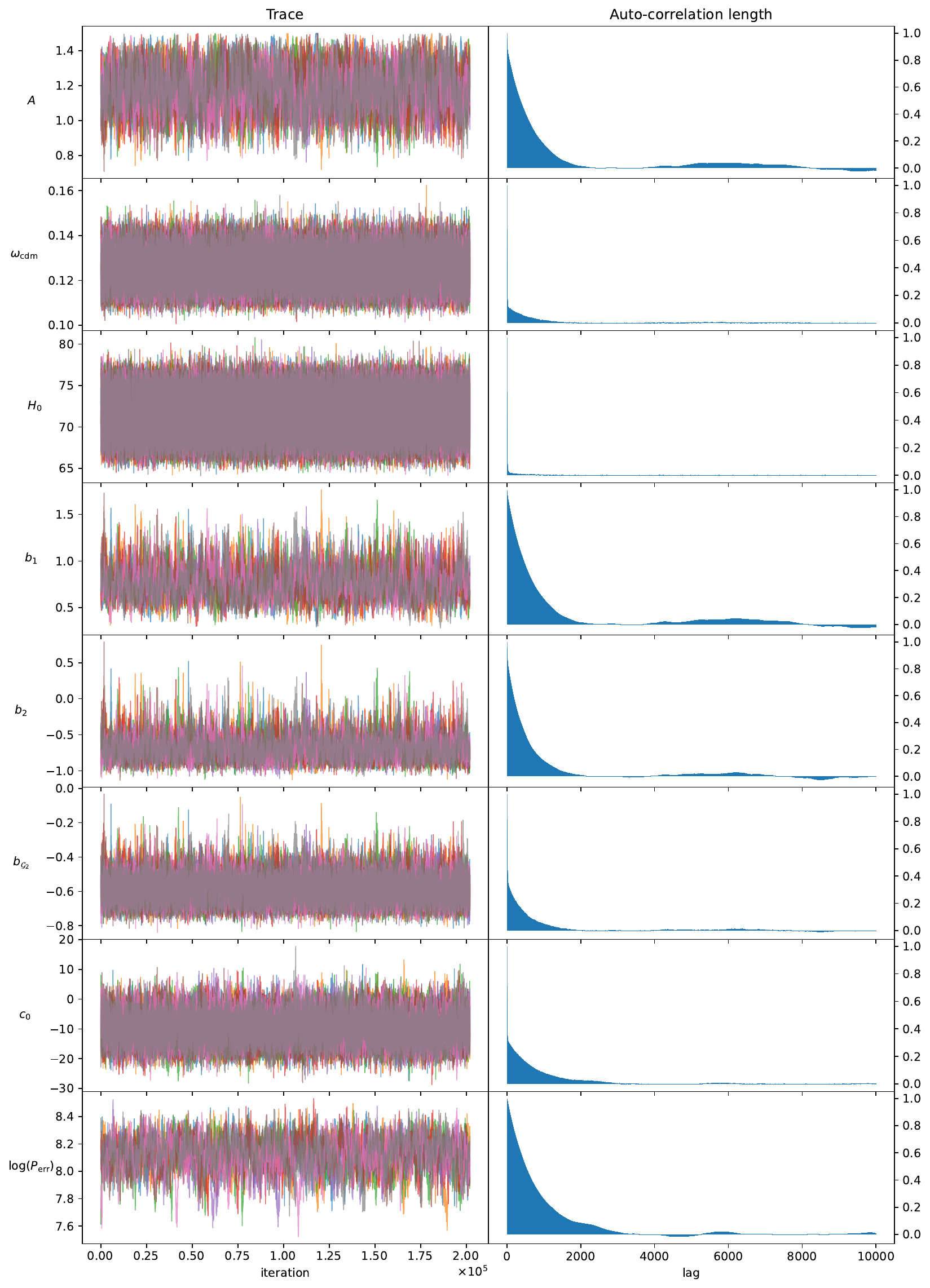}
    \caption{
    The trace plot (left) and the auto-correlation length (right) for the LPT-based model with the quadratic bias in real space at $k_{\rm max} = 0.1 ~h/{\rm Mpc}$.
    Different colors in the trace plot represent different MCMC chains.
    }
    \label{fig:trace_AC}
\end{figure}

\begin{table}[t]
    \centering
    \begin{tabular}{c|c|c}
        Parameters & Effective sample size & $\hat{R}$  \\ \hline
        $A$ & 1548 & 1.004\\ \hline
        $\omega_{\rm cdm}$ & 12931 & 1.0003 \\ \hline
        $H_0$ & 73834 & 1.0002\\  \hline
        $b_1$ & 1415 & 1.004 \\ \hline
        $b_2$ & 2063& 1.003  \\ \hline
        $b_{\mathcal G_2}$ & 5225& 1.001  \\ \hline
        $c_0$ & 3432& 1.003 \\ \hline
        $\log(P_{\rm err})$ & 1617& 1.007 \\ \hline
    \end{tabular}
    \caption{
    Effective sample sizes and Gelman-Rubin statistic $\hat{R}$ for the LPT-based model with the quadratic bias in real space at $k_{\rm max} = 0.1~h/{\rm Mpc}$,
    estimated from the eight MCMC chains with 200,000 samples each (see also Fig~\ref{fig:trace_AC}).
    }
    \label{tab:ESS}
\end{table}

To assess whether our MCMC chains efficiently explore the joint posterior, we here track several diagnostics.
As a representative example, we show the case of the LPT-based model with the quadratic bias in real space with $k_{\rm max}=0.1~h/{\rm Mpc}$.
Fig.~\ref{fig:trace_AC} presents the trace plots and the auto-correlation lengths for each parameter, 
and Table~\ref{tab:ESS} summarizes their estimated effective sample sizes.

Although the trace plots indicate that the chains are generally well mixed, 
it also suggests that some parameters, such as $A$, $b_1$ and ${\rm log}(P_{\rm err})$, exhibit slow mixing.
In fact, their auto-correlation lengths are larger and the effective sample sizes are correspondingly smaller than those of the other parameters.
In contrast, $\omega_{\rm cdm}$ and $H_0$ exhibit rapid mixing, 
probably because these parameters are well constrained by the shape of the linear power spectrum and thus less degenerate with the other parameters.
We also find that the sampling efficiency improves when we increase $k_{\rm max}$ or work in redshift space, 
where the parameter degeneracies are more easily broken.
In other words, this example corresponds to the lowest sampling efficiency among the cases we consider in this paper.

\section{Full 2D marginalized posteriors}
\label{app:full_posterior}
Here we present the full 2D marginalized posteriors for the LPT-based models with the quadratic bias as a reference.

\begin{figure}[t]
    \centering
    {$k_\text{max} = 0.1~h/\text{Mpc}$\par\vspace{1ex}}
    \begin{subfigure}{0.49\textwidth}
        \centering
        \includegraphics[width=\textwidth]{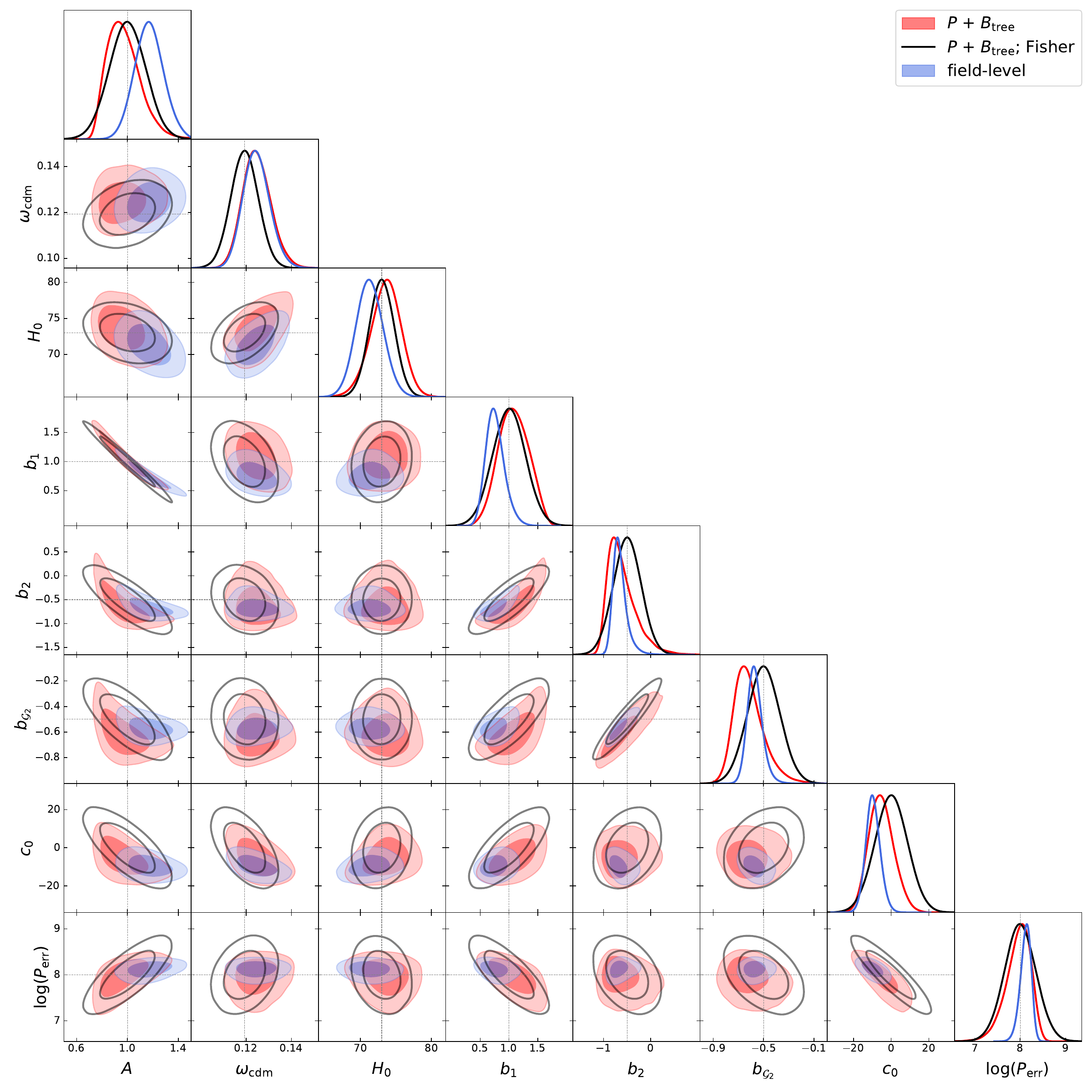}
    \end{subfigure}
    \begin{subfigure}{0.49\textwidth}
        \centering
        \includegraphics[width=\textwidth]{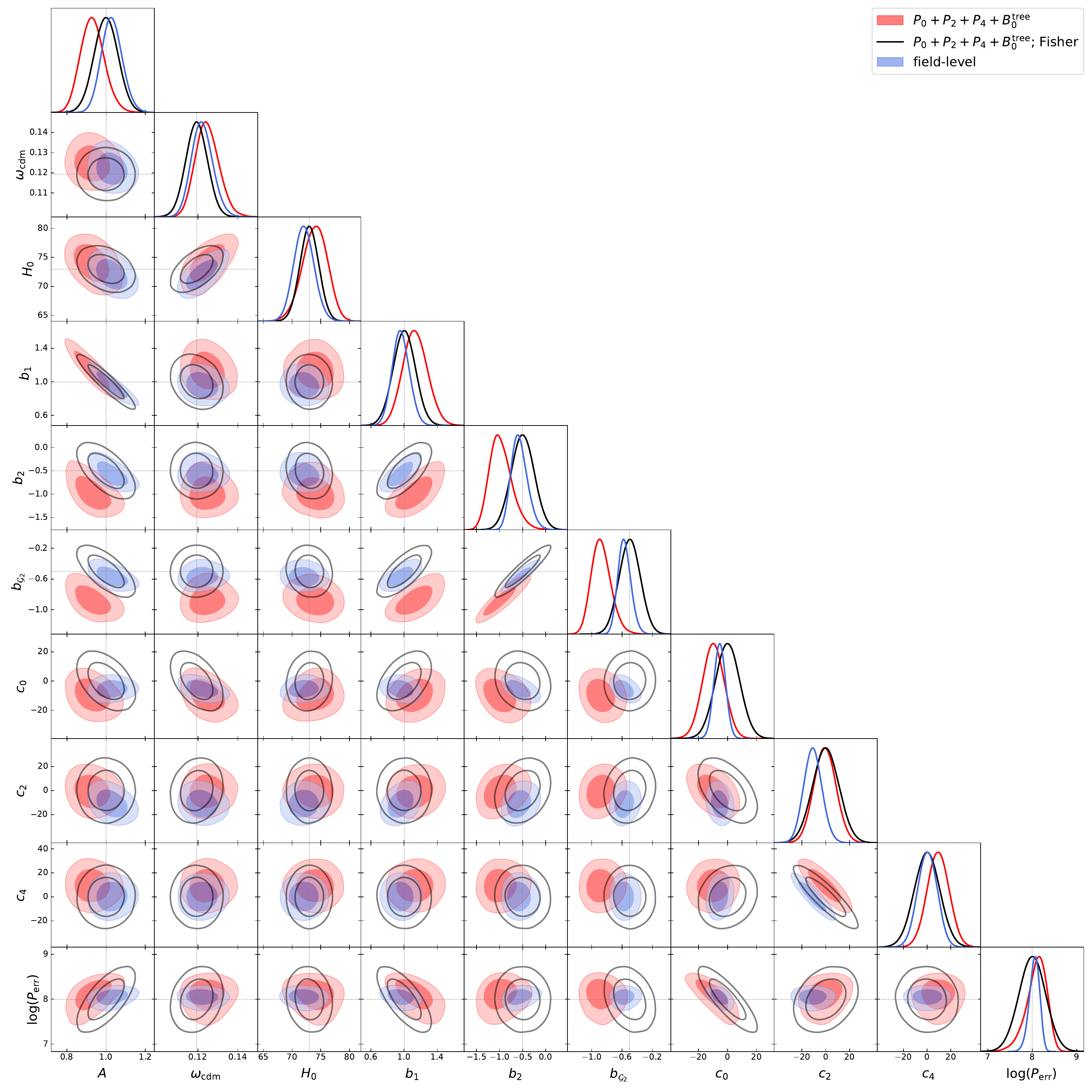}
    \end{subfigure} 
    \caption{The left and the right panels show the 2D marginalized posterior distributions with $k_{\rm max} = 0.1~h/{\rm Mpc}$ for the LPT-based model with the quadratic bias in real and redshift space, respectively.
    The constraints from the joint power spectrum and tree-level bispectrum are shown in red, and the field-level in blue.
    The black lines indicate the inverse Fisher matrix estimate for the error contours for the joint power spectrum and the bispectrum analysis.
    }
    \label{fig:LPT_quad_all_0.1}
\end{figure}

\begin{figure}[t]
    \centering
    {$k_\text{max} = 0.12~h/\text{Mpc}$\par\vspace{1ex}}
    \begin{subfigure}{0.49\textwidth}
        \centering
        \includegraphics[width=\textwidth]{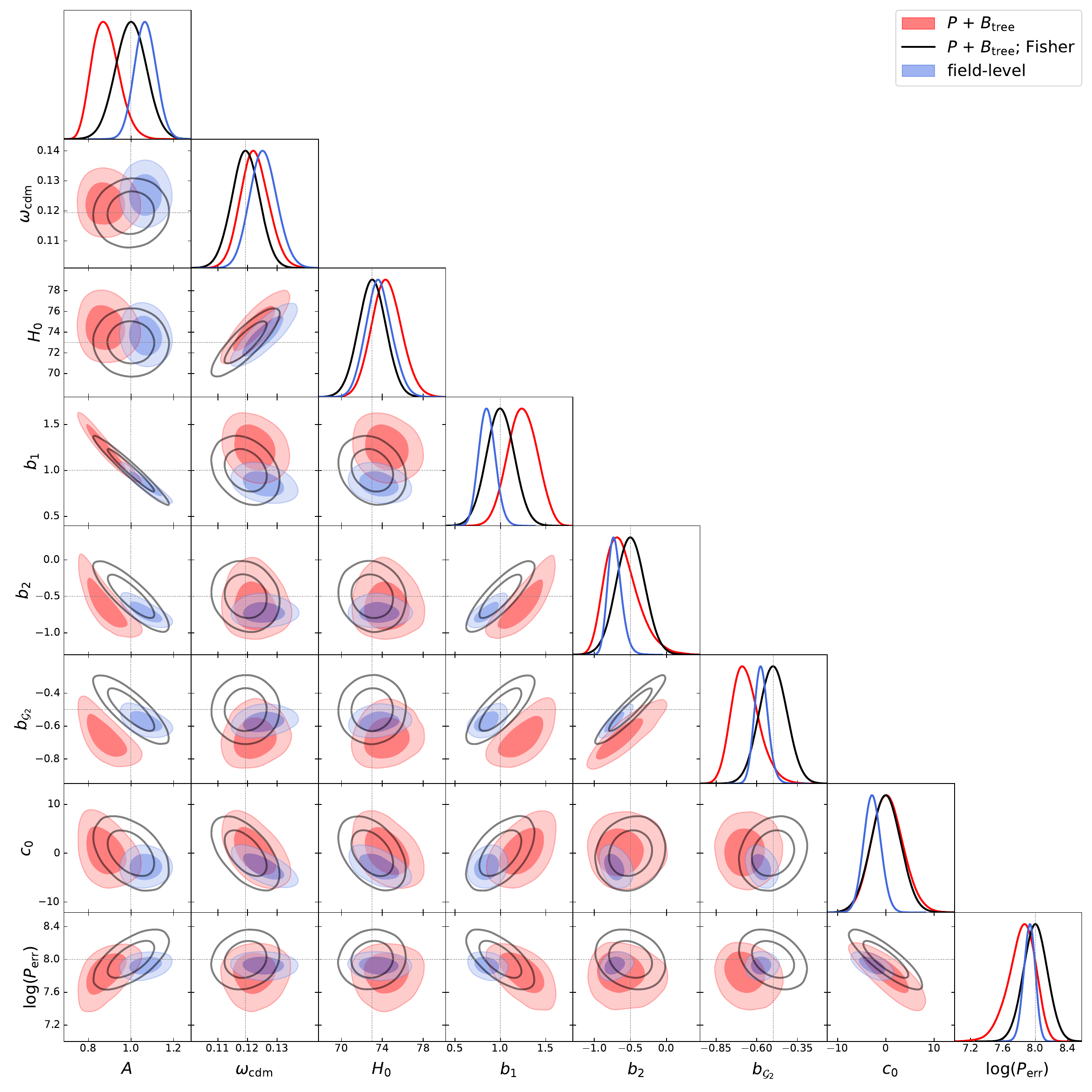}
    \end{subfigure}
    \begin{subfigure}{0.49\textwidth}
        \centering
        \includegraphics[width=\textwidth]{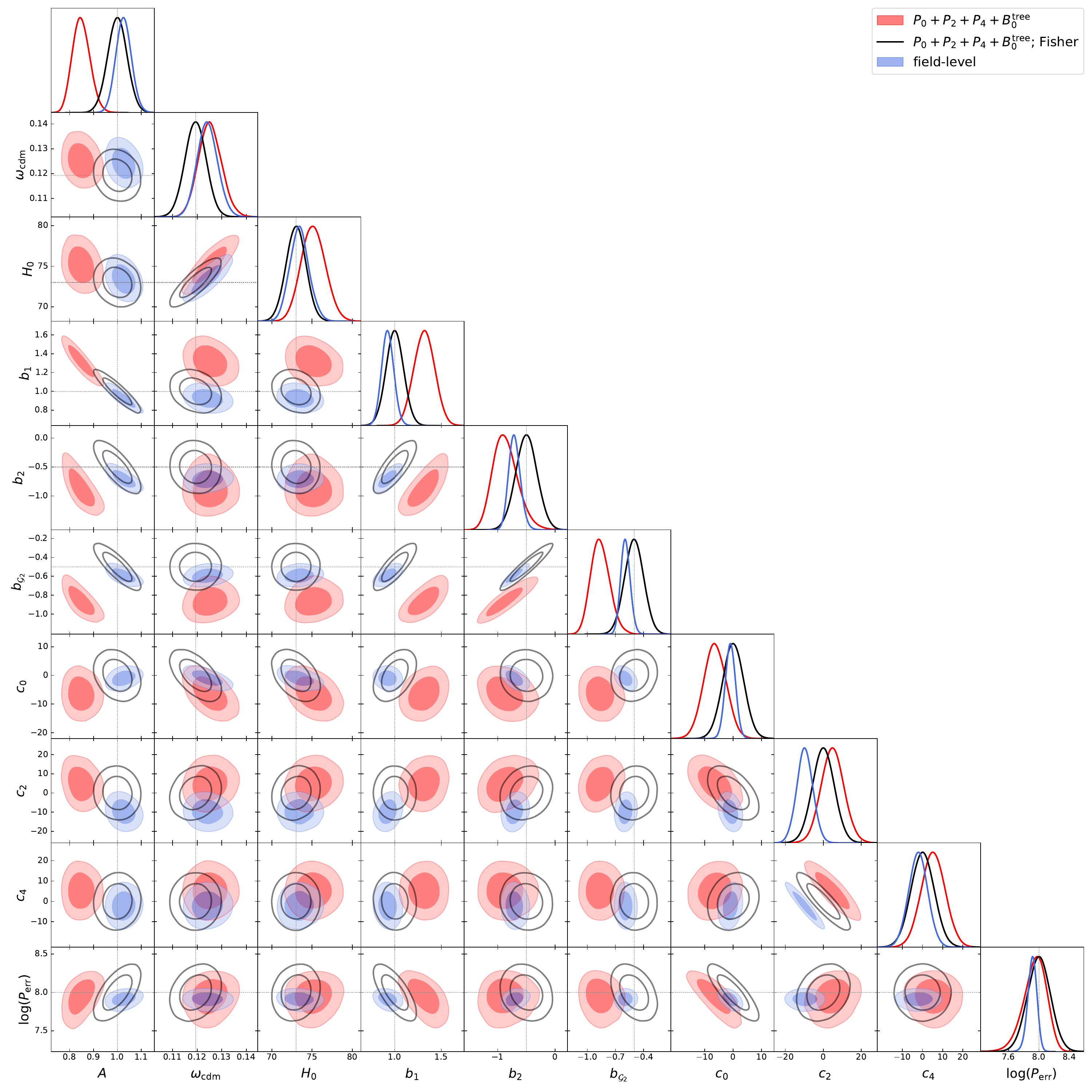}
    \end{subfigure} 
    \caption{The same as Fig.~\ref{fig:LPT_quad_all_0.1}, but with $k_{\rm max} = 0.12~h/{\rm Mpc}$.
    }
    \label{fig:LPT_quad_all_0.12}
\end{figure}

\bibliography{references}
\end{document}